\documentclass[12pt]{article}
\usepackage{graphicx}
\usepackage{amssymb}
\def\lsim{\mathrel{\raise.3ex\hbox{$<$\kern-.75em\lower1ex\hbox{$\sim$}}}}
\def\gsim{\mathrel{\raise.3ex\hbox{$>$\kern-.75em\lower1ex\hbox{$\sim$}}}}
\def\yd{$y_{\rm D}$~} 
\def\Nnu{$N_{\nu}$~}

\newcommand{\neb}{{\bar{\nu}_e}}
\newcommand{\tneb}{{T_{\bar{\nu}_e}}}
\newcommand{\tnx}{{T_{\nu_x}}}
\newcommand{\ene}{{\vev{E_{\nu_e}}}}
\newcommand{\eneb}{{\vev{E_{\bar{\nu}_e}}}}
\newcommand{\enx}{{\vev{E_{\nu_x}}}}
\newcommand{\lne}{{L_{\nu_e}}}
\newcommand{\lneb}{{L_{\bar{\nu}_e}}}
\newcommand{\lnx}{{L_{\nu_x}}}
\newcommand{\bneb}{{\beta_{\bar{\nu}_e}}}
\newcommand{\bnx}{{\beta_{\nu_x}}}
\def\vev#1{{\langle#1\rangle}}
\begin{document}

\title{Progress in the physics of massive neutrinos}
\author{V.~Barger$^{1,4}$, D.~Marfatia$^{2,4}$ and K.~Whisnant$^{3,4}$\\[2ex]
\small\it $^1$Department of Physics, University of Wisconsin, Madison, WI 53706\\
\small\it $^2$Department of Physics, Boston University, Boston, MA 02215\\
\small\it $^3$Department of Physics, Iowa State University, Ames, IA 50011\\
\small\it $^4$Kavli Institute for Theoretical Physics, University of
California, Santa Barbara, CA 93106}

\date{}

\maketitle

\begin{abstract}
The current status of the physics of massive
neutrinos is reviewed with a forward-looking emphasis. The article
begins with the general phenomenology of neutrino oscillations in vacuum
and matter and documents the experimental evidence for oscillations of
solar, reactor, atmospheric and accelerator neutrinos.  Both active and
sterile oscillation possibilities are considered. The impact of
cosmology (BBN, CMB, leptogenesis) and astrophysics (supernovae, highest 
energy cosmic rays) on
neutrino observables and vice versa, is evaluated.  The
predictions of grand unified, radiative and other models of neutrino mass are
discussed. Ways of determining the unknown
parameters of three-neutrino oscillations are assessed, taking into
account eight-fold degeneracies in parameters that yield the same
oscillation probabilities, as well as ways to determine the absolute
neutrino mass scale (from beta-decay, neutrinoless double-beta decay,
large scale structure and $Z$-bursts). Critical unknowns at present are the
amplitude of $\nu_\mu \to \nu_e$ oscillations and the hierarchy of the
neutrino mass spectrum; the detection of $CP$ violation in the
neutrino sector depends on these and on an unknown phase.  The estimated neutrino parameter sensitivities at future
facilities (reactors, superbeams, neutrino factories)
are given. The overall agenda of a future neutrino physics program to
construct a bottom-up understanding of the lepton sector is presented.

\end{abstract}

\newpage

\makeatletter
\def\numberline#1{\hbox to\@tempdima{\hfil#1}. }
\makeatother

\tableofcontents

\newpage
\section{Introduction}

After decades of immense experimental and theoretical effort, major
breakthroughs in our understanding of the properties of neutrinos have
occurred recently. Since the time that the neutrino was first proposed
by Pauli~\cite{pauli} and placed on a concrete theoretical foundation
by Fermi~\cite{fermi}, the question of whether neutrinos are massless
or massive has persisted\footnote{For a historical perspective and discussions
of early work on neutrinos, see Refs.~\cite{books, kayser}.}. The
standard technique of probing neutrino masses through studies of the
endpoints of decay spectra succeeded only in placing increasingly more
restrictive upper limits on neutrino masses, presently down to about
2~eV in the case of neutrinos emitted in the beta decays of
tritium~\cite{troitsk, mainz}.

A more sensitive measure of small neutrino masses was known since 1968
from the proposal of Gribov and Pontecorvo that a
neutrino of a given initial flavor could interchange its identity with
other flavors~\cite{mns}, with a probability that is dependent on 
the distance from the location of the source~\cite{gribov}. 
Early searches for evidence of neutrino
oscillations found only upper bounds on the oscillation
probabilities. A long-standing deficit of solar neutrinos in the
$^{37}$Cl radiochemical experiment of Davis~\cite{chlorine}, compared
to Standard Solar Model (SSM) expectations~\cite{ssmold,ssmnew}, was
attributed to oscillations, but this interpretation remained unproven
until recently. The observation of an electron to muon ratio in the
atmospheric neutrino events in the Kamiokande~\cite{kam} and
IMB~\cite{imb} experiments of about a factor of 2 above expectations
was interpreted as evidence for oscillations with neutrino
mass-squared difference $\delta m^2 \sim 10^{-2}\rm\,eV^2$ and near
maximal neutrino mixing~\cite{learned}. However, due to
the prevailing theoretical prejudice that neutrino mixings would be
small, based on the fact that quark mixings are small, this
interpretation of the atmospheric neutrino data did not receive
widespread acceptance.

The definitive evidence of atmospheric neutrino oscillations came ten
years later from the Super-Kamiokande (SuperK)
experiment~\cite{superK-atmos}. With the ability to measure the zenith
angular and energy distributions of both electron and muon events, the
SuperK experiment convincingly established that the muon-neutrino flux
has a deficit compared with the calculated flux that increased with
zenith angle (or equivalently the path distance), while the
electron-neutrino flux agreed with no-oscillation expectations. The
accumulation of high statistics data by SuperK eventually excluded
interpretations other than $\nu_\mu$ to $\nu_\tau$ oscillations with
nearly maximal mixing at a mass-squared-difference scale $\delta m^2
\sim 2.0\times10^{-3}\rm\,eV^2$. The energy and angular resolution of
the SuperK experiment are too coarse to allow the first minimum in
$\nu_\mu\to\nu_\mu$ to be resolved. That is the goal in the ongoing
K2K~\cite{k2k} and the forthcoming MINOS~\cite{minos} and
CNGS~\cite{icarus, opera} accelerator experiments.

The evidence for solar neutrino oscillations continued to build as
experiments with different detectors and energy sensitivity all found
deficits of 1/3 to 1/2 in rates compared to the SSM. The $^{71}$Ga
radiochemical experiments of SAGE~\cite{sage}, GALLEX~\cite{gallex}
and GNO~\cite{gno} found deficits in the flux of $pp$ neutrinos, which
is the dominant product of the $pp$ reaction chain that powers the
Sun{\footnote{The $pp$ neutrino flux is now determined experimentally
to $\pm2$\% and is in agreement with the SSM predictions to 1\% (the
theoretical uncertainty is also 1\%)~\cite{be7nogood}, thus confirming
the essential ingredients of the SSM.}}.  The SuperK water Cherenkov
detector~\cite{superK-solar} accurately measured the electron energy
spectrum from high-energy solar neutrinos (${\gsim}$5~MeV) originating
from $^8$B decays in the Sun and found it to be
flat~\cite{superK-flat,superK-spectrum} with respect to the SSM prediction. A
crucial theoretical aspect in the oscillations of solar neutrinos is
coherent forward $\nu_e e \to \nu_e e$ scattering, first discussed by
Wolfenstein~\cite{wolfenstein}, which affects electron neutrinos as
they propagate through the dense solar core. Matter effects can
produce large changes in the oscillation amplitude and wavelength
compared to vacuum oscillations, as first shown by Barger, Whisnant,
Pakvasa and Phillips~\cite{bppw}. A resonance
enhancement{\footnote{BWPP discovered this enhancement in a medium of
constant density by varying the neutrino energy; the matter effect
depends on the product of the neutrino energy and the electron
density. Mikheyev and Smirnov applied the enhancement at a given
neutrino energy to the propagation of solar neutrinos through the
varying electron density in the Sun.}} can be realized in matter for
one sign of $\delta m^2$. Because of the prevailing prejudice that
neutrino mixing would be small, there was a strong theoretical bias in
favor of a resonant solar solution, which was the original solution to
the solar neutrino problem proposed by Mikheyev and Smirnov (the
so-called  Mikheyev-Smirnov-Wolfenstein or MSW solution)~\cite{msw}. 
In addition to this small mixing
angle solution (known as SMA), other solutions with a large vacuum
mixing angle were later identified that could account for the solar
neutrino flux suppression~\cite{hata}. 

\begin{figure}[h]
\centering\leavevmode
\includegraphics[width=3in]{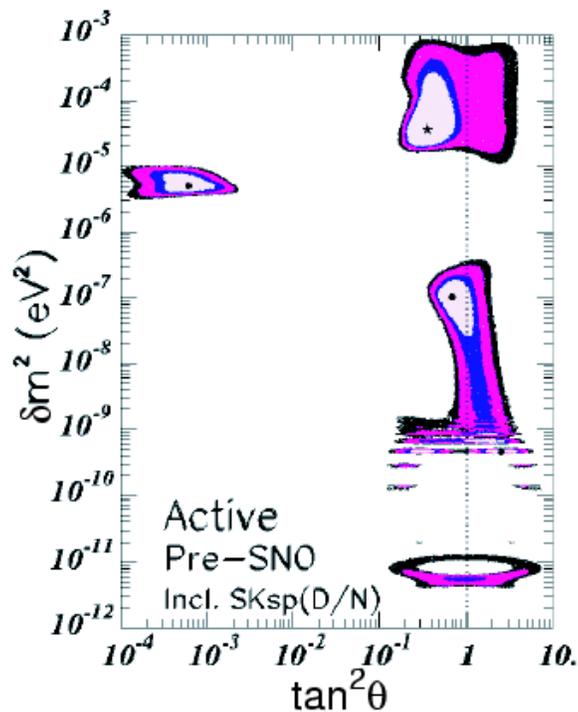}
\caption[]{90\%, 95\%, 99\% and 3$\sigma$ C.~L. allowed regions
 in the $(\delta m^2, \tan^2\theta)$
oscillation parameter space, before any SNO data. 
From Ref.~\cite{bahcall-concha}.
\label{fig:islands}}
\end{figure}

\begin{figure}[h]
\centering\leavevmode
\includegraphics[width=3in]{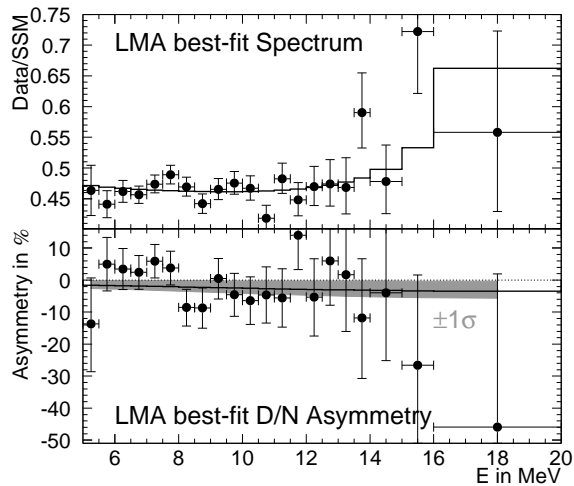}
\caption[]{The energy spectrum and the day-night asymmetry (day rate $-$ 
night rate divided by the average rate) as measured by SuperK. The solid lines
are the predictions for $\delta m^2_s= 6.3\times 10^{-5}$ eV$^2$ and 
$\tan^2\theta_s=0.55$.
From Ref.~\cite{superK-spectrum}.
\label{fig:superKflat}}
\end{figure}

The other solutions were named LMA (large mixing angle), LOW (low
$\delta m^2$, low probability)~\cite{krastev-petcov}, 
QVO (quasi-vacuum oscillations)~\cite{friedland} and 
VO (vacuum oscillations)~\cite{bpw-solar}. These islands in the
$(\delta m^2, \tan^2\theta)$ in the solar neutrino oscillation 
parameter space are illustrated in Fig.~\ref{fig:islands}. The
flat energy spectrum relative to the SSM and the absence of a
significant day/night difference (that can be caused by Earth-matter
effects), measured by SuperK (see Fig.~\ref{fig:superKflat}), 
led to a strong preference for the LMA solution. 
Subsequently, the Sudbury Neutrino Observatory (SNO)~\cite{sno} 
 measured the neutrino neutral currents as
well as the charged currents, and confirmed the oscillations independent
of the SSM normalization of the $^8$B flux. The recent SNO salt phase 
data~\cite{salt} in conjunction
with other solar neutrino data selects the
LMA solution uniquely at more than the 3$\sigma$ level. The
mass-squared difference indicated by the solar neutrino data is
$\sim 6 \times 10^{-5}\rm\,eV^2$ and the mixing is large but not maximal,
$\tan^2\theta\sim0.4$; see Section~4.

The definitive confirmation of the LMA solar solution has 
come from the KamLAND experiment~\cite{kamland1}. If
$CPT$ invariance is assumed, the probabilities of $\nu_e\to\nu_e$ and
$\bar\nu_e\to\bar\nu_e$ oscillations should be equal at the same
values of $L/E$. At the average distance $L\sim180$~km of the reactors
from the KamLAND detector and the typical energies of a few MeV of the
reactor $\bar\nu_e$, the experiment has near optimal sensitivity to
the $\delta m^2$ value of the LMA solar solution. The first year of
data from KamLAND~\cite{kamland2} shows the rate suppression of
$P(\bar\nu_e \to\bar\nu_e)$ expected from the solar LMA solution,
vindicating the oscillation interpretation of the solar neutrino
problem. The solar $\nu_e$ are oscillating to a combination of
$\nu_\mu$ and $\nu_\tau$.
 
With three neutrinos ($\nu_e, \nu_\mu, \nu_\tau$) there are two
distinct $\delta m^2$ that can account for the atmospheric and solar
neutrino oscillations. However, the LSND accelerator
experiment~\cite{lsnd-dar} found evidence at 3.3$\sigma$ significance
for $\bar\nu_\mu \to \bar\nu_e$ oscillations at a higher $\delta m^2
\sim 0.2$ to 1~eV$^2$ with small mixing, $\sin^22\theta\sim 0.003$ to
0.04. Evidence for oscillations of $\nu_\mu \to \nu_e$ was also
observed, although at lesser significance~\cite{lsnd-dif}.  Such
oscillations are constrained but not excluded by the KARMEN
experiment~\cite{karmen}. A possible way to explain the LSND effect is
to invoke oscillations involving a fourth, sterile neutrino with no Standard
Model
interactions~\cite{caldwell}. Two possible scenarios are
considered, with 2~+~2 and 3~+~1 spectra of $\delta m^2$~\cite{okada,
bpww, bklww,peres}, where the large mass-squared
difference of the LSND oscillations connects the upper and lower
levels. However, there is a tension in the global fits between the
solar and atmospheric data on one hand and the reactor and accelerator
data on the other, with the conclusion that both schemes are 
highly disfavored~\cite{tortola}. The MiniBooNE~\cite{miniboone}
experiment, now in progress at Fermilab, is designed to test the
$\nu_\mu \to \nu_e$ oscillation hypothesis for the LSND effect. The
source of the antineutrino beam at LSND is muon decays, while at
MiniBooNE the source is pion decays. An alternative nonoscillation
interpretation of the LSND effect as a nonstandard muon
decay~\cite{babu-pakvasa} will not be tested by MiniBooNE. Since the
LSND effect is fragile, this review emphasizes three neutrino
oscillation phenomena.

The phenomenology of three neutrino mixing, like that of three
quarks~\cite{km}, involves three mixing angles ($\theta_{23} \equiv
\theta_a,\ \theta_{12} \equiv \theta_s,\ \theta_{13} \equiv
\theta_x$), two mass-squared differences ($\delta m_{31}^2 \equiv
\delta m_a^2,\ \delta m_{21}^2 \equiv \delta m_s^2$), and one
$CP$-violating phase ($\delta$)~\cite{mns, cabibbo, bpw-three, bpw-cp,pdg}. 
In order that $\delta$ be measurable, both $\delta m^2$ 
scales must contribute to the oscillation~\cite{bpw-three}. 
If neutrinos are Majorana~\cite{majorana}, 
two further $CP$-violating phases ($\phi_2,\ \phi_3$)
enter in the calculation of neutrinoless 
double-beta decay~\cite{wolfen} but not oscillations~\cite{doi}. 
We already have approximate
knowledge of $\theta_a, \theta_s, \delta m_a^2$ (but not the sign of
$\delta m_a^2$) and $\delta m_s^2$ (its sign is known from solar matter
effects). The major challenge before us now is the measurement of
$\theta_x$, for which we have only an upper bound, $\sin^22\theta_x
\lesssim 0.2$ (for $\delta m^2_a=2.0\times 10^{-3}$ eV$^2$) at the 95\% C.~L. 
from the CHOOZ~\cite{chooz} and Palo
Verde~\cite{paloverde} reactor experiments.

The amplitude of the oscillations $\nu_\mu\to\nu_e$ and
$\nu_e\to\nu_\tau$ are governed by the size of
$\theta_x$. Long-baseline and reactor experiments are planned with 
improved sensitivity to $\sin^22\theta_x$~\cite{jhfsk}; see Section~10. The
$CP$-violating phase enters oscillations via a factor $\sin\theta_x
e^{-i\delta}$ and thus it is essential to establish a nonzero value of
$\theta_x$ in order to pursue the measurement of $\delta$. The size of
$CP$ violation in long-baseline experiments also depends on the value
of $\delta m^2_s$. Oscillation parameter ambiguities exist~\cite{bmw}
that must be resolved by the experiments.

The anticipated steps in the long-baseline program are off-axis
beams~\cite{jhfsk,bnl-e889, para}, superbeams~\cite{richter,
superbeams}, detectors with larger fiducial volumes and sophistication
~\cite{jhfsk,uno,ar}, 
and neutrino
factories~\cite{geer, autin}. The
ultimate sensitivities will be derived from neutrino factories, where
the neutrino beams are obtained from the decays of muons that are stored
in a ring with straight sections.

Although oscillations have allowed us to establish that neutrinos have
mass, they do not probe the absolute neutrino mass scale. In particle
and nuclear physics, the only avenues for this purpose are tritium beta
decay and neutrinoless double-beta decay, and the latter only if
neutrinos are Majorana particles. Experiments are
beginning to probe the interesting region of $m_\nu$. Another route to
the absolute mass is the power spectrum of galaxies, which gets
modified on small scales when the sum of neutrino masses is
nonzero~\cite{power}. The WMAP analysis~\cite{wmap} of the cosmic
microwave background (CMB) and large-scale structure data have already
given an upper limit on $\sum m_\nu$ of about 1~eV.

The field of neutrino physics is progressing at a rapid rate. The
purpose of this review is to summarize the present status of the field
and to discuss ways that progress can be made in experimentally
answering the outstanding questions. For other recent reviews see 
Refs.~\cite{reviews,barr-dorsner, altarelli-feruglio,
 mohapatra, mcchen}.
We have provided an extensive bibliography, but due to the large number of papers on
the subject, it may not be comprehensive. 

\section{Neutrino counting: $Z$-decays, CMB and BBN}

Studies of $e^+e^-$ annihilation at the $Z$-resonance pole at the
Large Electron Positron collider have determined 
the invisible width of the $Z$ boson.  
The experimental value $N_\nu = 2.984 \pm 0.008$
is close to the number expected from 3 active light neutrinos, though
the value is 2$\sigma$ low~\cite{pdg}.

The cosmic microwave background (CMB) anisotropies and Big Bang Nucleosynthesis (BBN) probe the effective number of neutrinos $(N_\nu = 3 + \Delta N_\nu)$ that were present in the early universe. The extra relativistic energy density due to sterile neutrinos, or other possible light particles, is normalized to that of an equivalent neutrino flavor as~\cite{ssg}
\begin{equation}
\rho_x \equiv \Delta N_\nu \rho_\nu = {7\over8}\Delta N_\nu \rho_\gamma \,,
\end{equation}
where $\rho_\gamma$ is the energy density in photons. Sterile neutrinos would contribute to $\Delta N_\nu$, but so could other new physics sources.

\begin{figure}[ht]
\centering\leavevmode
\includegraphics[width=3.75in]{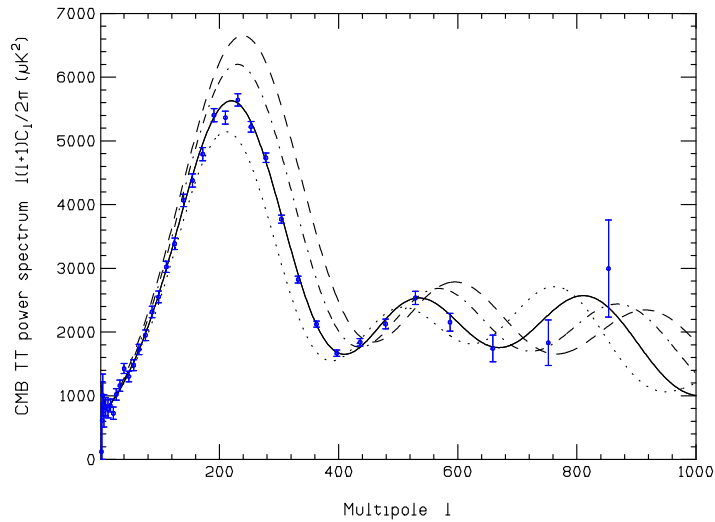}
\caption[]{The power spectrum for the best-fit ($N_\nu=2.75$) to 
the WMAP data~\cite{wmap} is the solid line. With all other parameters
and the overall normalization of the primordial spectrum fixed, 
the spectra for $N_\nu=1$, $N_\nu=5$ and $N_\nu=7$ are the dotted,
dot-dashed and dashed lines, respectively. The data points represent
the binned TT power spectrum from WMAP. From Ref.~\cite{nupap}. 
\label{fig:wmapNnu}}
\end{figure}

\begin{figure}[ht]
\centering\leavevmode
\includegraphics[width=3.25in]{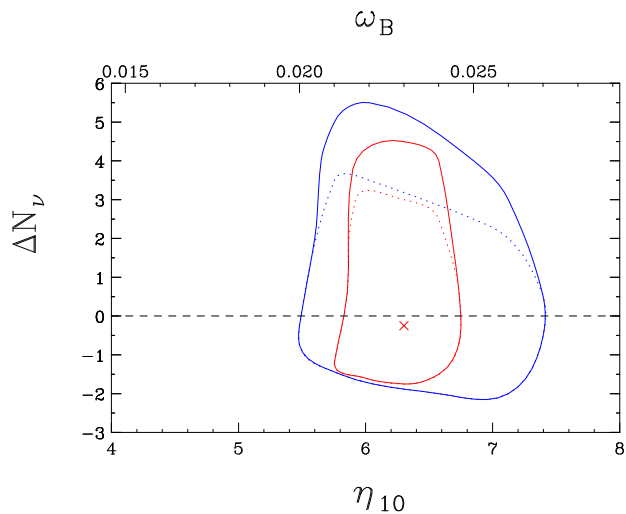}
\caption[]{The $1\sigma$ and $2\sigma$ contours in the 
$\eta_{10}-\Delta N_\nu$ plane from an analysis~\cite{nupap} of WMAP
data. The solid (dotted) lines correspond to $t_0 \ge 11$ (12) Gyr. The
cross marks the best-fit at $\omega_{\rm B}=0.023$ and $\Delta N_\nu=-0.25$.
\label{fig:DNnu-constraints}}
\end{figure}

The precise WMAP measurements~\cite{wmap} of the CMB have been
analyzed to constrain $\Delta N_\nu$~\cite{nupap,others}. In
Ref.~\cite{nupap}, a flat universe with a
cosmological constant $\Lambda$ as dark energy is assumed. 
The parameters varied are the
reduced Hubble constant $h$, the baryon density $\omega_B = \Omega_B
h^2$, the total matter density $\omega_M = \Omega_M h^2$ (comprised of
baryons and cold dark matter), the optical depth $\tau$, the spectral
index $n_s$ and amplitude $A_s$ of the primordial power spectrum
$\left( P = A_s (k/k_*)^{n_s-1}\right)$ and $\Delta N_\nu$. The HST
measurement of $h=0.72\pm0.08$~\cite{HST} is imposed as a top-hat
distribution. This strong $h$ prior helps to break the degeneracy
between $\omega_M$ and $\Delta N_\nu$. The universe is also required
to be older than 11~Gyr, as inferred from globular
clusters~\cite{chaboyer}.  The effects of $N_\nu$ on the CMB are
illustrated in Fig.~\ref{fig:wmapNnu}, for a fixed normalization of
the power spectrum.  The resulting $\Delta N_\nu$ constraints are
shown in Fig.~\ref{fig:DNnu-constraints}, where $\eta_{10} \equiv
10^{10} n_B/n_\gamma = 274\omega_B$. The best-fit is $\Delta N_\nu
= -0.25$, but at $2\sigma$, $\Delta N_\nu \leq 5.3$ is allowed; see Table
\ref{tab1}~\cite{nupap}.

\begin{table}[t] 
\caption{The 2$\sigma$ ranges (for 1 degree of freedom) of \Nnu and 
$\eta_{10}$ from analyses~\cite{nupap} of WMAP data, deuterium and
helium abundances and their combinations. The WMAP analysis involves
the assumption of a flat universe, along with the strong HST prior on
$h$ and the age constraint $t_0\ge 11$ Gyr. For BBN the adopted
primordial abundances are: \yd $\equiv 10^{5}$(D/H)$=2.6 \pm
0.4$~\cite{k}, Y = $0.238 \pm 0.005$~\cite{osw}.} 
\label{tab1} 
\begin{center} 
\begin{tabular}{|l|c|c|} 
\hline 
  &\Nnu (2$\sigma$ range)  & $\eta_{10}$ (2$\sigma$ range) \\ \hline 
  WMAP& 0.9 -- 8.3 & 5.58 -- 7.26\\ \hline 
  \yd + Y & 1.7 -- 3.0 & 4.84 -- 7.11\\ \hline 
  WMAP + \yd + Y&1.7 -- 3.0 & 5.53 -- 6.76\\ \hline
\end{tabular} 
\end{center} 
\end{table} 

BBN is a much better probe of $\Delta N_\nu$ than the CMB. The
prediction of the primordial abundance of $^4$He depends sensitively
on the early expansion rate, while the prediction of the D abundance
is most sensitive to the baryon density
$\eta_{10}$~\cite{gary}.
The best-fit value of $\Delta N_\nu$
from BBN is $\Delta N_\nu = -0.7$ but at $2\sigma$, \Nnu$=3$ is
allowed. The D-inferred baryon density is in excellent agreement with
the baryon density determined from the CMB and Large Scale Structure.

\begin{figure}[h]
\centering\leavevmode
\includegraphics[width=3.25in]{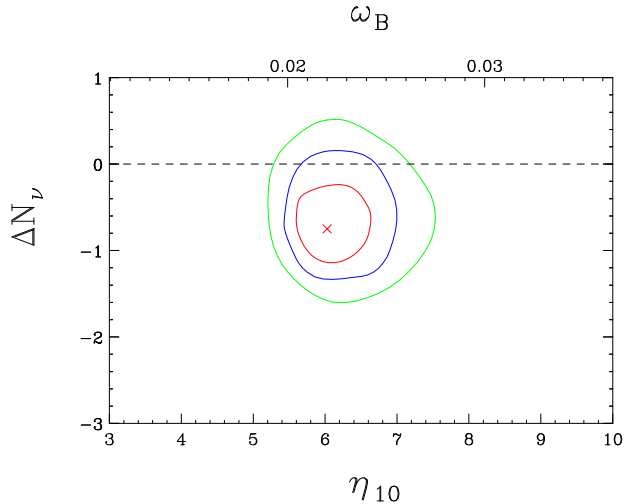}
\caption[]{The $1\sigma$, $2\sigma$ and $3\sigma$ contours in the 
               $\eta_{10}-\Delta N_\nu$ plane from a combination of WMAP data 
               and the adopted D and $^4$He abundances. From Ref.~\cite{nupap}.
\label{fig:comb-allowed}}
\end{figure}

A combined analysis of the BBN and the WMAP data yields the
allowed regions in Fig.~\ref{fig:comb-allowed}. As with the BBN
analysis above, the best-fit value is $\Delta N_\nu = -0.7$ and
\Nnu$=3$ is allowed at 2$\sigma$.

The BBN analysis is consistent with 3 neutrinos, but gives no support to the possible existence of extra neutrinos. Even one extra, fully thermalized neutrino $(\Delta N_\nu = 1)$ is strongly disfavored~\cite{nupap}. We discuss the implications of this result for the sterile neutrino interpretation of the LSND experiment in Section~\ref{sec:LSND}.

\section{Neutrino mixing and oscillations}

\subsection{Vacuum oscillations}

The dramatic increase in our knowledge of neutrino properties
has come from observational evidence of neutrino oscillations.  These
neutrino flavor changes require that the neutrino flavor states,
$\nu_\alpha$ are not the same as the neutrino mass eigenstates, $\nu_i$.
The eigenstates are related by a unitary matrix $V$~\cite{mns},
\begin{equation}
   \nu_\alpha = \sum V_{\alpha i}^* \nu_i \,.
\end{equation}
$V$ is often denoted as $V_{MNS}$, where MNS
 represents the authors of Ref.~\cite{mns}\footnote{
Sometimes it is denoted as $V_{PMNS}$ or $V_{MNSP}$ to acknowledge the
contributions of Pontecorvo.}.
For 3 neutrinos, the mixing matrix $V$ is specified by three rotation
angles $\theta_a$, $\theta_x$, $\theta_s$ ($0\leq \theta_i \leq \pi/2$)
and three $CP$-violating phases
$\delta$, $\phi_2$ and $\phi_3$ 
($0\leq \delta, \phi_i \leq 2\pi$).  $V$ can be conveniently written as the
matrix product
\begin{equation}
V =  \left[ \begin{array}{ccc}1&0&0\\ 0& c_a & s_a\\ 0 & -s_a & c_a
        \end{array}\right]
       \left[\begin{array}{ccc}c_x & 0 & s_x e^{-i\delta}\\ 0 & 1 & 0\\
                                              -s_xe^{i\delta} & 0 & c_x
        \end{array}\right]
        \left[\begin{array}{ccc}c_s & s_s & 0\\ -s_s & c_s & 0\\ 0 & 0 & 1
        \end{array}\right]
       \left[\begin{array}{ccc}1 & 0 & 0\\ 0 & e^{i\left({1\over2}\phi_2\right)} & 0\\
                                               0 & 0 & e^{i\left({1\over2}\phi_3 + \delta\right)}
      \end{array}\right]
\label{eq:mixmat}
\end{equation}
where $c_i$ denotes $\cos\theta_i$ and $s_i$ denotes  
$\sin\theta_i$.
The angle $\theta_a$, customarily denoted as $\theta_{23}$, governs
the oscillations of atmospheric neutrinos, the angle $\theta_s$
($\theta_{12}$) describes solar neutrino oscillations, and the angle
$\theta_x$ ($\theta_{13}$) is an unknown angle that is bounded by
reactor neutrino experiments at short distances ($L \simeq 1$~km).  
 The
oscillation probabilities are independent of the Majorana phases
$\phi_2$ and $\phi_3$.  Vacuum neutrino oscillations are given by
\begin{equation}
P( \nu_\alpha \to\nu_\beta) =
\left| \sum_j V_{\beta j} e^{-{im_j^2L\over 2E_\nu}} V^*_{\alpha j} \right|^2 \,,
\end{equation}
where the $m_j$ are the neutrino eigenmasses.  The oscillation
probabilities depend only on differences of squared neutrino masses.
The oscillation arguments for the atmospheric and solar phenomena are
\begin{equation}
\Delta_a \equiv  {\delta m_a^2 L\over 4E_\nu} \,,  \qquad   
\Delta_s \equiv  {\delta m_s^2 L\over 4E_\nu} \,,
\label{eq:delta}
\end{equation}
respectively, where
\begin{equation}
\delta m_a^2 = m_3^2 - m_1^2 \,, \qquad \delta m_s^2 = m_2^2 - m_1^2 \,.
\end{equation}
The vacuum oscillation probabilities are
\begin{eqnarray}
P(\nu_e\to\nu_e) &=& 1 - \sin^22\theta_x \sin^2\Delta_a
- (c_x^4\sin^22\theta_s + s_s^2\sin^22\theta_x)\sin^2\Delta_s
\nonumber\\
&&+s_s^2\sin^22\theta_x({1\over2}\sin2\Delta_s\sin2\Delta_a +
2\sin^2\Delta_a\sin^2\Delta_s) \,,
\label{eq:Pee}
\\
P(\nu_e\to\nu_\mu) &=& s_a^2\sin^22\theta_x\sin^2\Delta_a
+4 J (\sin2\Delta_s\sin^2\Delta_a - \sin2\Delta_a\sin^2\Delta_s)
\nonumber\\
&&-(s_a^2 s_s^2 \sin^22\theta_x - 4K)
[{1\over2}\sin2\Delta_s\sin2\Delta_a + 2\sin^2\Delta_s\sin^2\Delta_a]
\nonumber\\
&&+ [c_x^2(c_a^2-s_x^2 s_a^2)\sin^22\theta_s + s_a^2 s_s^2\sin^22\theta_x
- 8 K s_a^2]\sin^2\Delta_s \,,
\label{eq:Pem}
\\
P(\nu_\mu\to\nu_\mu) &=& 1 - (c_x^4\sin^22\theta_a+s_a^2\sin^22\theta_x)
\sin^2\Delta_a
\nonumber\\
&&+[c_x^2(c_s^2-s_x^2s_a^2)\sin^22\theta_a+s_s^2s_a^2\sin^22\theta_x-8Ks_a^2]
\nonumber
\\
&&\quad\times[{1\over2}\sin2\Delta_s\sin2\Delta_a + 2\sin^2\Delta_s\sin^2\Delta_a]
\nonumber\\
&&-[\sin^22\theta_s(c_a^2-s_x^2s_a^2)^2
+ s_x^2\sin^22\theta_a(1-c_\delta^2\sin^22\theta_s)
\nonumber\\
&&\quad +2s_x\sin2\theta_s\cos2\theta_s\sin2\theta_a\cos2\theta_a c_\delta
- 16Ks_a^2s_s^2
\nonumber\\
&&\quad +\sin^22\theta_ac_x^2(c_s^2-s_x^2s_s^2)+s_x^2s_a^2\sin^22\theta_x]
\sin^2\Delta_s
\label{eq:Pmm}
\\
P(\nu_\mu\to\nu_\tau) &=& c_x^4 \sin^22\theta_a \sin^2\Delta_a
+ 4 J (\sin2\Delta_s\sin^2\Delta_a - \sin2\Delta_a\sin^2\Delta_s)
\nonumber\\
&&-[c_x^2\sin^22\theta_a(c_s^2 - s_x^2 s_s^2) + 4K\cos2\theta_a]
\nonumber\\
&&\quad\times[{1\over2}\sin2\Delta_s\sin2\Delta_a + 2\sin^2\Delta_s\sin^2\Delta_a]
\nonumber\\
&&+[\sin^22\theta_a(c_s^2-s_x^2s_s^2)^2
+s_x^2\sin^22\theta_s(1-\sin^22\theta_a c_\delta^2)
+4K\cos2\theta_a\nonumber\\
&&\quad
+s_x\sin2\theta_s\cos2\theta_s\sin2\theta_a\cos2\theta_a(1+s_x^2)c_\delta]
\sin^2\Delta_s \,,
\label{eq:Pmt}
\end{eqnarray}
where the $CP$-violating quantity $J$~\cite{jarlskog} is
\begin{equation}
J = {1\over8}c_x\sin2\theta_x\sin2\theta_a\sin2\theta_s s_\delta \,,
\label{eq:J}
\end{equation}
and for convenience we have defined
\begin{equation}
K = {1\over8}c_x\sin2\theta_x\sin2\theta_a\sin2\theta_s c_\delta \,.
\label{eq:K}
\end{equation}
Oscillation probabilities for other neutrino channels may be obtained
by probability conservation, {\it i.e.}, $\sum_\alpha
P(\nu_\alpha\to\nu_\beta) = \sum_\beta P(\nu_\alpha\to\nu_\beta) =
1$. Probabilities for antineutrino channels are obtained by replacing
$\delta$ by $-\delta$ in the corresponding neutrino formulae. Also,
since $CPT$ is conserved for ordinary neutrino
oscillations, the antineutrino probabilities are given by
$P(\bar\nu_\alpha \to \bar\nu_\beta) = P(\nu_\beta \to \nu_\alpha)$,
and
$P(\nu_\alpha \to \nu_\beta)$ is obtained from $P(\nu_\beta \to
\nu_\alpha)$ by replacing $\delta$ by $-\delta$.

For oscillations of atmospheric and long-baseline neutrinos, the
oscillation argument $\Delta_a$ is dominant, and the vacuum oscillation
probabilities in the leading oscillation approximation (where only the $\delta m_a^2$ oscillations are appreciable) are
\begin{eqnarray}
P(\nu_e\to\nu_e) &\simeq& 1 - \sin^22\theta_x \sin^2\Delta_a
\\
P(\nu_e\to\nu_\mu) &\simeq& s_a^2 \sin^22\theta_x \sin^2\Delta_a
\\
P(\nu_\mu\to\nu_\mu) &\simeq& 1 - (c_x^4 \sin^22\theta_a +
s_a^2 \sin^22\theta_x) \sin^2\Delta_a \,.
\\
P(\nu_\mu\to\nu_\tau) &\simeq& c_x^4 \sin^22\theta_a\sin^2\Delta_a\,.
\end{eqnarray}

For vacuum oscillations of solar neutrinos, where $|\Delta_a| \gg 1$,
the terms involving $\Delta_a$ approach their average values over a
complete cycle and
\begin{equation}
P(\nu_e\to\nu_e) \simeq 1 - {1\over2}\sin^22\theta_x
-c_x^4\sin^22\theta_s\sin^2\Delta_s \,.
\end{equation}
Note that in the limit $\theta_x \to 0$, $\nu_\mu \to \nu_\tau$
oscillations of atmospheric neutrinos and $\nu_e \to \nu_e$
oscillations of solar neutrinos completely decouple, {\it i.e.}, they are
determined by independent parameters, and each has the form of
two-neutrino oscillations.

\subsection{Matter effects on oscillations}

The scattering of $\nu_e$ on electrons in matter can modify the vacuum
oscillation probabilities~\cite{wolfenstein, bppw, langacker}. 
For the two-neutrino case, with mixing angle $\theta$
and mass-squared difference $\delta m^2$, the oscillation probability
amplitude $\sin^22\theta_m$ in matter is
\begin{equation}
\sin^2 2\theta_m = {\sin^22\theta \over
\left( {A\over \delta m^2} - \cos2\theta \right)^2 + \sin^22\theta} \,,
\label{eq:matteramp}
\end{equation}
where
\begin{equation}
A = 2\sqrt2\,G_F\,N_e\,E_\nu = 1.54\times10^{-4}{\rm~eV}^2\, Y_e\,
\rho(\rm{\rm g/cm}^3)\, E_\nu({\rm GeV}) \,,
\end{equation}
and $N_e$ is the electron number density, which is the product of the
electron fraction $Y_e$ and matter density $\rho$. The oscillation
amplitude in matter is enhanced if $\delta m^2 > 0$, and a resonance
occurs ({\it i.e.}, the amplitude reaches its maximal value of unity) at the
critical density $N_e^c = \delta m^2
\cos2\theta/(2\sqrt2\,G_F\,E_\nu)$.  For antineutrinos, $A \to -A$ in
Eq.~(\ref{eq:matteramp}), and the oscillation amplitude in matter is
enhanced if $\delta m^2 < 0$. For $N_e$ much larger than the critical
density, the oscillation amplitude in matter is strongly suppressed
for both neutrinos and antineutrinos. The effective value of $\delta
m^2$, and hence the oscillation wavelength, is also changed in matter:
\begin{equation}
\delta m^2_m = \delta m^2
\sqrt{\left( {A\over \delta m^2} - \cos2\theta \right)^2
+ \sin^22\theta} \,.
\end{equation}

The angle $\theta_m$ represents the mixing between the flavor
eigenstates $\nu_\alpha$ and the instantaneous eigenstates in matter $\nu_{im}$:
\begin{equation}
\left[\begin{array}{c} \nu_{1m}\\ \nu_{2m} \end{array} \right] =
\left[\begin{array}{ccc} \cos\theta_m & -\sin\theta_m \\
                    \sin\theta_m & \cos\theta_m \end{array} \right]
\left[\begin{array}{c} \nu_e\\ \nu_\mu \end{array} \right].
\label{eq:eigen}
\end{equation}
If the electron density is $N_e^0$ when the neutrino is created, the
initial value of $\theta_m$ is given by
\begin{equation}
\cos2\theta_m^0 = - {{A^0\over \delta m^2} - \cos2\theta \over
\sqrt{({A^0\over \delta m^2} - \cos2\theta)^2 + \sin^22\theta}} \,,
\end{equation}
where $A^0 = 2\sqrt2\,G_F\,N_e^0\,E_\nu$.  A neutrino originally
created as a $\nu_e$ can be expressed in terms of the lower and upper
eigenstates as $\nu_e = \cos\theta_m^0\,\nu_{1m} +
\sin\theta_m^0\,\nu_{2m}$. 

\subsection{Solar neutrino oscillations}

For matter of varying density, such as for
neutrinos propagating through the Sun, the instantaneous eigenstates
change as the neutrinos propagate. Once a solar neutrino reaches the
vacuum, the lower eigenstate is $\nu_1 = \cos\theta\, \nu_e -
\sin\theta\, \nu_\mu$ and the upper eigenstate is $\nu_2 =
\sin\theta\, \nu_e + \cos\theta\, \nu_\mu$.

\begin{figure}[t]
\centering\leavevmode
\includegraphics[width=2.5in]{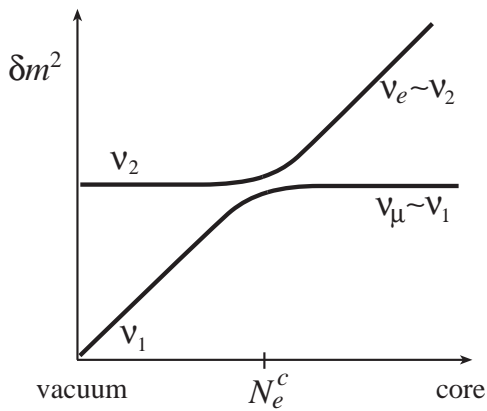}
\caption[]{The level-crossing diagram for solar neutrinos.
\label{fig:levelcross}}
\end{figure}

A $\nu_e$ created far above the resonance density is predominantly in
the upper eigenstate. Then if the neutrino propagation is adiabatic,
the neutrino will remain in the upper eigenstate, and if $\theta$ is
small it will be predominantly $\nu_\mu$ once it reaches the
vacuum~\cite{bethe, haxton} (see Fig.~\ref{fig:levelcross}). For
nonadiabatic propagation, if the probability of jumping from one
eigenstate to another is $P_x$, then averaging over the
oscillations~\cite{parke-walker}
\begin{equation}
\langle P(\nu_e \to \nu_e) \rangle = {1\over2} \left[ 1 +
(1-2P_x)\cos2\theta\cos2\theta_m^0 \right] \,,
\label{eq:LZS1}
\end{equation}
where~\cite{jump}
\begin{equation}
P_x = {\rm{exp}\big(-{\pi \over 2}\,\gamma\, F \big)-
\rm{exp}\big(-{\pi \over 2}\,\gamma\, {F \over {\rm{sin}}^2\theta} \big) \over
 1-
\rm{exp}\big(-{\pi \over 2}\,\gamma\, {F \over {\rm{sin}}^2\theta} \big)} \,,
\label{eq:Px}
\end{equation}
is the transition probability with $F=1-{\rm{tan}}^2\theta$ for the
exponentially varying matter density in the Sun,
and~\cite{haxton,parke-walker}
\begin{equation}
\gamma = {(\delta m^2)^2 {\rm{sin}}^2 2\theta \over 4\sqrt{2}G_F E_{\nu}^2\
|dN_e/dL|_c} \,,
\label{eq:adiabaticity}
\end{equation}
measures the degree of adiabaticity of the transition. In
Eq.~(\ref{eq:adiabaticity}), $|d N_e/d L|_c$ is the density gradient at
the critical density. This process is analogous to level crossings in
atoms~\cite{lzs}. For an electron neutrino that is created well above the
resonance density ($\theta^0_m \simeq \pi/2$) and which undergoes a
perfectly adiabatic transition ($\gamma \to \infty, P_x \to 0$), the
oscillation probability is $P(\nu_e\to\nu_e) = \sin^2\theta$. Thus a
very large depletion of solar $\nu_e$'s is possible even for small
vacuum mixing angles. This is known as the
MSW effect, and was first studied
numerically in Ref.~\cite{msw}. In the extreme nonadiabatic limit
($\gamma\to0$)~\cite{rosen}, $P_x\to \cos^2 \theta$ and the
oscillation probability approaches $1-{1\over2}\sin^22\theta$, the
expected value for two-neutrino vacuum oscillations averaged over the
oscillations.

The range over which $P(\nu_e\to\nu_e) < {1\over2}$ is
\begin{equation}
{\delta m^2 \cos2\theta \over 2\sqrt2 G_F N_e^0} < E_\nu <
\delta m^2\sin2\theta \sqrt{\pi\over 8\sqrt2\ln2 G_F|dN_e/dL|_c}\,.
\label{eq:range}
\end{equation}
For neutrino energies below this range, the initial density is below
the critical density and the neutrino starts with a large fraction in
the lower eigenstate. For neutrino energies above this range, the
transition becomes very nonadiabatic and the neutrino has a high
probability of hopping from the upper eigenstate to the lower
eigenstate. In either case, a large component of the neutrino ends up
in the lower eigenstate, in which case the survival probability is
greater than ${1\over2}$. For $N_e^0 = 100 N_A/{\rm cm}^3$ (the
approximate number density at the center of the Sun), neutrinos
with energies in the range $2{\rm~MeV} \lsim E_\nu \lsim 20{\rm~GeV}$
will have survival probabilities smaller than $1\over2$, assuming the
best-fit oscillation parameters of the LMA solution ($\delta m^2_s =
7\times10^{-5}$~eV$^2$ and $\sin^22\theta_s = 0.83$). 

Exact formulae that include cases when the neutrino is created near
resonance are presented in Ref.~\cite{haxton2}. More discussions of
exact formulae for the transition probability are given in
Ref.~\cite{petcov2}. A semi-classical treatment for
an arbitrary density profile is given in
Ref.~\cite{balantekin}. Formulae for the MSW effect in a
three-neutrino model are presented in Ref.~\cite{kuo-pantaleone2}.

\subsection{Long-baseline oscillations through the Earth}

Oscillations of long-baseline neutrinos are affected
by electrons in the Earth if the path length is an appreciable
fraction of the Earth's diameter. The full propagation equations for three
neutrinos in matter are
\begin{equation}
i{d\nu_\alpha\over dL} = {1\over2E_\nu} \sum_\beta
\left( A \delta_{\alpha e} \delta_{\beta e} +
\sum_i V_{\beta i}^{*} \delta m_{i1}^2 V_{\alpha i} \right) \nu_\beta \,.
\label{eq:prop3}
\end{equation}
A constant density approximation often provides a good
representation of the neutrino propagation over long baselines through the Earth. Since
$\delta m^2_s \ll |\delta m^2_a|$ and $\theta_x$ is small, the probabilities 
can be expanded 
to second order in terms of the small parameters $\theta_x$ and
$|\delta m^2_s/\delta m^2_a|$~\cite{b-c, freund}. The following useful
approximations for $\delta m^2_a > 0$ are obtained~\cite{bmw}:
\begin{eqnarray}
P(\nu_\mu \to \nu_e) &=&
x^2f^2 + 2xyfg(\cos\delta\cos\Delta - \sin\delta\sin\Delta)
+ y^2g^2 \,,
\label{eq:Pme}
\\
P(\bar\nu_\mu \to \bar\nu_e) &=&
x^2\bar f^2 + 2xy\bar fg(\cos\delta\cos\Delta + \sin\delta\sin\Delta)
+ y^2g^2 \,,
\label{eq:Pmbeb}
\end{eqnarray}
where
\begin{eqnarray}
x &=& \sin\theta_a \sin2\theta_x \,,
\\
y &=& \alpha \cos\theta_a \sin2\theta_s \,,
\\
f,\bar f &=& \sin\left[(1\mp\hat A)\Delta\right]/(1\mp\hat A) \,,
\\
g &=& \sin(\hat A\Delta)/\hat A \,,
\end{eqnarray}
and
\begin{equation}
\Delta = |\Delta_a| \,,\qquad \hat A = |A/\delta m^2_a| \,,\qquad
\alpha = |\delta m^2_s/\delta m^2_a| \,.
\end{equation}
For $\delta m^2_a < 0$, the corresponding formulae are
\begin{eqnarray}
P(\nu_\mu \to \nu_e) &=&
x^2\bar f^2 - 2xy\bar fg(\cos\delta\cos\Delta + \sin\delta\sin\Delta)
+ y^2g^2 \,,
\label{eq:Pme2}
\\
P(\bar\nu_\mu \to \bar\nu_e) &=&
x^2f^2 - 2xyfg(\cos\delta\cos\Delta - \sin\delta\sin\Delta)
+ y^2g^2 \,.
\label{eq:Pmbeb2}
\end{eqnarray}
Oscillation probabilities for an initial $\nu_e$ and final $\nu_\mu$ 
can be found by
changing the sign of the $\sin\delta$ term in Eqs.~(\ref{eq:Pme},
\ref{eq:Pmbeb}, \ref{eq:Pme2}), and (\ref{eq:Pmbeb2}).  These expansions
are nearly exact for distances less than 4000~km when $E_\nu \gsim
0.5$~GeV~\cite{bmw}. For more accurate results at longer distances,
Eq.~(\ref{eq:prop3}) may be integrated numerically over the density
profile of the neutrino path.

Other approximate solutions can also be useful in certain
situations. At relatively short distances where the matter effect is
not as large, an expansion can be made in $\alpha$ and $A/\delta
m^2_a$ for the constant density
solution~\cite{koike-sato}. Relationships between the vacuum and
matter oscillation parameters for three-neutrino oscillations are
given in Ref.~\cite{harrison-scott}. Exact results for the
three-neutrino case with constant density are given in
Refs.~\cite{bppw, zaglauer}. Several properties of the
general three neutrino solution with a nonconstant density profile 
are discussed in
Ref.~\cite{yokomakura2}. Consequences of random density fluctuations are
discussed in Ref.~\cite{baha}; they are not expected to play an important role
in most situations.

The evolution in Eq.~(\ref{eq:prop3}) is modified if a sterile neutrino
$\nu_s$, is involved~\cite{sterileprop}:
\begin{equation}
A \delta_{\alpha e} \delta_{\beta e} \to 2 \sqrt2 G_F E_\nu
\left[ N_e \delta_{\alpha e} \delta_{\beta e} - {1\over2}N_n
(\delta_{\alpha\beta} - \delta_{\alpha s} \delta_{\beta s}) \right] \,,
\label{eq:sterile}
\end{equation}
where $N_n$ is the neutron number density. In two-neutrino
oscillations between $\nu_e$ and a sterile neutrino, the electron
number density $N_e$ is changed to an effective number density
$N_{eff} = N_e - {1\over2}N_n$, which changes the critical density for a
resonance in oscillations of solar neutrinos. Also, in two-neutrino
oscillations between a $\nu_\mu$ or $\nu_\tau$ and a sterile neutrino,
$N_{eff} = - {1\over2} N_n$; consequently there can be substantial
matter effects in $\nu_\mu \to \nu_s$ oscillations of atmospheric neutrinos.


\section{The solution to the solar neutrino problem}

\begin{figure}[ht]
\centering\leavevmode
\includegraphics[width=3in]{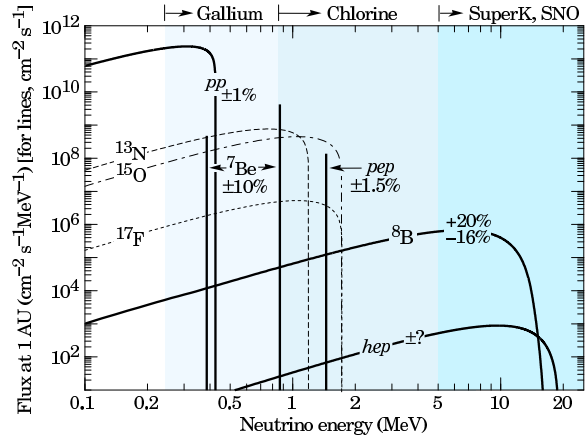}
\caption[]{The neutrino flux predictions of the Standard Solar 
Model~\cite{ssmnew}. From Ref.~\cite{pdg}.
\label{fig:SM-flux}}
\end{figure}

Decades of study of neutrinos from the Sun~\cite{chlorine,
sage,gallex,gno,superK-flat,superK-spectrum,sno,salt} have convincingly
established that neutrino oscillations are the cause of the deficits
of 1/3 to 1/2 in the measured electron-neutrino flux relative to the
Standard Solar Model expectations~\cite{ssmnew}.  
The water Cherenkov experiments
of SuperK and SNO measure the high energy neutrinos ($E
\gtrsim 5$~MeV) from the $^8$B chain, the Chlorine experiment also detects
the intermediate energy neutrinos from $^7$Be and 
$pep$, and the GALLEX, GNO
and SAGE experiments have dominant contributions from the $pp$ neutrinos~\cite{ssmold}; see
Fig.~\ref{fig:SM-flux}~\cite{ssmnew}. Until recently, the interpretation of the
deficits depended on comparisons with SSM predictions of the flux.
With the SNO experiment, which directly measures the total active
neutrino flux via neutral currents, the evidence for flavor conversion becomes robust.

\begin{figure}[h]
\centering\leavevmode
\includegraphics[width=3in]{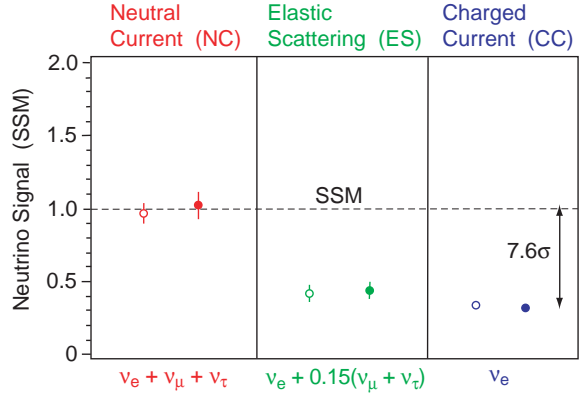}
\caption[]{Evidence for neutrino flavor
change seen by SNO. The open (filled) circles represent the 2003 SNO flux 
results, relative to the SSM, under the assumption of an undistorted 
(unconstrained) $^8$B neutrino energy spectrum.
\label{fig:heeger-5sigma}}
\end{figure}

The SNO experiment utilizes a heavy water target and measures the following
processes~\cite{sno,salt}:
\begin{eqnarray}
\mbox{Charged-Current (CC):} &&  \nu_e + d \to e^- + p + p\\
\mbox{Neutral-Current (NC):} && \nu_x + d \to \nu_x + n + p\\
\mbox{Elastic-Scattering (ES):} && \nu_x + e^- \to \nu_x + e^-
\end{eqnarray}
The CC/NC ratio establishes the oscillations of $\nu_e$ to $\nu_\mu$ and $\nu_\tau$
flavors,
\begin{equation}
{{\rm CC} \over {\rm NC}} =
{{\rm flux}(\nu_e) \over {\rm flux}(\nu_e + \nu_\mu + \nu_\tau)} \,.
\end{equation}
Only $\nu_e$ are produced in the Sun; the $\nu_\mu$ and $\nu_\tau$
fluxes are a consequence of oscillations. The charged-current signal was
found to be suppressed by 7.6$\sigma$ from the neutral-current signal;
see Fig.~\ref{fig:heeger-5sigma}.

\begin{figure}[ht]
\centering\leavevmode
\includegraphics[width=3.5in]{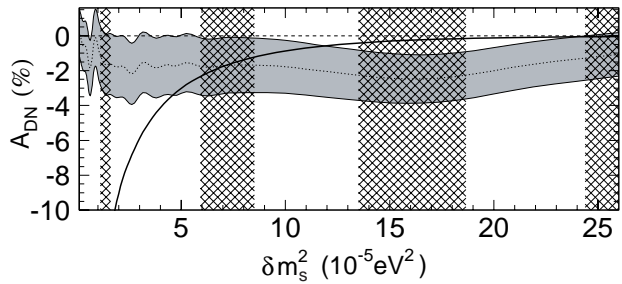}
\caption[]{The SuperK day-night asymmetry as a function of $\delta m^2_s$, with
the $1\sigma$ band shaded. The solid line is the prediction 
for $\tan^2 \theta_s =0.55$, and the cross-hatched bands are the ranges 
of $\delta m^2_s$ allowed by KamLAND.
From Ref.~\cite{superK-spectrum}. 
\label{fig:superdn}}
\end{figure}

\begin{figure}[ht]
\centering\leavevmode
\includegraphics[width=3in]{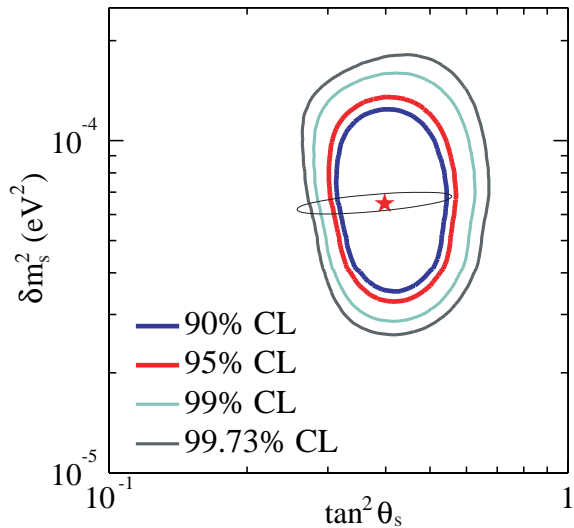}
\caption[]{90\%, 95\%, 99\% and 3$\sigma$ C.~L. allowed regions from a fit 
to the Homestake, GALLEX+GNO
and SAGE rates, and the SuperK and SNO spectra (with NC sensitivity enhanced
by salt). 
The ellipse is a projection of the 3$\sigma$ region
from three years of KamLAND data assuming  
the best-fit LMA parameters ($\delta m_s^2=6.5 \times 10^{-5}$ eV$^2$, 
$\tan^2 \theta_s=0.40$). 
Adapted from Ref.~\cite{salt} and Ref.~\cite{bmw1}. 
\label{fig:global}}
\end{figure}

\begin{figure}[ht]
\centering\leavevmode
\includegraphics[width=3in]{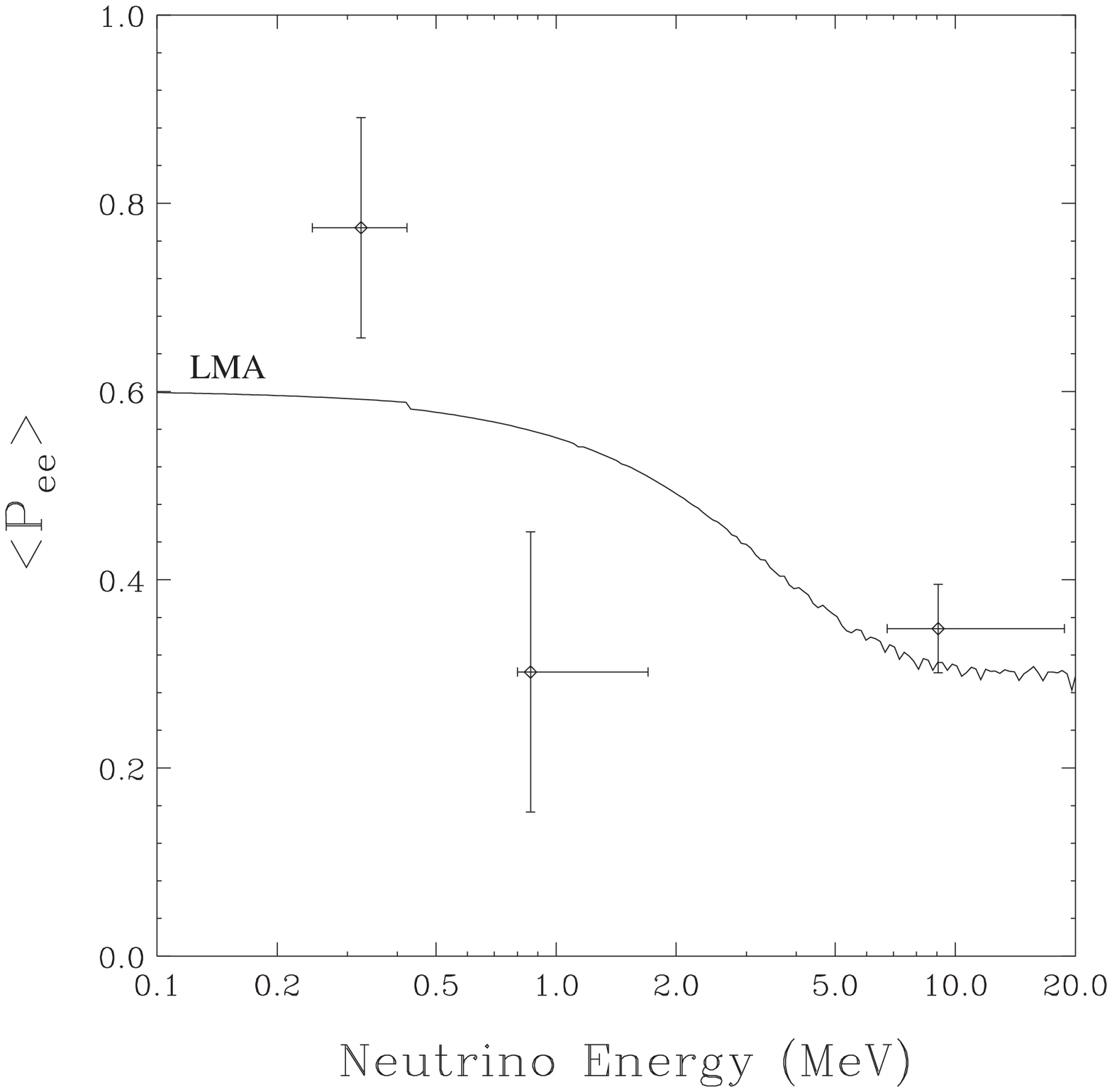}
\caption[]{The flux-weighted survival probability for an 
LMA  
solution. Adapted from Ref.~\cite{ourfit}.
\label{fig:surv-prob-vs-E}}
\end{figure}

The day and night energy spectra of charged-current events are
potentially sensitive to matter effects on oscillations that occur
when the neutrinos travel through the Earth~\cite{balwen}, though a
significant day-night asymmetry is yet to be established. SuperK (SNO)
 measured the day minus night $\nu_e$ rate to be  $-2.1\% \pm 2.0\%^{+1.3}_{-1.2}\%$
 ($-7.0\% \pm 4.9\%^{+1.2}_{-1.3}\%$) 
of the average rate~\cite{superK-flat,superK-spectrum} (\cite{sno}); 
see Fig.~\ref{fig:superdn}. 
From a global fit
to neutrino data,   
regions of the solar oscillation parameters have
been determined, as shown in Fig.~\ref{fig:global}~\cite{salt}.  
The Large Mixing
Angle (LMA) solution is preferred at more than 3$\sigma$.
The best fit to the
solar data is $\delta m_s^2 = 6.5\times10^{-5}\rm\, eV^2$ and
$\tan^2\theta_s = 0.40$~\cite{salt}. 
The survival probability
versus neutrino energy for LMA parameters is shown in
Fig.~\ref{fig:surv-prob-vs-E}.

\begin{figure}[t]
\centering\leavevmode
\includegraphics[width=2.75in]{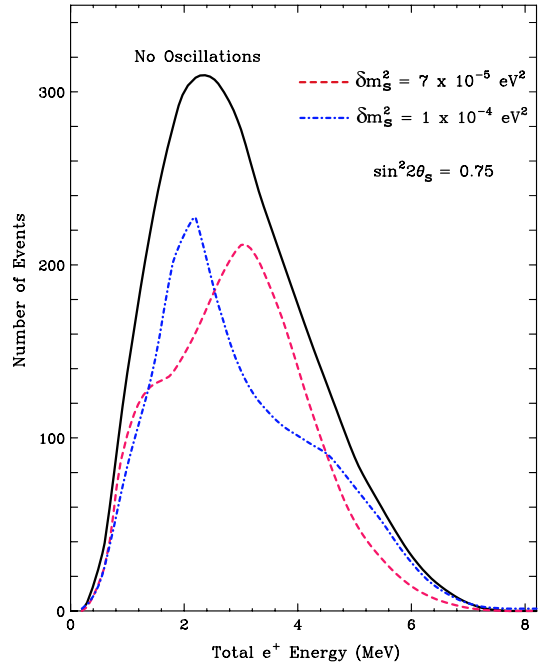}
\caption[]{KamLAND is expected to detect the spectral distortion resulting 
from oscillations. Adapted from Ref~\cite{bmw1}. \label{fig:kam-dm2}}
\end{figure}

The KamLAND experiment~\cite{kamland1} measures the electron
antineutrino flux at the Kamiokande site from surrounding reactors.
The dominant reactor is at $L=160$~km and the average distance from
the sources is $L\sim180$~km.  The measured reaction is $\bar\nu_e + p
\to e^+ + n$.  If $CPT$ invariance holds, which is expected in quantum
field theory, then $P(\bar\nu_e \to \bar\nu_e) =P(\nu_e \to \nu_e)$.
Therefore, for the LMA solar solution, reactor antineutrinos should also
disappear due to oscillations.  For any other solar oscillation
solution, no disappearance would be observed at KamLAND.  The
KamLAND expectation~\cite{bmw1, oth} for 3 years data, assuming
the LMA oscillation parameters, is shown in Fig.~\ref{fig:global} by
the narrow ellipse superimposed on the LMA region from solar data.
With sufficient data, the KamLAND experiment should ``see'' the oscillations
in the positron energy spectra, as illustrated in Fig.~\ref{fig:kam-dm2}~\cite{bmw1}.

\begin{figure}[ht]
\centering\leavevmode
\includegraphics[width=3.25in]{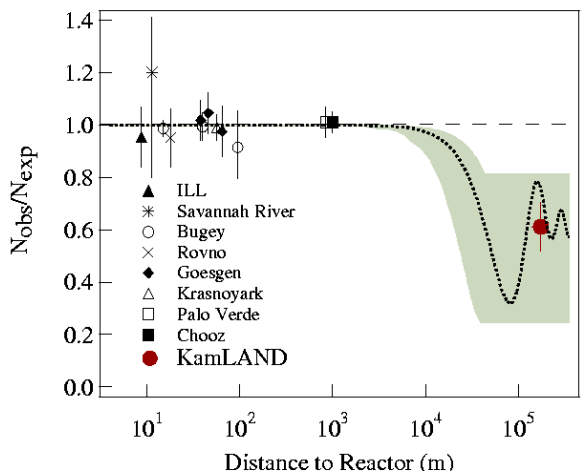}
\caption[]{The ratio of the measured to expected $\bar\nu_e$ flux from
reactor experiments. The shaded region encompasses the fluxes predicted for 
oscillation parameters in the 95\% C.~L. region 
from an analysis of pre-SNO salt phase data~\cite{foglifit}.  
The dotted curve corresponds to $\delta
m^2_s=5.5\times 10^{-5}$ eV$^2$ and $\tan^2 \theta_s=0.42$. 
 From Ref.~\cite{kamland2}. 
\label{fig:kamresults}}
\end{figure}

\begin{figure}[ht]
\centering\leavevmode
\includegraphics[width=3in]{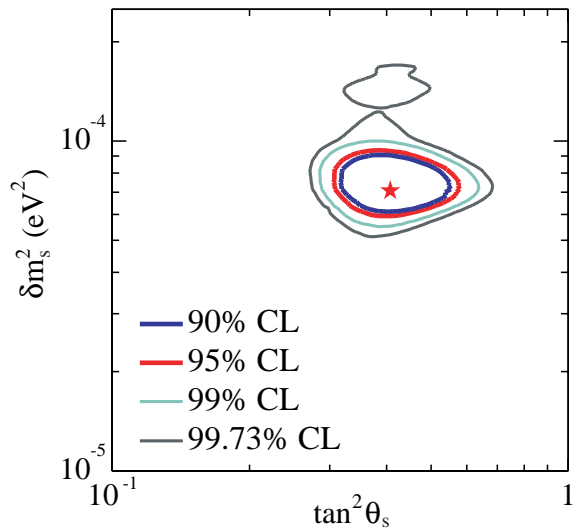}
\caption[]{90\%, 95\%, 99\% and 3$\sigma$ C.~L. allowed regions from a 
combined fit to KamLAND and solar neutrino data. 
The best-fit point is at $\delta
m^2_s=7.1\times 10^{-5}$ eV$^2$
and $\tan^2 \theta_s=0.41$. From Ref.~\cite{salt}.\label{fig:combined}}
\end{figure}

The first KamLAND results~\cite{kamland2} are based on 145 days of
operation.  The data give spectacular confirmation of the solar oscillation
analysis predictions; see Fig.~\ref{fig:kamresults}.  The fractional number of
events $(N({\rm observed})-N({\rm bkg}))/N({\rm expected}) = 0.611\pm
0.085 \rm (stat) \pm 0.041(syst)$ excludes no oscillations at 99.95\%
C.~L. and eliminates all solar solutions but LMA~\cite{kamland2}. A
combined analysis of solar and KamLAND data select the LMA solution
uniquely at the $4-5 \sigma$ C.~L~\cite{kamanalysis}. Maximal mixing is 
excluded at the 5.4$\sigma$ C.~L. and  $\delta m^2_s< 10^{-4}$ eV$^2$ at 
greater than the 99\% C.~L.~\cite{salt}; see Fig.~\ref{fig:combined}.

KamLAND has not only eliminated all oscillation solutions other than LMA, but 
has relegated nonoscillation solutions to the solar neutrino problem to
be at most subleading  effects. Non-standard neutrino interactions 
(NSNI)~\cite{nsni}, 
which lead to
energy-independent conversion probabilities, were consistent with the flat energy 
spectra seen by SuperK and SNO~\cite{nsni2}. 
With KamLAND data, 
NSNI are generically rejected as the leading cause of $\bar\nu_e$-disappearance 
at about the 3$\sigma$ C.~L.
Also, the resonant and nonresonant
spin-flavor precession solutions~\cite{sfp} are allowed only at the 99.86\% and 
99.88\% C.~L., respectively~\cite{sfprej}. Yet another excluded alternative  
invokes the violation of the equivalence
principle to induce oscillations even for massless neutrinos~\cite{vep}.
These three solutions fail because 
the KamLAND baseline is too short for any significant disappearance 
to occur.  

The continuation of the KamLAND reactor experiment will provide a
measurement of $\delta m_s^2$ that will be precise to about 10\%.  
Data from solar neutrinos require
that the sign of $\delta m_s^2$ is positive and that the mixing angle
$\theta_s$ is nonmaximal. A more precise determination of $\theta_s$
is important to test models of neutrino mass.  
Future SNO data (from the $^3$He proportional counter
phase) should reduce the presently
allowed range of $\theta_s$ by measuring the CC/NC ratio and the 
day-night asymmetry more precisely.
As far as the parameters responsible for the mixing of 
solar neutrinos are concerned, the goal of any new 
solar or reactor experiment should be to pin down $\theta_s$ more accurately
than a combination of future SNO and KamLAND data. 
Borexino~\cite{borexino} or any other
$^7$Be solar neutrino 
experiment will not improve on the accuracy with which three years of KamLAND data
and solar data will determine $\theta_s$~\cite{be7nogood}. A future $pp$
solar neutrino experiment with better than 3\% precision can lead to 
significant improvement. Proposals for measuring $pp$ neutrinos include
LENS~\cite{lens}, MOON~\cite{moon}, SIREN~\cite{siren}, XMASS~\cite{xmass}, 
CLEAN~\cite{clean}, HERON~\cite{heron} and GENIUS~\cite{genius}. Another
avenue for an improved $\theta_s$ measurement is a lithium-based 
radiochemical detector that detects neutrinos from the 
CNO cycle~\cite{kopylov}. 
A reactor neutrino experiment with baseline
such that the measured 
survival probability is a minimum can lead to a more precise  measurement
of $\theta_s$~\cite{futreac}. Since the reactor
neutrino spectrum is accurately known and KamLAND will determine 
$\delta m_s^2$ precisely, such an experiment is conceivable. For 
$\delta m_s^2= 7\times 10^{-5}$ eV$^2$, the required baseline is 70 km.

\section{Atmospheric neutrinos}

The first compelling evidence for neutrino oscillations came from the
measurement of atmospheric neutrinos.  Interactions of cosmic rays
with the atmosphere produce pions and kaons that decay to
muon neutrinos, electron neutrinos, and their antineutrinos:
\begin{eqnarray}
\pi^+, K^+ \to \nu_\mu \mu^+ \to \nu_\mu e^+ \nu_e \bar\nu_\mu \,,
\\
\pi^-, K^- \to \bar\nu_\mu \mu^- \to \bar\nu_\mu e^- \bar\nu_e \nu_\mu \,.
\end{eqnarray}
On average there are twice as many muon neutrinos as electron
neutrinos at energies of about 1~GeV, although the electron neutrinos tend to be at somewhat
lower energies since they are produced only in a secondary decay. The
atmospheric neutrino flux is well understood: the normalizations are
known to 20\% or better and ratios of fluxes are known to
5\%~\cite{gaisser-honda}. The flux falls off rapidly with neutrino
energy for $E_\nu \gsim 1$~GeV.

\begin{figure}[ht]
\centering\leavevmode
\includegraphics[width=3.65in]{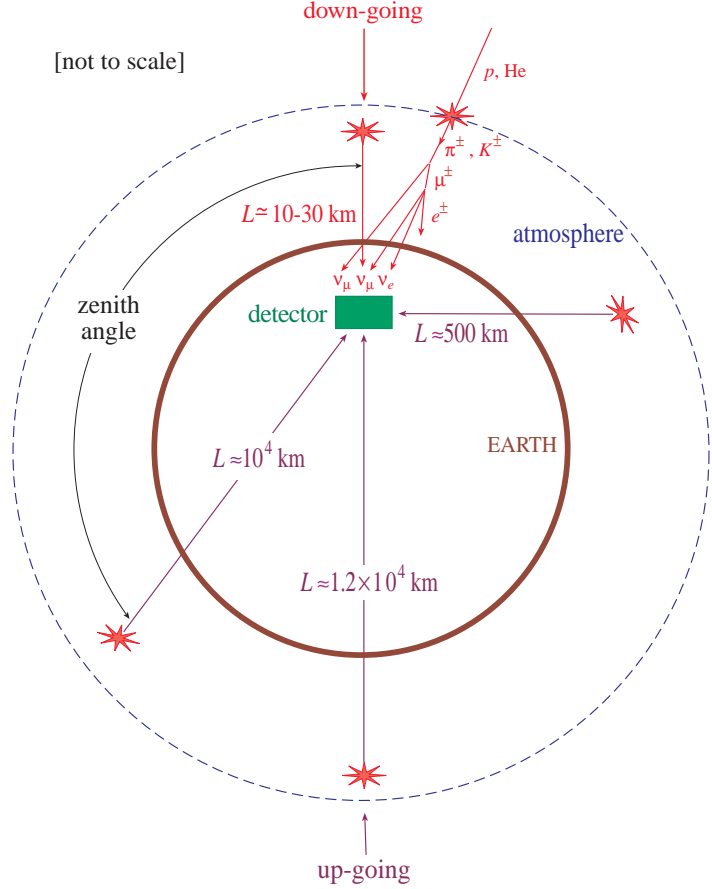}
\caption[]{A schematic view (not to scale) of the different zenith angles of
atmospheric neutrinos and distances they travel before detection.
\label{fig:atmo-nu}}
\end{figure}

\begin{figure}[t]
\centering\leavevmode
\includegraphics[width=2.8in]{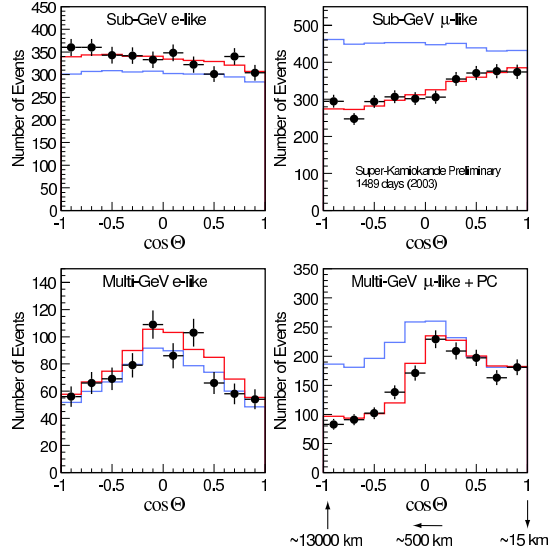}
\caption[]{Zenith angle distributions for $e$-like and $\mu$-like
contained atmospheric neutrino events in SuperK~\cite{hayato};
$\cos\Theta = 1$ corresponds to downward events
with $L \sim 15$~km and $\cos\Theta = -1$ corresponds to upward events
with $L \sim 13000$~km. The lines show the best
fits with and without oscillations; the best-fit is $\delta m^2_a =
2.0\times10^{-3}$~eV$^2$ and $\sin^22\theta_a = 1.00$. 
\label{fig:zenith}}
\end{figure}

\begin{figure}[ht]
\centering\leavevmode
\includegraphics[width=2.8in]{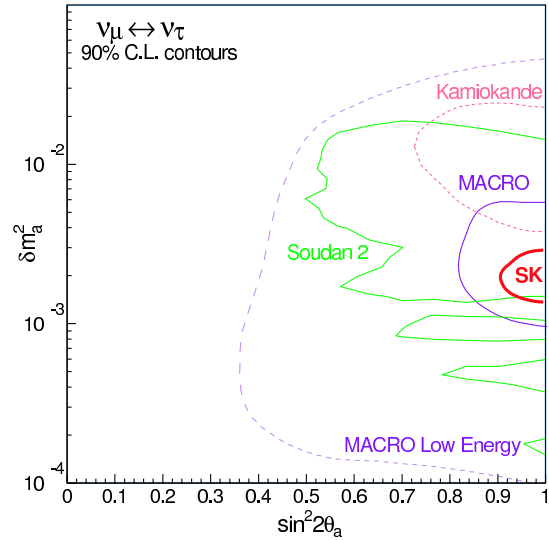}
\caption[]{90\% C.~L. allowed regions for $\nu_{\mu} \rightarrow \nu_{\tau}$ 
oscillations of atmospheric neutrinos for Kamiokande, SuperK, Soudan-2
and MACRO. From Ref.~\cite{kearns}.
\label{fig:macro}}
\end{figure}

Neutrinos observed at different zenith angles have path distances that
vary from $L\sim 10-30$~km for downward neutrinos to $L\sim 10^4$\,km
for upward neutrinos, as illustrated in Fig.~\ref{fig:atmo-nu}. The
ratio of observed to expected neutrino events provides a sensitive
measure of neutrino oscillations, especially since different values of
neutrino baselines and energies can be studied.  Initial evidence for
atmospheric neutrino oscillations was an overall depletion of muon
neutrinos~\cite{kam, imb} compared to the theoretical expectation. The
SuperK experiment~\cite{superK-atmos} has studied CC events
in four categories: fully contained ($E_\nu \sim 1$~GeV), partially
contained ($E_\nu \sim 10$~GeV), upward-going stopped ($E_\nu \sim
10$~GeV) and through-going ($E_\nu \sim 100$~GeV). The
contained events have the highest statistics and give the more precise
measurement, but all of the data samples are fully consistent with the
same oscillation parameters. Zenith angle distributions for the
$e$-like and $\mu$-like contained events are shown in
Fig.~\ref{fig:zenith} along with the no oscillation expectation and
the best fit assuming oscillations. Atmospheric neutrinos are
primarily sensitive to the leading oscillation which involves
$\theta_a$ and $\theta_x$. The angle $\theta_x$ is constrained to be smaller
than 13$^\circ$ (for $\delta m^2_a=2.0\times 10^{-3}$ eV$^2$) 
at the 95\% C.~L. Assuming $\theta_x=0$
in Eqs.~(\ref{eq:Pmm}) and (\ref{eq:Pmt}) ({\it i.e.}, atmospheric muon
neutrinos oscillate exclusively to tau neutrinos), the latest (preliminary) 
analysis by the SuperK collaboration of their data
yields best-fit values $\sin^22\theta_a = 1.00$ (maximal mixing) and $\delta m_a^2 =
2.0\times10^{-3}\rm\,eV^2$~\cite{hayato}. The 90\% C.~L. ranges for these
oscillation parameters are $\sin^22\theta_a \gsim 0.90$ and $\delta m_a^2 \simeq
(1.3-3.0)\times 10^{-3}\rm\,eV^2$. Both the Soudan-2~\cite{soudan2}
and MACRO~\cite{macro} experiments have also measured atmospheric
neutrinos and find allowed regions consistent with the SuperK result
(see Fig.~\ref{fig:macro}).

\begin{figure}[ht]
\centering\leavevmode
\includegraphics[width=3in]{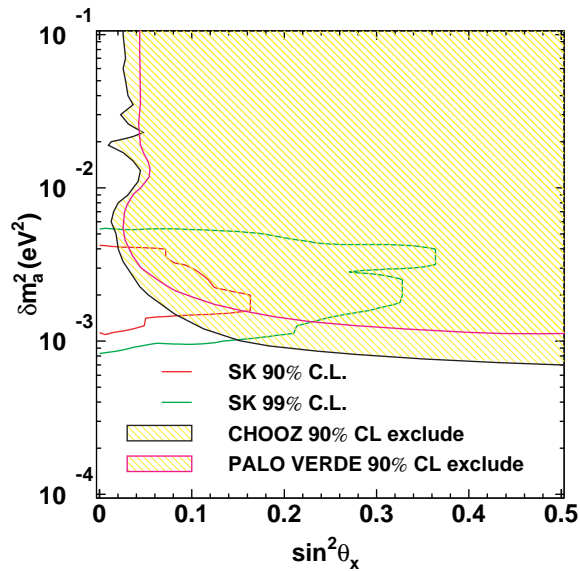}
\caption[]{Allowed regions in $\sin^2\theta_x-\delta m^2_a$ plane
for three-neutrino oscillations from SuperK atmospheric data. From
Ref.~\cite{nakaya}.
\label{fig:three-nu}}
\end{figure}

If $\theta_x\ne0$, then $\nu_e$ would participate in the oscillations
of atmospheric neutrinos with amplitude $\sin^2 2\theta_x$.  In the
SuperK data the number of observed electron neutrinos is consistent
with $\theta_x=0$ (see Fig.~\ref{fig:zenith}), whereas an enhancement
would be expected if there were $\nu_\mu \leftrightarrow \nu_e$
oscillations (due to the 2:1 ratio of $\nu_\mu$ to $\nu_e$ in the
flux). Furthermore, the zenith angle distribution of the SuperK muon
sample is inconsistent with oscillations involving
$\nu_e$. Figure~\ref{fig:three-nu} shows the SuperK allowed region in
$\sin^2\theta_x-\delta m^2_a$ plane. 
There is also slightly greater than 
2$\sigma$ evidence in the SuperK data of hadronic showers from $\tau$
decays~\cite{kearns,toshito}, 
consistent with the hypothesis that the primary
oscillation of atmospheric neutrinos is $\nu_\mu \to \nu_\tau$.

Another consequence of $\theta_x\ne0$ is that reactor
$\bar\nu_e$ fluxes should exhibit disappearance due to the leading
oscillation when $L/E_\nu \ge 40$~m/MeV. Data from the CHOOZ reactor
experiment~\cite{chooz} ($L \sim 1000$~m, $E_\nu \sim 3$~MeV) place
upper limits on $\theta_x$ for the values of $\delta m^2_a$ indicated
by the atmospheric neutrino data; similar limits have been obtained
from the Palo Verde reactor experiment~\cite{paloverde} (see
Fig.~\ref{fig:three-nu}). For the $\delta m^2_a$ values obtained from the SuperK
collaboration's latest two-neutrino analysis (slightly lower than from previous analyses), 
bounds on $\theta_x$ are quite sensitive to $\delta m^2_a$. 
While for $\delta m^2_a=2.0\times 10^{-3}$ eV$^2$, the 95\% C.~L. 
upper bound on
$\sin^2 2 \theta_x$ is 0.2, for $\delta m^2_a=1.3\times 10^{-3}$ eV$^2$, 
the corresponding bound is 0.36. Thus, it has become necessary to specify the 
$\delta m^2_a$ for which a bound on $\theta_x$ is quoted.

\begin{figure}[ht]
\centering\leavevmode
\includegraphics[width=3in]{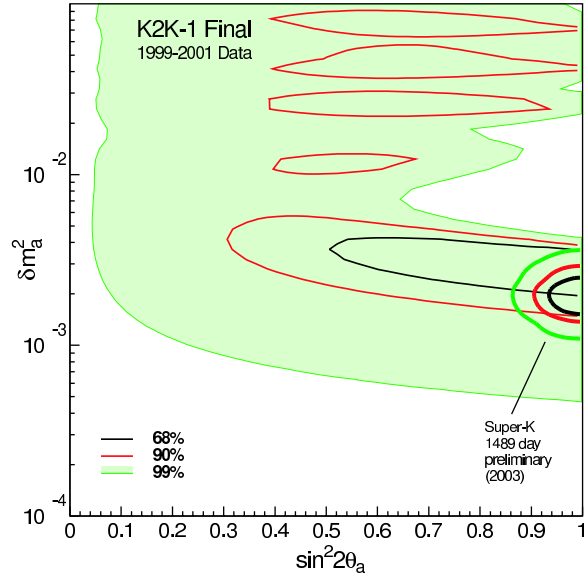}
\caption[]{Allowed regions in $\sin^2\theta_a-\delta m^2_a$ plane
from K2K, compared with the allowed regions from the SuperK atmospheric
data~\cite{kearns}.
\label{fig:k2k}}
\end{figure}

The K2K experiment, in which $\nu_\mu$ with energies typically
$\sim1.4$~GeV are directed from KEK to SuperK ($L = 250$~km), has
measured a $\nu_\mu$ survival probability consistent with the
atmospheric neutrino results, $P(\nu_\mu \to \nu_\mu) =
0.70^{+0.11}_{-0.10}$~\cite{k2k}. The K2K allowed region, from the
number of events and the spectrum shape combined, is consistent with
the allowed region from the atmospheric neutrino data (see
Fig.~\ref{fig:k2k}). An approximate two-neutrino
analysis that adopted the SuperK allowed regions in Fig.~\ref{fig:k2k} 
in combination with K2K data found 
$\delta m_a^2=(2.0^{+0.8}_{-0.6})\times 10^{-3}$ eV$^2$ at the
 95\% C.~L.~\cite{foglisub}. Imposing this $\delta m_a^2$ range 
as a prior in a three-neutrino analysis of 
CHOOZ, KamLAND and pre-SNO salt phase solar data yielded
$\sin^2 2 \theta_x \leq 0.17$ at the 95\% C.~L. after marginalizing
over $\delta m^2_a$~\cite{foglisub}.

The MINOS experiment~\cite{minos} expects to detect atmospheric
neutrinos via the $\nu_\mu$ and $\bar\nu_\mu$ charged-current reactions
 {\it and} to determine the
signs of the resulting charged leptons with a magnetic field, thereby
separately testing oscillations of $\nu_\mu \to \nu_\mu$ and
$\bar\nu_\mu \to \bar\nu_\mu$.

Since upward-going atmospheric neutrinos traverse a large fraction of
the Earth's diameter, matter
effects could be relevant. The dominant oscillation of atmospheric neutrinos
appears to be $\nu_\mu \to \nu_\tau$; for two-neutrino $\nu_\mu \to \nu_\tau$
oscillations there would be no matter effects. However, 
 probability conservation in three-neutrino
oscillations involves large matter effects which
change the $\nu_\mu \to \nu_\tau$ oscillation probability to a small
extent. 

Specific relationships between the changes in the oscillation
phase and the matter density can lead to an enhancement of the
oscillation probability, analogous to the classical phenomenon of
parametric resonance~\cite{ermilova}. Also,
constructive quantum mechanical interference between the probability
amplitudes for different density layers can give total neutrino flavor
conversion~\cite{chizhov}. Such enhancement phenomena generally
require passage of the neutrino through the Earth's core. Conditions
for observing matter effects in the Earth's mantle and core in
atmospheric neutrino experiments are discussed in
Ref.~\cite{bernabeu}.


\section{Absolute neutrino mass}

Neutrino oscillations tell us nothing about the absolute scale of
neutrino masses.  The standard technique for probing the absolute mass
is to study the end-point region of the electron spectrum in tritium
beta-decay.  The effect of a nonzero neutrino mass is to suppress and
cut off the electron distribution at the highest energies.  The
effective neutrino mass that could be determined in beta-decay
is~\cite{dbeta}
\begin{equation}
m_\beta^2 = \sum \left| V_{ei} \right|^2 m_{i}^2 \,. 
\end{equation}
The present limit from the Troitsk~\cite{troitsk} and Mainz~\cite{mainz} experiments is $m_\beta \leq 2.2$~eV at 2$\sigma$.  Future sensitivity down to $m_\beta = 0.35$~eV is expected in the
KATRIN experiment~\cite{katrin}, which will begin in 2007.

The sum of neutrino masses $\Sigma \equiv \sum m_\nu$, can be probed in cosmology by measuring 
$\omega_\nu = \Sigma/(94$ eV) using the
large scale structure (LSS) of the universe~\cite{power}. Relativistic 
neutrinos do not cluster
on scales smaller than they can travel in a Hubble time. Thus, 
the power spectrum is suppressed on scales smaller than the horizon when the 
neutrinos become nonrelativistic. (For eV neutrinos, this is the horizon at 
matter-radiation equality).
 The effect is subtle. Lighter neutrinos
freestream out of larger scales and cause the power spectrum 
suppression to begin at smaller wavenumbers~\cite{morehu}:
\begin{equation}
k_{nr} \approx 0.026 \bigg({m_\nu \omega_M \over 1\, {\rm{eV}}}\bigg)^{1/2} {\rm{Mpc}}^{-1}\,,
\end{equation}
assuming almost degenerate neutrinos.
On the other hand, heavier neutrinos
constitute a larger fraction of the matter budget and suppress power on 
smaller scales more strongly than lighter neutrinos~\cite{power}:
\begin{equation}
{\Delta P \over P} \approx -8 {\Omega_\nu \over \Omega_M}\approx 
-0.8 \bigg({\Sigma \over 1\, {\rm{eV}}}\bigg) \bigg({0.1 \over \omega_M}\bigg)\,.
\end{equation}
 The galaxy power
spectrum is influenced by the sum of neutrino
masses, even down to 0.1~eV~\cite{power}.    
Analysis of the 2dF Galaxy Redshift Survey (2dFGRS) data
found a limit of $\Sigma\leq 2.2$~eV on the masses of degenerate
neutrinos, or about 0.7~eV for each neutrino~\cite{elgaroy}.  An
improved limit of $\Sigma \leq 0.71$~eV was obtained by the WMAP
collaboration in an analysis of CMB data in conjunction with the
2dFGRS and the Lyman alpha forest power
spectrum data~\cite{wmap}. Since there are questions about the treatment 
of the uncertainties~\cite{seljak} in the Lyman alpha forest data, the WMAP
collaboration performed an analysis without this data and find the
bound is {\it strengthened} to $\Sigma \leq 0.63$~eV~\cite{wmap}.
With only CMB and 2dFGRS data, other analyses found the limit
on the summed neutrino masses to be about 1~eV~\cite{elg2}.
In connection with the above limits, it is interesting that 
an argument relying on anthropic selection concluded that 
$\Sigma \sim 1$ eV so that neutrinos cause a small but nonnegligible suppression of
galaxy formation~\cite{vilenkin}. In the future, lensing measurements
 of galaxies and the CMB by large scale structure 
may also provide a sensitive probe of $\Sigma$~\cite{kaplinghat}.

The $Z$-burst mechanism~\cite{zburst} provides
another astrophysical 
probe of the absolute scale of neutrino masses~\cite{weilerpas}. 
The Greisen-Zatsepin-Kuzmin (GZK) cutoff energy 
($\sim 5\times 10^{19}$ eV)~\cite{gzk}
expected in the cosmic ray spectrum 
is absent in data from the AGASA experiment~\cite{agasa}, although the
GZK cutoff may be respected by the HiRes data~\cite{bergman}.
A possible explanation for ultra high energy cosmic ray events above the
GZK cutoff is resonant annihilation of ultra high energy 
neutrinos on the cosmic neutrino background to
produce $Z$ bosons which decay in a burst of about 20 photons and
2 super-GZK nucleons.
The average energy of the secondary nucleon arriving at Earth after travelling a
distance $D$ ($\sim$ 50--100 Mpc) is~\cite{weilersong} 
\begin{equation} 
E \approx {10^{21}\, \rm{eV} \times 0.8^{D/6 \rm{Mpc}} \over m_\nu/0.1\,\rm{eV}}\,.
\end{equation} 
The neutrino mass scale constrained by atmospheric data from below and by 
cosmology from above is suitable to initiate
$\sim 10^{20}$ eV air-showers. Higher $m_\nu$ would lower $E$, thus precluding
$Z$-bursts as an explanation of the super-GZK events. Lower $m_\nu$ would
necessitate an unrealistically large neutrino flux at the resonant energy.
The speculative assumption of the $Z$-burst mechanism is that a substantial
cosmic neutrino flux exists at $\sim 10^{22}$ eV. 
If the existence of this flux is confirmed at teraton 
neutrino detectors~\cite{euso}, $Z$-bursts must occur.

\begin{figure}[ht]
\centering\leavevmode
\includegraphics[width=2.15in]{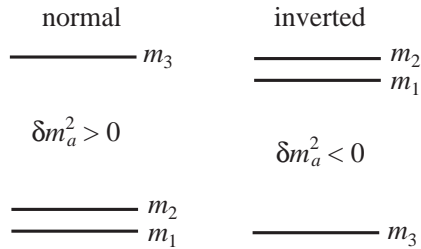}
\caption[]{The patterns of relative mass differences in normal (left) and inverted (right) neutrino mass hierarchies. \label{fig:hierarchies}}
\end{figure}

All neutrino masses are linked to the lightest mass by the values of
$\delta m^2_a$ and $\delta m^2_s$ determined by the neutrino
oscillation studies~\cite{bww}. If the scale of the lightest mass is small, then
the heaviest mass is approximately $\sqrt{|\delta m^2_a|} \simeq
0.05$~eV and a neutrino mass hierarchy exists. Since the sign of
$\delta m^2_a$ is unknown, there are two possible hierarchies, as
illustrated in Fig.~\ref{fig:hierarchies}. The mass hierarchy is an important
discriminant of neutrino mass models. If the scale of the
lightest mass is larger than 0.05~eV, then the neutrino masses are
approximately degenerate.

\begin{figure}[t]
\centering\leavevmode
\includegraphics[width=2.5in]{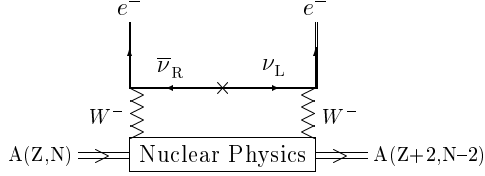}
\caption[]{Neutrinoless double-beta decay mediated by
Majorana neutrinos. \label{fig:00bb}}
\end{figure}

\begin{figure}[h]
\centering\leavevmode
\includegraphics[width=2.55in]{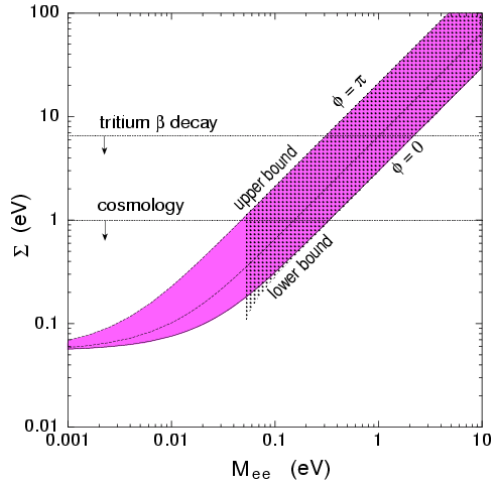}    
\caption[]{$\Sigma$ vs.\ $M_{ee}$ for the normal (shaded) and inverted (dotted) hierarchies. For the inverted hierarchy, $M_{ee} \ge \sqrt{|\delta m_a^2|}$. 
(Here, $|\delta m_a^2|$ was taken to be $3\times 10^{-3}$ eV$^2$).
The 95\% C.~L. bounds from tritium $\beta$ decay and cosmology are shown. Adapted from Ref.~\cite{our1}.\label{fig:bounds} }
\end{figure}

In the Standard Model with massive neutrinos and no other new physics,
neutrinoless double-beta decay ($0\nu\beta\beta$) probes the absolute
mass, provided that neutrinos are Majorana particles; see Fig.~\ref{fig:00bb}. 
Numerous
theoretical analyses have been made of what can be learned about the
neutrino sector from
$0\nu\beta\beta$~\cite{0nubb,our1,speculate,nogo}.  The decay rate
depends on the $\nu_e\mbox{--}\nu_e$ element of the mass
matrix~\cite{wolfen}:
\begin{equation}
  M_{ee} = \left| \sum V_{ei}^2 m_i \right| \,.
\label{eq:Mee}
\end{equation}
The prediction is insensitive to $\theta_x$ and $\delta m^2_s$ because
they are small. Setting $\theta_x = 0=\delta m_s^2$, the following
relation between $M_{ee}$ and $\Sigma$ is
obtained for both hierarchies~\cite{our1}:
\begin{eqnarray}
 M_{ee} = \bigg({2} \Sigma- \sqrt{\Sigma^2+3 \delta m_a^2}\bigg)\left| c_s^2 + s_s^2 e^{i\phi} \right|/3\,,
\end{eqnarray}
where $\phi$ is a Majorana phase.
For a given measured value of $M_{ee}$ both upper (since $\theta_s
\neq \pi/4$) and lower bounds are implied for $\Sigma$.  These bounds
are displayed in Fig.~\ref{fig:bounds}. 
The present upper limit on $M_{ee}$ is 0.35~eV at the 90\% C.~L.~\cite{klap}, 
with an overall factor
of 3 uncertainty associated with the $0\nu\beta\beta$ 
nuclear matrix elements~\cite{elliott}.  
A detection
of neutrinoless double beta decay, corresponding to $M_{ee} =0.39$~eV, 
has been reported~\cite{klapdor}, but this experimental
result is highly controversial~\cite{controversy}.

An extensive review of past and proposed $0\nu\beta\beta$ experiments has been 
made in Ref.~\cite{elliott}. Future experiments include 
CUORE ($^{130}$Te)~\cite{cuore}, 
EXO ($^{136}$Xe)~\cite{exo}, XMASS ($^{136}$Xe)~\cite{xmass},
 GENIUS ($^{76}$Ge)~\cite{genius}, 
Majorana ($^{76}$Ge)~\cite{majoranaexp} and MOON 
($^{100}$Mo)~\cite{moon}.  The upcoming experiments are expected to have 
sensitivity better than 50 meV, which is the critical mass scale of 
$\sqrt{|\delta m_a^2|}$.

There has been speculation about detecting $CP$ violation using
$0\nu\beta\beta$~\cite{speculate}. However, it has been shown in 
Ref.~\cite{nogo} that  this is impossible
in the foreseeable future since there is no reliable method for estimating
the uncertainty in the nuclear matrix elements.
Moreover, the further the solar amplitude is constrained away
from unity, the more stringent will be the precision requirement on the matrix
elements for such a detection to be made even in principle~\cite{nogo}. 
Even under extremely optimistic assumptions, at best it may be possible to
determine whether $\phi$ is closer to 0 or to $\pi$, corresponding to 
$CP$ conservation.


\section{Supernova neutrinos}

\begin{figure}[ht]
\centering\leavevmode
\includegraphics[width=4.75in]{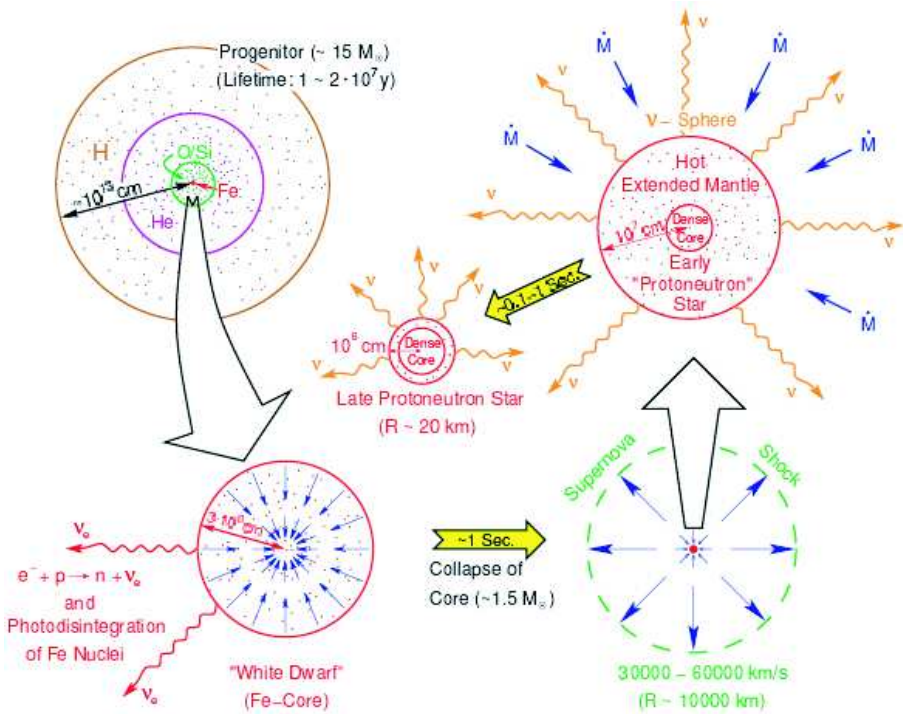}
\caption[]{Schematic illustration of a supernova explosion. The dense Fe core
collapses in a fraction of a second and gets neutronized (lower-left). The 
inner core rebounds and gives rise to a shock-wave (lower-right). The 
protoneutron star cools by the emission of neutrinos. From Ref.~\cite{snexpl}.
\label{fig:snfig}}
\end{figure}

 Stars with masses above eight solar masses undergo collapse. Once the core of the
star becomes constituted primarily of iron, further compression of the core
does not ignite nuclear fusion and the star is unable to
thermodynamically support its outer envelope. As the surrounding
matter falls inward under gravity, the temperature of the core rises
and iron dissociates into $\alpha$ particles and nucleons.  Electron
capture on protons becomes heavily favored and electron neutrinos are
produced as the core gets neutronized (a process known as 
neutronization).  When the core reaches densities above $10^{12}$
g/cm$^3$, neutrinos become trapped (in the so-called
neutrinosphere). The collapse continues until $3-4$ times nuclear
density is reached, after which the inner core rebounds, sending a
shock-wave across the outer core and into the 
mantle; see Fig.~\ref{fig:snfig}. This shock-wave
loses energy as it heats the matter it traverses and incites further
electron capture on the free protons left in the wake of the shock.
During the few milliseconds in which the shock-wave travels from
the inner core to the neutrinosphere, electron neutrinos are released
in a pulse. This neutronization burst carries away
approximately $10^{51}$ ergs of energy. However, 99\% of the binding energy
$E_b$, of the protoneutron star
is released in the following $\sim 10$ seconds primarily 
via $\beta$-decay (providing a
source of electron antineutrinos), $\nu_e\neb$~\cite{newmodel5} and
$e^+e^-$ annihilation and nucleon bremsstrahlung~\cite{suzuki}
(sources for all flavors of neutrinos including $\nu_\mu$,
$\bar{\nu}_\mu$, $\nu_\tau$ and $\bar{\nu}_\tau$), in addition to
electron capture.  The neutrinos following the neutronization burst
are the ones of interest in the following discussion.

We focus on charged current $\nu_e$ and $\neb$ interactions. 
We therefore cannot distinguish between the
different nonelectron species, and denote them collectively as  
 $\nu_x$ ($x=\mu,\tau, \bar{\mu}, \bar{\tau}$){\footnote{However, note that weak magnetism can cause the 
luminosities and temperatures of the $\nu_{\mu,\tau}$ and  $\bar{\nu}_{\mu,\tau}$ to
differ by about 10\%~\cite{newmodel3}. 
Practically speaking, this effect is too small to be detectable.}}.
The various cooling processes result in a state of approximate 
(within a factor of two or so) equipartition of energy with
the luminosity of the electron neutrinos $\lne$ 
being up to $10\%$ larger than the that of 
the electron 
antineutrinos and $50-100\%$ larger than that of $\nu_x$~\cite{newmodel5}.
The degree to which equipartition is violated can be quantified 
through constants $\bneb$ and $\bnx$ which are defined by
\begin{equation}
\lne = \bneb \lneb = \bnx \lnx\,,
\label{lumin}
\end{equation}
where $1 \leq \bneb \lsim 1.1$ and $1\leq \bnx \lsim 2$. Perfect equipartition 
corresponds to $\bneb=\bnx=1$.

Since the protoneutron star is opaque to neutrinos, it takes a few tens of
seconds for them to diffuse out. The $\nu_e$ and $\neb$ interact with nuclear matter 
via both charged and neutral current reactions (with a smaller cross-section 
for $\neb$), while the $\nu_x$  experience only neutral current
scattering. Consequently, the different species have neutrinospheres such that their radii
obey $R_{\nu_e} > R_{\neb} > R_{\nu_x}$. Each neutrino species decouples at a temperature 
characterized by the temperature at the surface of 
its neutrinosphere. Based on this simple argument,
which relies only on the well-known interaction strengths of neutrinos with matter,
the hierarchy of average energies is
\begin{equation}
\ene < \eneb < \enx\,.
\end{equation}
Early supernova (SN) models typically predicted~\cite{models1}
\begin{eqnarray}
\ene &=& 10-12\ {\rm MeV} \,,  \nonumber \\
\eneb &=& 14-17\  {\rm MeV}\,,  \nonumber \\
\enx &=& 24-27 \  {\rm MeV}\,, \ \ \ \ \ \ \  \nonumber \\
E_b &=& 1.5-4.5 \times 10^{53}\  {\rm ergs} \label{models} \,.
\end{eqnarray}   
The inclusion of additional energy transfer processes in modern 
SN codes indicates that the hierarchy of average energies is 
likely smaller than originally expected.
There is evidence that nuclear recoils can lower $\enx$ by as much as 
20\%~\cite{newmodel2}. Also, 
nucleon bremsstrahlung softened the spectra in the simulations of
Ref.~\cite{newmodel1}. 
The cumulative effect is that $\enx/\eneb$ is typically expected to be
1.1, and unlikely
to be greater than 1.2~\cite{raffsn}.
 
The energy spectra of neutrinos can be modeled 
by pinched Fermi-Dirac distributions. The 
unoscillated differential flux (flux per unit energy) at 
a distance $D$ can be written as
\begin{equation}
F_{\alpha} = \frac{L_{\alpha}}{24\pi D^2 \, T^4_{\alpha} |Li_4(-e^{\eta_{\alpha}})|} \: \frac{E^2}{e^{E/T_{\alpha}-\eta_\alpha} \, + \, 1} \:,
\label{fermi}
\end{equation}
where $\alpha=\nu_e, \neb, \nu_x$,  $Li_n(z)$ is the 
polylogarithm function and $\eta_{\alpha}$ is the 
degeneracy parameter.  
The temperature of the neutrinos, $T_{\alpha}$, is related to 
$\vev{E_{\alpha}}$ via $\vev{E_{\alpha}}=
3 {Li_4(-e^{\eta_{\alpha}}) \over Li_3(-e^{\eta_{\alpha}})}T_{\alpha}$.

\begin{figure}[ht]
\centering\leavevmode
\includegraphics[width=2.8in]{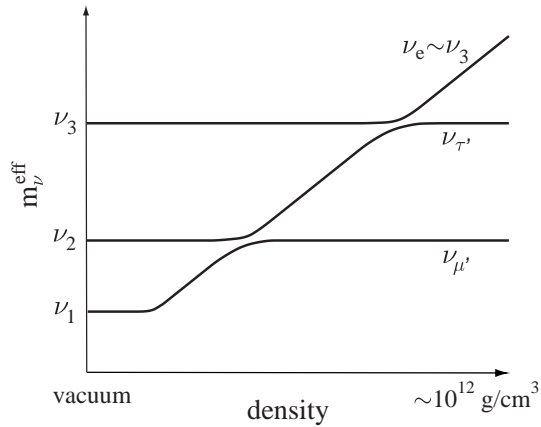}
\caption[]{Schematic level-crossing diagram for neutrinos emitted by a SN in the
case of a normal mass hierarchy. $\nu_{\mu'}$ and $\nu_{\tau'}$ are basis states which
diagonalize the ($\nu_\mu,\nu_\tau$) submatrix of the 
Hamiltonian governing the neutrino evolution. 
\label{sncross}}
\end{figure}

Because of the extremely high density of the matter in the neutrino
production region of a SN, all flavors of neutrinos start out in pure
mass eigenstates.  As the neutrinos stream out from the production
region, they pass through a density profile that is well-represented by $V_0
(R_{\odot}/r)^3$~\cite{dense}, where $R_\odot$ is the solar radius and $V_0$ is
a constant.
Due to the wider range of densities that the neutrinos encounter, both
the solar and atmospheric scales contribute to the oscillation
dynamics.  The hierarchical nature of the two scales ($|\delta
m^2_{s}| \ll |\delta m^2_{a}|$) and the smallness of the mixing
parameter $\sin^2 2 \theta_{x}$ imply that the dynamics can be
approximately factored so that oscillations are governed by $\delta
m^2_{a}$ and $\sin^2 2\theta_{x}$ at high densities ($10^3-10^4$
g/cm$^3$), and by $\delta m^2_{s}$ and $\sin^2 2\theta_{s}$ at low
densities ($\sim 20$ g/cm$^3$ for the LMA solution)~\cite{dighe}; 
see Fig.~\ref{sncross}.
Transitions in the latter region are adiabatic. In the high density region,
neutrinos (antineutrinos) pass through a resonance if  $\delta m^2_{a}>0$ 
($\delta m^2_{a}<0$). The jumping
probability is the same for both neutrinos
and antineutrinos~\cite{foglisn} and is of the form $P_H \sim
e^{-\sin^2 \theta_{x} (|\delta m^2_{a}|/E_{\nu})^{2/3} V_0^{1/3}
}$~\cite{kuosn}. Note the exponential dependence of $P_H$ on
$\sin^2\theta_x$. That SN neutrinos provide a handle on the sign of $\delta
m^2_a$ can be seen as follows.

For the normal
and inverted hierarchies, the survival probability of electron 
antineutrinos is given by~\cite{dighe}
\begin{equation}
\bar{p} = \bar{P}_{1e}\,,
\label{norprob}
\end{equation} 
and
\begin{equation}
\bar{p}=P_H\bar{P}_{1e} + (1-P_H) \sin^2 \theta_{x}\,,
\label{invprob}
\end{equation}
respectively. Here, 
$\bar{P}_{1e} = \bar{P}_{1e}(E_{\nu},\delta m^2_{s},\sin^2 2 \theta_{s})$ 
is the probability that an 
antineutrino reaching the Earth in the $\bar{\nu}_1$ mass eigenstate will interact with the detector as a $\neb$.
If $\sin^2 2 \theta_{x} \ll 10^{-3}$, then $P_H \approx 1$ and the 
survival probabilities 
for the two hierarchies are the same. Thus, the normal and inverted mass 
hierarchies are indistinguishable for $\sin^2 2 \theta_{x} \ll 10^{-3}$. 
If $\sin^2 2 \theta_{x} \gsim 10^{-3}$, 
for the inverted hierarchy $\bar{p} \approx
\sin^2 \theta_{x}\lsim 0.05$ and the original
electron antineutrinos have all been swapped for the more energetic $\mu$ and 
$\tau$
antineutrinos by the time they exit the supernova envelope, 
resulting in a harder incident
spectrum.
Thus, the initial $\neb$ spectrum would have to be softer for the 
inverted hierarchy than for the normal hierarchy.

The detection of neutrinos from SN 1987A was momentous. The 11 events
at Kamiokande II~\cite{k2sn} and 8 events at the Irvine Michigan
Brookhaven~\cite{imbsn} detectors have lent strong support to the
generic model of core collapse supernovae~\cite{loredo}. The
significance of these few events provides a tantalizing glimpse into
the physics potential offered by a future galactic SN event. Despite
the fact that only a few galactic SN are expected per
century, the potential payoff is so huge that experiments dedicated
to SN neutrino detection have been proposed~\cite{uno}.

Attempts have been made to extract neutrino oscillation parameters
from the 19 SN 1987A events (see Refs.~\cite{snanalysis,minnun,us} for
recent analyses).  However, conclusions drawn from these analyses
depend crucially on the assumed neutrino temperatures and spectra.
For example, it was claimed that the data favor the normal hierarchy
over the inverted hierarchy provided $\sin^2 \theta_{x} \gsim
10^{-4}$~\cite{minnun}, but this conclusion was
contradicted~\cite{us}. Table~\ref{tab2par} shows the results of
two-parameter fits in $E_b$, and $\tneb$, for the normal
hierarchy and the inverted hierarchy (with $\sin^2 2
\theta_{x}=0.01$). In all cases the likelihoods are
comparable, so the data do not favor one neutrino mass hierarchy over
the other.  While SN 1987A was of great astrophysical significance, it
did not allow firm deductions about neutrino mixing.

\begin{table}[ht]
\caption{Best fit values for $E_b$ and $\tneb$ from two-parameter fits
to all the KII and IMB data. Results are for the cases of no oscillations,
the inverted hierarchy with $\sin^2 2\theta_x=0.01$ and 
the normal hierarchy. From Ref.~\cite{us}. }
\label{tab2par}
\begin{center}
\begin{tabular}{|l|c|c|c|} 
\hline 
 & $E_b$ ($10^{53}$ ergs) & $\tneb$ (MeV) & $ln({\cal L}_{max})$  \\ \hline
 no oscillations           & 3.2 & 3.6 & -42.0\\ \hline
 $\enx/\eneb=1.25 \;,\; \sin^2 2 \theta_x=0.01 $ & 3.1 & 2.9 & -42.0\\ \hline
 $\enx/\eneb=1.25 \;,\;$ normal   & 3.2 & 3.4 & -41.9\\ \hline
 $\enx/\eneb=1.4\  \;,\; \sin^2 2 \theta_x=0.01 $ & 3.1 & 2.6 & -42.0\\ \hline
 $\enx/\eneb=1.4\  \;,\;$ normal   & 3.4 & 3.2 & -41.6\\ \hline
 $\enx/\eneb=1.7\  \;,\; \sin^2 2 \theta_x=0.01 $ & 3.2 & 2.1 & -42.0\\ \hline
 $\enx/\eneb=1.7\  \;,\;$ normal   & 4.2 & 2.7 & -41.2\\ \hline
 $\enx/\eneb=2.0\  \;,\; \sin^2 2 \theta_x=0.01 $ & 3.2 & 1.8 & -42.0\\ \hline 
 $\enx/\eneb=2.0\  \;,\;$ normal   & 5.8 & 2.2 & -40.6\\ \hline
\end{tabular}
\end{center}
\end{table}

Neutrinos from a galactic SN incident on a large water or heavy water 
detector could in principle provide much information on neutrino
oscillations. A determination of $\theta_{x}$ and the neutrino mass 
hierarchy from SN neutrinos is special in that degeneracies arising 
from the unknown $CP$ phase
$\delta$ and whether $\theta_{a}$ is above or below 
$\pi/4$ do not contaminate it, {\it i.e.,} 
the eight-fold parameter degeneracies that are 
inherent in long baseline experiments~\cite{bmw} are absent. 
This cleanness  results because (i)
nonelectron fluxes do not depend on the $CP$ phase 
$\delta$~\cite{newmodel4},  
and so SN neutrinos directly probe $\theta_{x}$, and (ii)
whether $\theta_{a}$ is above or below $\pi/4$ is immaterial since
this parameter does not affect the oscillation dynamics.  

Investigations of the effect of neutrino oscillations on SN neutrinos have been
made in Refs.~\cite{dighe,futuresn,us2}.  
Whether or not the mass hierarchy can
be determined and $\theta_x$ be constrained depends strongly on how
much $\enx/\eneb$ is greater than unity~\cite{dighe}. 
The higher the value of $\enx/\eneb$, the
better the possible determinations.  As noted earlier, $\enx/\eneb$ is 
expected to be about 1.1, and no larger than 1.2~\cite{raffsn}. 
Then, assuming that the $\eta_\alpha$ values predicted by SN models 
are accurate, a safe deduction is that
observations of a galactic SN at SuperK or Hyper-Kamiokande (HyperK)
could place either a lower or upper bound on $\theta_x$ if the
neutrino mass hierarchy is inverted~\cite{us2}. The hierarchy can be determined
if it is inverted and
$\sin^2 2\theta_x \gsim 10^{-3}$; for $\sin^2 2\theta_x
\lsim 10^{-4}$, the survival probabilities of the electron
antineutrinos are similar for both hierarchies rendering them
indistinguishable even in principle. On the other hand, if the
hierarchy is normal, neither can $\theta_x$ be constrained nor can the
hierarchy be determined~\cite{us2}. Nonetheless, the
importance of the detection of a galactic SN should not be understated 
because of its major
impact on the understanding of the SN explosion mechanism. 

\begin{figure}[b]
\centering\leavevmode
\includegraphics[width=2.8in]{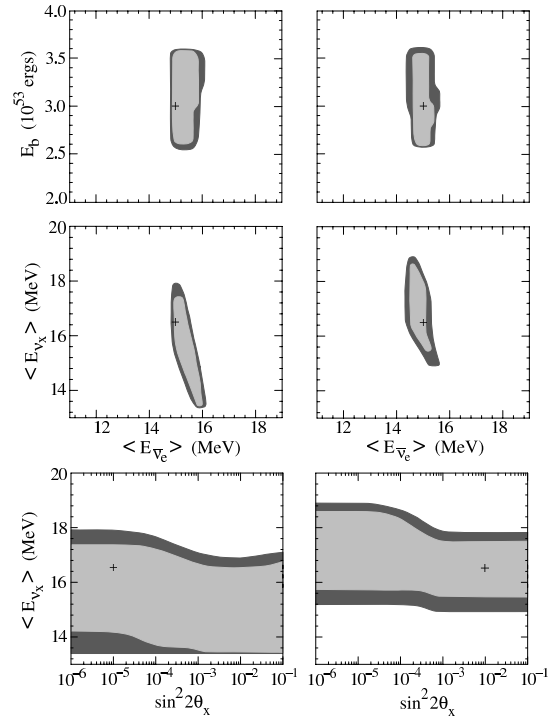}
\caption[]{Determination of 
 the binding energy $E_b$, the supernova
neutrino mean energies (temperatures) and $\sin^22\theta_x$ 
for the normal mass hierarchy. The left-hand
and right-hand panels correspond to data simulated at 
$\sin^22\theta_x = 10^{-5}$
and $\sin^22\theta_x = 10^{-2}$, respectively.
The cross-hairs 
mark the theoretical inputs, and the $90$\% and 
$99\%$ C.~L. regions are light and dark shadings, respectively. 
$\vev{E_{{\nu}_x}}$ is
the mean energy of the nonelectron neutrinos. From Ref.~\cite{us2}. \label{fig:sn}}
\end{figure}

Broadly speaking, the binding energy of the star and the 
temperatures of the different flavors of
neutrinos and antineutrinos are determinable; see Fig.~\ref{fig:sn}~\cite{us2}. 
Precise measurements
of the temperatures can provide an unique window into the dominant microphysics
of neutrino transport in addition to confirming the expectation that $\tneb<\tnx$. With very large detectors it may even be 
possible to estimate how equipartitioned the energy is among the neutrino flavors. This knowledge would help refine SN codes that predict different degrees to which equipartitioning
is violated. For example, in Ref.~\cite{totani} an almost perfect equipartitioning 
is obtained while according to Refs.~\cite{newmodel5,mezzacappa}, 
equipartitioning holds only to within a factor of 2.

Since it is difficult to determine $\theta_x$ and the mass hierarchy
simultaneously from galactic SN data, an interesting prospect is
to consider what can be learned from a future SN 
if $\sin^2 2\theta_x$ is already known to be
larger than $0.01$ from a future reactor or accelerator neutrino
experiment. With this information, SuperK would easily discriminate between
the two hierarchies if the hierarchy is inverted.  If the hierarchy is
normal, Earth-matter effects on the SN neutrino flux 
could be used to verify that this is the case, with
either a high-statistics detector like the proposed half-megaton
HyperK~\cite{jhfsk} or at a high-resolution scintillation
detector~\cite{raffsn2}.
 

\section{Model building}

\subsection{Patterns of neutrino masses and mixings}

One of the important challenges in particle physics is to understand
the spectrum of fermion masses. The mixing matrix in the quark sector,
$V_{CKM}$, is given by the product $V_u^\dagger V_d$, where $V_u$ and
$V_d$ are the unitary transformations applied to the left-handed up
and down quarks to diagonalize the up and down quark mass
matrices. Similarly, the mixing matrix that enters into neutrino
oscillations is $V_{MNS} = V_L^\dagger V_\nu$, where $V_L$ and $V_\nu$
are the unitary transformations applied to the left-handed charged
leptons and neutrinos to diagonalize the charged lepton and
neutrino mass matrices. In the quark sector, all mixing angles in
$V_{CKM}$ are small and there is a mass hierarchy among the
generations, whereas in the lepton sector (although not
necessarily in the neutrino sector) a mass hierarchy exists with two
large mixing angles and one small mixing angle in $V_{MNS}$. A
remarkable property of neutrino masses is that they are so much
lighter than the charged leptons. Any theory of fermion mass must
reconcile the extreme differences between quark and lepton masses and
mixings. A plethora of papers has addressed the problem of
neutrino mass over the years in the context of different models, which
we are unable to exhaustively cover here; for recent comprehensive reviews and
more references, see Refs.~\cite{barr-dorsner, altarelli-feruglio,
 mohapatra}. 

Since absolute neutrino masses are not yet known, there are three
possible mass patterns for neutrinos: (i) normal hierarchy ($m_1 \ll
m_2 \ll m_3$), (ii) inverted hierarchy ($m_2 \gsim m_1 \gg m_3$), and
(iii) quasi-degenerate ($m_1 \simeq m_2 \simeq m_3$). Because
$V_{MNS}$ is a product of the mixing matrices for the charged leptons
and neutrinos, the observed mixing in neutrino oscillations can originate
from $V_L$, $V_\nu$, or a combination of the two. Viable models exist
with different combinations of mass pattern and origins of the mixing
angles.

In models where where the charged lepton mixing matrix is
approximately diagonal, $V_{MNS}$ derives directly from $V_\nu$. If
there are three Majorana neutrinos, then there are nine independent
parameters in the mass matrix: three absolute masses and six mixing
matrix parameters (see Eq.~\ref{eq:mixmat}). Six of these may be
measured in neutrino oscillations (three mixing angles, the Dirac
phase, and two mass-squared differences). The absolute mass scale may
be determined by measuring tritium beta-decay, $0\nu\beta\beta$ decay or 
the suppression of the matter power spectrum. 
The magnitude of the $\nu_e-\nu_e$ mass matrix element
(which depends on the three mixing angles, absolute masses, and two
Majorana phases) may be determined from $0\nu\beta\beta$
experiments (although the value of the associated nuclear matrix elements
makes this measurement less than precise). Therefore the complete
$3\times3$ mass matrix for Majorana neutrinos cannot be fully
determined by experiment in the near future, and in practice neither
of the Majorana phases will be well-measured. In models where the
charged lepton mass matrix is nondiagonal, there are even more
independent parameters. However, symmetry arguments can help to
reduce the number of parameters.

Soon after the initial SuperK discovery that atmospheric neutrinos
oscillate with maximal or nearly maximal amplitude, it was noted that
the neutrino sector might exhibit bimaximal mixing~\cite{bimaximal}, 
{\it i.e.}, maximal or nearly maximal mixing of solar {\it
and} atmospheric neutrinos. Then the mixing matrix has the unique form
(up to state redefinitions)
\begin{equation}
V_{MNS} = \left( \begin{array}{ccc} {1\over\sqrt2} & {1\over\sqrt2} & 0 \\
-{1\over2} & {1\over2} & {1\over\sqrt2}\\ 
{1\over2} & -{1\over2} & {1\over\sqrt2} \end{array} \right) \,.
\label{eq:bimaximal}
\end{equation}
Perturbations on this basic form can yield mixing that is not quite
maximal, and can make $V_{e3}$ ($=s_x e^{-i\delta}$) nonzero. More
recently, solar neutrino and KamLAND data disfavor maximal mixing
in the solar sector ($\sin^22\theta_s \le 0.95$ at the 3$\sigma$ level),
and the emphasis is now on finding models that can give bi-large
mixing.

\subsection{The seesaw mechanism}

A popular model for understanding the smallness of neutrino masses is
the seesaw mechanism~\cite{seesaw1}, which
is particularly well-motivated in a grand unified theory (GUT). In the
one-generation version, the neutrino mass matrix in the
$\nu_L$-$\nu_R$ basis is
\begin{equation}
M_\nu = \left( \begin{array}{cc} 0 & m_D\\ m_D & m_R \end{array} \right) \,,
\label{eq:seesaw}
\end{equation}
where $m_D$ is a Dirac mass and $m_R$ a right-handed Majorana mass.
The eigenmasses are then approximately $-m_D^2/m_R$ and $m_R$ (the
negative value of the lighter state can be made positive by a
redefinition of the phases of the neutrino fields). If the heaviest
light neutrino has mass of order $\sqrt{|\delta m^2_a|} \simeq
0.05$~eV, and the Dirac mass is the $\tau$ lepton mass then $m_R \sim
10^{11}$~GeV. Other interesting possibilities for $m_D$ are the
electroweak vacuum expectation value or the top quark mass which would
imply $m_R = 10^{15}$~GeV (close to the GUT scale).  Since heavy
right-handed neutrinos exist in most grand unified models, a
GUT/seesaw model is very attractive.

In practice, there are three generations of neutrinos and the neutrino
mass matrix for the light neutrinos is $-M_D^TM_R^{-1}M_D$, where
$M_D$ is the $3\times3$ matrix describing the Dirac neutrino masses
(presumably related by symmetries to the charged lepton or quark
masses) and $M_R$ is the $3\times3$ matrix describing the right-handed
Majorana neutrino masses. If $M_D$ has a hierarchical form (similar to
the charged lepton mass spectrum), then the neutrino mixing angles in
$V_{MNS}$ tend to be small unless there is an unnatural conspiracy
between $M_D$ and $M_R$~\cite{albright,jezabek}. 
The choice of particular forms for $M_D$ and
$M_R$ can avoid this problem~\cite{jezabek,costa}. 
A more detailed discussion of such models and relevant
references can be found in Ref.~~\cite{barr-dorsner}.

\subsection{GUT models}

\begin{figure}[ht]
\centering\leavevmode
\includegraphics[width=3.5in]{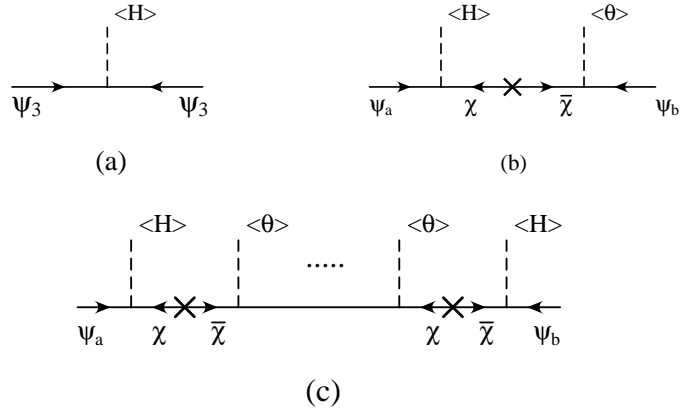}
\caption[]{Froggatt-Nielsen diagrams. Here $a$ and $b$ are 
family-indices and 
$(\chi, \bar\chi)$ are vector-like fields of mass $M$ 
and $\vev{\theta}$ is the vacuum expectation value of the
flavor Higgses or flavons. The tree-level diagram (a) generates the mass of the
third family and the lighter masses are obtained by 
${\cal{O}}(\vev{\theta}/M)^n$ suppressions from diagrams (b) and (c). From 
Ref.~\cite{mcchen}.
\label{fig:frogniel}}
\end{figure}

Since GUT models relate quarks and leptons, the fact that
quark mixing angles are large and lepton mixing angles are small is
potentially a problem for these models. In fact, many GUT models predicted
small solar neutrino mixing, now ruled out by data from
SNO and KamLAND. However, there is a way around this difficulty, due
to the fact that GUT theories relate leptons and quarks of opposite
chiralities, {\it e.g.}, right-handed down quarks are in the same fermion
multiplets as left-handed charged leptons, and vice versa. Therefore
the mixing we observe in the lepton sector is connected to the
right-handed quark mixing, which is unknown and not constrained.
So-called lopsided models~\cite{barr} take advantage of this fact. The
three-generation Dirac mass matrices for the charged leptons and down
quarks have the forms
\begin{equation}
M_L \propto \left( \begin{array}{ccc} x & x & x \\ x & 0 & \epsilon \\
x & \sigma & 1 \end{array} \right) \,, \qquad
M_d \propto \left( \begin{array}{ccc} x & x & x \\ x & 0 & \sigma \\
x & \epsilon & 1 \end{array} \right) \,,
\label{eq:lopsided}
\end{equation}
where $\epsilon \ll \sigma \sim 1$. The entries involving only the
second and third generations come from specific Higgs Yukawa
interactions, with no contribution to the middle diagonal. The entries
with an ``$x$'' involve the first generation and are very small, and
in many (although not all) models are generated by Froggatt-Nielsen
diagrams mediated by exotic vector-like matter fields~\cite{froggatt};
see Fig.~\ref{fig:frogniel}.  The up-quark and neutrino mass matrices
are approximately diagonal.  Then the quark mixing element $V_{bu}$ is
small, but the leptonic mixing element $V_{\mu3}$, which relates to
the mixing of atmospheric neutrinos, is large. In lopsided models, the
large atmospheric neutrino mixing comes from the diagonalization of
the charged lepton mass matrix; the solar neutrino mixing angle arises
from the structure of the three-generation right-handed Majorana
neutrino mass matrix, also determined in some models by
Froggatt-Nielsen diagrams. Most lopsided models are embedded in a
GUT~\cite{sato-yanagida}, but some are not~\cite{irges}. In many cases,
horizontal ({\it i.e.}, family) symmetries determine the textures of the
mass matrices. A Monte Carlo study suggests that lopsided textures are
favored by the data~\cite{antonelli}. In any GUT framework, proper
comparison with data can only be made after allowing for the
renormalization group running of the neutrino mass terms in the
Lagrangian~\cite{babuleung}; for recent discussions, see
Ref.~\cite{albright-geer}. Of particular importance are how
zeroes in the mass matrices behave under renormalization and the
stability of the large mixing angles~\cite{frigerio}. For a more complete
discussion and further references on GUT models, see
Refs.~\cite{barr-dorsner, albrighttalk}.

Lopsided models can yield either large or small solar neutrino
mixing~\cite{albright-barr}; for a discussion of which models can
naturally yield the LMA solution, see Ref.~\cite{dorsner}. The known
SO(10) models that satisfy all experimental data favor a normal
hierarchy~\cite{albrighttalk}, {\it i.e.}, $m_3 > m_1, m_2$ in the neutrino
sector, and therefore $\delta m^2_a > 0$. Predictions for $V_{e3}$
vary~\cite{barr-dorsner}.

Another interesting possibility is to assume that the unification
group is replicated at the Planck scale, {\it i.e.}, there is one copy for
each generation. This leads to family-dependent $U(1)$ symmetries as
the theory breaks down to become the Standard Model at low
energies. Some consequences of such models are discussed in
Ref.~\cite{ling}. Models that utilize family symmetries can have
either (i) both large mixing angles in the neutrino sector deriving
from the neutrino mass matrix~\cite{ross}, or (ii) the large
atmospheric neutrino mixing angle deriving from the charged lepton
mass matrix and the large solar neutrino mixing angle deriving from
the neutrino mass matrix~\cite{datta-ling-ramond}. SUSY GUT models
with $R$-parity violation are discussed in Ref.~\cite{koide}.

\subsection{Non-GUT models}

There are many alternatives to explicit GUT models (although it is
often assumed that they could emerge from an unspecified grand unified
theory). Some popular possibilities include: (i) the Zee model, (ii)
models with low-energy new physics in which neutrino masses are
generated as loops, such as supersymmetry with R-parity violation,
(iii) so-called ``democratic'' models, (iv) models with triplet Higgs
bosons, (v) models with specific textures for the neutrino mass matrix
or special relationships involving the entries in the mass matrix,
(vi) models with dynamical electroweak symmetry breaking, and (vii)
models with large extra dimensions.  In the non-GUT models, $M_\nu$ is
unrelated to the charged fermion mass matrices, and hence {\it a
priori} there are few constraints on its structure. In many cases,
horizontal (family) symmetries can be used to provide constraints
and produce a phenomenologically compelling model.

\subsubsection*{Zee model}

\begin{figure}[h]
\centering\leavevmode
\includegraphics[width=2.2in]{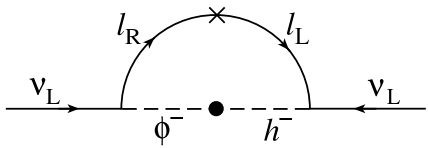}
\caption[]{Neutrino mass is generated at one loop in the Zee model.
\label{fig:zee}}
\end{figure}

In the Zee model~\cite{zee}, which invokes radiative neutrino masses
via a charged $SU(2)_L$ singlet and a Higgs field in the loop (see Fig.~\ref{fig:zee}), 
the neutrino mass matrix has the approximate form
\begin{equation}
M_\nu = \left( \begin{array}{ccc} 0 & $A$ & $B$ \\ $A$ & 0 & 0 \\
$B$ & 0 & 0 \end{array} \right) \,,
\label{eq:zee}
\end{equation}
where $A \sim B$. In Zee-type models, the diagonal elements of $M_\nu$
are zero, and the remaining off-diagonal elements may be nonzero (but
small compared to $A$ and $B$). Such a texture can also result from an
approximate $L_e - L_\mu - L_\tau$ symmetry~\cite{glashow}; 
for some other possibilities see Ref.~\cite{matsuda}. 
The mass matrix in Eq.~(\ref{eq:zee}) yields large mixing for
both solar and atmospheric neutrinos (although some specific models
yield vacuum solar neutrino oscillations that are now excluded). 
The mass hierarchy for the Zee-type mass matrix is
inverted, {\it i.e.}, $m_3 \ll m_1, m_2$ and $\delta m^2_a < 0$. The
prediction of the Zee model that the solar neutrino mixing is nearly
maximal mixing is now excluded by data.

\begin{figure}[b]
\centering\leavevmode
\includegraphics[width=2.2in]{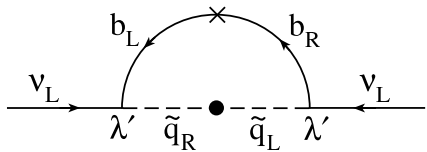}
\caption[]{The dominant one-loop diagram which generates Majorana
neutrino masses for left-handed neutrinos in $R$-parity violating models.
The coupling $\lambda^\prime$ violates lepton number as well as
$R$-parity.
\label{fig:rpvloop}}
\end{figure}

\subsubsection*{New physics at low energy}

In models with low-energy new physics, neutrinos couple to a heavy
fermion in the theory. Mass terms for the light neutrinos are
generated by loop diagrams involving the neutrino and the heavy
fermion.  If the heavy fermion coupling to the second and third
generation neutrinos is larger than to the first generation, a normal
mass hierarchy and large mixing for atmospheric neutrinos results. In
the minimal supersymmetric extension of the Standard Model (MSSM)
radiative neutrino mass generation is a direct consequence of R-parity
violation~\cite{rparity} (see Fig.~\ref{fig:rpvloop}).  For specific
realizations to explain the neutrino anomalies see
Ref.~\cite{drees}. R-parity violating SUSY models that reproduce the
neutrino mass and mixing parameters can have specific signatures in
future collider experiments, such as lepton-number violating final
states~\cite{hirsch}, neutralino decay within the
detector~\cite{porod,chun2}, neutralino decay
branching ratios~\cite{chun2,sierra}, multi-$b$-jet events with an
isolated charged lepton~\cite{hesselbach}, and multi-lepton
events~\cite{magro}. It is also possible that neutrino masses are 
generated at the two-loop level~\cite{babu-model}.

\subsubsection*{Flavor democracy}

In models with flavor democracy~\cite{democratic}, the quark and
charged lepton mass matrices have the form
\begin{equation}
M \propto \left( \begin{array}{ccc} 1 & 1 & 1 \\ 1 & 1 & 1 \\
1 & 1 & 1 \end{array} \right) \,,
\label{eq:democratic}
\end{equation}
while the neutrino mass matrix is approximately diagonal. This
scenario also leads to large mixing for solar and atmospheric
neutrinos and a normal mass hierarchy. 
The democratic model (and many other non-seesaw models) predicts that the solar
neutrino mixing angle is maximal and the atmospheric neutrino mixing
angle is large but not necessarily maximal, whereas the data indicate
the opposite.

\subsubsection*{Triplet Higgs bosons}

In models with triplet Higgs bosons, horizontal symmetries are used to
constrain the texture of the neutrino mass matrix~\cite{frampoh}. An
example that uses an $S_2\times S_2$ permutation symmetry in a
four-neutrino theory (with one neutrino becoming heavy) is given in
Ref.~\cite{mohapatra-nussinov}.

\subsubsection*{Textures or special relationships}

Since it unlikely that all nine parameters of the neutrino mass matrix
can be determined, many studies have examined simpler structures with
fewer independent parameters (with the assumption that the charged
lepton mass matrix is diagonal). Some recent examples:

\begin{enumerate}

\item[(i)] It has been shown that $3\times3$ Majorana mass matrices with three
or more independent zero entries are excluded by current neutrino data
but there are seven distinct textures with exactly two
independent zeroes that are acceptable~\cite{fgm}. Note that since a Majorana
mass matrix is symmetric, a reflected off-diagonal zero is not counted
as independent. Two of these textures lead to a normal mass hierarchy and
the other five to a quasi-degenerate mass spectrum. In fact, 
it is possible to fully determine the neutrino mass spectra corresponding to 
these textures~\cite{zzxing}. Several aspects of these
seven matrices have been studied in Ref.~\cite{followup}.
Some of these
textures can be realized in a seesaw model with or without extra
$U(1)$ flavor symmetries~\cite{kageyama}. These textures can also be
obtained in models with three Higgs triplets 
and a sufficiently massive triplet Majoron~\cite{frampoh}.

\item[(ii)] The weak-basis independent condition det($M_\nu = 0$)
(which would be approximately true if the lightest neutrino was nearly
massless) can also lead to a complete determination of the
neutrino mass matrix~\cite{branco3}.

\item[(iii)] Another possible condition that the neutrino mass matrix
might obey is form invariance, $U M U^T = M$, where $U$ is a specific
unitary matrix such that $U^N$ represents a well-defined discrete
symmetry in the neutrino flavor basis~\cite{ma}. This condition leads
to a variety of possible mass matrices, including all three of the
allowed mass patterns~\cite{ma2}. If the discrete symmetry is the
non-Abelian group $A_4$, the mass pattern is
quasi-degenerate~\cite{ma3}.

\item[(iv)] If the sum of the neutrino masses is zero~\cite{nasri}, 
which can occur in models whose neutrino mass matrix can be expressed as the
commutator of two matrices, only the inverted hierarchy and
quasi-degenerate mass patterns are allowed by current neutrino 
data~\cite{he-zee}.

\end{enumerate}

If the charged lepton mass matrix is not assumed to be diagonal, then
there are more independent parameters. A simplifying
ansatz may be used to reduce the number of parameters. For example,
the Fritzsch ansatz~\cite{fritzsch} assumes $M_{11} = M_{22} = M_{13}
= M_{31} = 0$ for both the charged leptons and neutrinos, which can
lead to acceptable phenomenology. The large mixing of
atmospheric neutrinos can come from $V_L$~\cite{datta}.

\subsubsection*{Dynamical electroweak symmetry breaking}

The seesaw mechanism can be realized in models with dynamical
electroweak symmetry breaking (extended technicolor, or ETC,
models). By suppressing the Dirac mass terms $m_R$ in
Eq.~(\ref{eq:seesaw}), the heavy Majorana scale need not be so high.
(typically the $m_D$ terms can be much smaller than the ETC
scale)~\cite{appelquist}. Dynamical electroweak symmetry breaking due
to a neutrino condensate has also been considered~\cite{antusch3}.

\subsubsection*{Extra dimensions}

Theories with large extra dimensions~\cite{extra} have been postulated
to avoid the hierarchy problem. In such theories, there is no very
large scale ({\it e.g.}, the GUT or Planck scale), and so the smallness of
the neutrino masses cannot be obtained from a conventional seesaw
mechanism; instead, it is a consequence of the suppressed coupling
between the active neutrinos on the brane (the usual four-dimensional
world) and sterile neutrinos in the bulk (Kaluza-Klein modes) or on
other branes, associated with the small overlap of their
wavefunctions. According to the particular model and coupling
mechanisms, the neutrino masses can be either Dirac or Majorana. Some
examples of models of neutrino mass in extra dimensions are given in
Refs.~\cite{dienes}.  However, no evidence of Kaluza-Klein modes, whose
effects are like those of sterile neutrinos, has been found in the
oscillation data.  For a more complete discussion of theories with
large extra dimensions, see Ref.~\cite{mohapatra}.

Alternatively, extra dimensions can be generated dynamically at low
energies from a theory which is four-dimensional and renormalizable at
high energies (a process called dimensional
deconstruction~\cite{cohen}).  Acceptable neutrino phenomenology
appears to be possible in such a scenario~\cite{balaji}.

\subsubsection*{Neutrino anarchy}

Finally, even though it is aesthetically pleasing to think that
symmetries in one form or another account for the structure seen in
the neutrino masses and mixings, it could be that an essentially
random three-neutrino mass matrix can give the appropriate
phenomenology~\cite{hall}. In models with neutrino anarchy, it was
originally thought that large mixing angles are quite natural, and the
value of $\theta_x$ could lie just below the current experimental
bound. However, a recent study suggests that large mixing angles are
not preferred if the mass matrix elements are truly random in a
basis-independent way~\cite{espinosa}. Statistical analyses of
nonrandom structures have also been performed~\cite{antonelli,
altarelli4}.


\section{Leptogenesis}


The range of neutrino masses that the data suggest lends credibility to
the seesaw mechanism. A direct consequence of the seesaw mechanism is
{\it leptogenesis}~\cite{leptogen}.  This process generates a net
lepton asymmetry $Y_L \equiv (n_L- n_{\bar L})/s$ because all of
Sakharov's conditions~\cite{sakharov} are met: (i) the heavy
right-handed neutrinos $N_i$ decay into a lepton-Higgs pair ($l H$)
and into the $CP$ conjugate pair with different partial widths,
thereby violating lepton number; (ii) $CP$ violation results from
phases in the Yukawa couplings and neutrino mass matrices; (iii) the
cosmological expansion yields the departure from thermal equilibrium.

As the universe cools and the $N_i$ drop out
of equilibrium, their decays lead to a $CP$
asymmetry~\cite{sarkar},
\begin{equation}
\epsilon_i={\Gamma(N_i \to l H)- \Gamma(N_i \to \bar{l} H^*) \over 
\Gamma(N_i \to l H)+ \Gamma(N_i \to \bar{l} H^*)}\,.
\end{equation}
The lepton asymmetry generated will be $Y_L \sim \sum \epsilon_i/g_*$,
where $g_*$ is the number of degrees of freedom.

Since sphaleron~\cite{klink} interactions preserve $B-L$ but violate
$B$ and $L$~\cite{thooft}, the lepton asymmetry is partially converted
to a baryon asymmetry $Y_B$. In terms of the initial
$B-L$~\cite{harvey},
\begin{equation}
Y_B= a Y_{B-L} = {a \over a-1} Y_L\,,
\label{eq:b-l}
\end{equation}  
where $a$ depends on the processes in equilibrium. In the seesaw
extended SM (MSSM), $a=28/79$ ($a=8/23$). Note that Eq.~(\ref{eq:b-l})
is valid only for temperatures far above the weak scale (for a review
see Ref.~\cite{buchmuller}).

An interesting aspect of leptogenesis is that it requires the low
energy neutrino masses to be sufficiently light that the baryon
asymmetry is not washed out by neutrino-mediated $L$-violating
scatterings~\cite{fukugita}{\footnote{
If leptons couple to a heavy SU$(2)_L$ triplet in addition to $N_i$, neutrino masses
do not necessarily induce asymmetry washout effects and an upper bound on the
neutrino masses cannot be placed~\cite{hambye}.}}.
This has been further explored in
Refs.~\cite{pludi1,pludi2}. Reference~\cite{pludi2} finds that for the
mass-squared differences relevant to solar and atmospheric neutrino
oscillations, leptogenesis is the unique source of the baryon
asymmetry provided the lightest of the heavy neutrinos is ${\cal{O}}(10^{5})$ 
times lighter than the other
heavy neutrinos.  

Although a very appealing idea, leptogenesis is difficult to test~\cite{hambye2}.
The $\epsilon_i$ can be expressed independently of the neutrino mixing
matrix $V$ of Eq.~(\ref{eq:mixmat})~\cite{branco2}. Any connection
between the $CP$ phase $\delta$ and the $CP$ violation required for
leptogenesis requires assumption about the texture of the Yukawa
matrix, and is therefore model-dependent~\cite{leptomodels}.

The only way to test leptogenesis directly is to constrain the Yukawa
matrix and the masses of right-handed neutrinos. 
This can be achieved by searching for lepton flavor
violating decays and the electric dipole moments of the charged
leptons to which orders of magnitude improved sensitivity 
are expected in the near
future~\cite{meco}: $\mu \to e \gamma$, $\tau \to \mu \gamma$, $\tau
\to e \gamma$ and $\mu \to 3e$.  Even so, any determination of the
Yukawa matrix will depend on how precisely the Higgs sector is
known and on assumptions about the GUT model.

\section{Future long-baseline experiments}

A summary of present knowledge of neutrino parameters is given in
Table~\ref{tab:parameters}, along with the near future projects that
will improve this knowledge. The mixing angle $\theta_x$, the Dirac
$CP$ phase $\delta$, and the sign of $\delta m^2_a$ are as yet
undetermined; their measurement will be the main goal of future
long-baseline neutrino experiments. In this section, we discuss the
the next-generation long-baseline experiments that are being
considered after MINOS, ICARUS, and OPERA.

\begin{table}[ht]
\caption[]{Present knowledge of neutrino parameters and future ways of 
improving this knowledge.}
\label{tab:parameters}
\smallskip
\centering\leavevmode
\begin{tabular}{|c|c|c|}
\hline
3-neutrino& Present knowledge& \\[-1.5ex]
observables& ($\sim$ 95\% C.~L.)& Near future\\[.5ex]
\hline
$\theta_a$& $45^\circ\pm10^\circ$& $P(\nu_\mu\to\nu_\mu)$ MINOS, CNGS\\ \hline
$\theta_s$& $32.5^\circ\pm 3.6^\circ$& SNO NC, KamLAND\\ \hline
$\theta_x$& ${\leq}13^\circ$ (for $|\delta m^2_a|=2.0\times 10^{-3}$ eV$^2$)& $P(\bar\nu_e\to\bar\nu_e)$ 
Reactor, $P(\nu_\mu\to\nu_e)$ LBL\\ \hline
$|\delta m_a^2|$& $(2.0^{+1.2}_{-0.8})\times10^{-3}\rm\,eV^2$& $P(\nu_\mu\to\nu_\mu)$ MINOS, CNGS\\ \hline
sgn$(\delta m_a^2)$& unknown& $P(\nu_\mu\to\nu_e), P(\bar\nu_\mu\to\bar\nu_e)$ 
LBL\\ \hline
$|\delta m_s^2|$& $(7.1^{+1.8}_{-1.1})\times10^{-5}\rm\,eV^2$& $P(\bar\nu_e\to\bar\nu_e)$ KamLAND\\ \hline 
sgn$(\delta m_s^2)$& + (MSW)& done \\ \hline
$\delta$& unknown& $P(\nu_\mu\to\nu_e), P(\bar\nu_\mu\to\bar\nu_e)$ LBL\\ \hline
Majorana& unknown& $0\nu\beta\beta$\\ \hline
$\phi_2$ & unknown& $0\nu\beta\beta$ (if $\simeq0,\pi$)\\ \hline
$\phi_3$& unknown& hopeless\\ \hline
 $m_\nu$ &$\sum m_\nu <1$ eV& LSS, $0\nu\beta\beta$, $\beta$-decay\\ 
\hline
\end{tabular}
\end{table}

\subsection{Conventional neutrino beams and superbeams}

The near term agenda is to confirm atmospheric neutrino oscillations
in accelerator experiments and improve the accuracy with which $|\delta m^2_a|$
and $\sin^22\theta_a$ are determined. Experiments that measure $\nu_\mu$
disappearance will establish the first oscillation minimum in
$P(\nu_\mu\to\nu_\mu)$.  The K2K experiment from KEK to
SuperK~\cite{k2k}, a distance of $L=250$~km, has begun taking data
again following the restoration of the SuperK detector. The MINOS
experiment from Fermilab to the Soudan mine~\cite{minos}, at a
distance of $L=730$~km, will begin in 2005.  It is expected to obtain
10\% precision on $|\delta m^2_a|$ and $\sin^22\theta_a$ in 3 years
running. The CERN to Gran Sasso (CNGS) experiments,
ICARUS~\cite{icarus} and OPERA~\cite{opera}, also at a distance
$L=730$~km but with higher neutrino energy, are
expected to begin in 2007. The appearance of $\nu_\tau$ should be
observed in the CNGS experiments, which would confirm that the primary
oscillation of atmospheric neutrinos is $\nu_\mu \to \nu_\tau$.

The three parameters that are not determined by solar and atmospheric
neutrino experiments are $\theta_x$, sgn$(\delta m_a^2)$, which fixes
the hierarchy of neutrino masses, and the $CP$-violating phase
$\delta$. The appearance of $\nu_e$ in $\nu_\mu \to \nu_e$
oscillations is the most critical measurement, since the probability
is proportional to $\sin^22\theta_x$ in the leading oscillation, for
which there is currently only an upper bound 
(0.2 for $\delta m^2_a=2.0\times 10^{-3}$ eV$^2$ at the 95\% C.~L., from the CHOOZ
reactor experiment~\cite{chooz}). By combining ICARUS/MINOS/OPERA
data, it may be possible to establish whether
$\sin^22\theta_x>0.01$ at 95\% C.~L.~\cite{bgmtw}.

The study of $\nu_\mu \to \nu_e$ oscillations also allows one to test
for $CP$ violation in the lepton sector~\cite{bpw-cp}. 
Intrinsic $CP$ violation in the Standard Model requires
both $\delta \ne 0$ or $\pi$ {\it and} $\theta_x \ne 0$. In vacuum, the
$CP$ asymmetry in the $\nu_\mu \to \nu_e$ channel, to leading order in the
$\delta m^2$'s, is
\begin{equation}
{P(\nu_\mu \to \nu_e) - P(\bar\nu_\mu \to \bar\nu_e) \over
P(\nu_\mu \to \nu_e) + P(\bar\nu_\mu \to \bar\nu_e)} \simeq -
\left({\sin2\theta_s \sin2\theta_a \over 2 \sin^2\theta_a} \right)
\left({\sin2\Delta_s \over \sin2\theta_x}\right) \sin \delta \,.
\label{eq:cpasy}
\end{equation}
For large-angle solar and atmospheric neutrino mixing the first factor
on the right-hand side of Eq.~(\ref{eq:cpasy}) is of order unity. The
existence of $CP$ violation requires that the contribution of the
sub-leading scale, $\Delta_s$, is nonnegligible, so large $L/E_\nu$
values are essential. In practice, the $CP$ conserving and $CP$
violating contributions may have similar size~\cite{shrock}, depending
on the values of $L/E_\nu$ and $\theta_x$. Furthermore,
Earth-matter effects can induce fake $CP$ violation, which must be
folded into any deduction of $\delta$ (on the other hand, matter
effects are essential in determining the sign of $\delta m_a^2$).

The standard proposed method for measuring $CP$ violation is to
compare event rates in two charge conjugate oscillations channels,
such as $\nu_\mu \to \nu_e$ and $\bar\nu_\mu \to \bar\nu_e$. However,
there are three two-fold parameter degeneracies that are present when
two such measurements are made, which may result in an overall
eight-fold degeneracy (a parameter degeneracy occurs when two or more
parameter sets are consistent with the same data):

\begin{enumerate}

\item[(i)] The $(\delta,\theta_x)$ ambiguity~\cite{bmw, b-c, peak,
minakata2, huber}, in which two different parameter pairs, $(\delta,
\theta_x)$ and $(\delta^\prime, \theta_x^\prime)$, lead to the same
values for $P(\nu_\mu \to \nu_e)$ and $P(\bar\nu_\mu \to \bar\nu_e)$.

\item[(ii)] The sgn($\delta m^2_a$) ambiguity~\cite{bmw, peak, lipari,
leading}, where
$(\delta,\theta_x)$ for one sign of $\delta m^2_a$ gives the same
values for the oscillation probabilities as $(\delta^\prime,
\theta_x^\prime)$ with the opposite sign of $\delta m^2_a$.

\item[(iii)] The $(\theta_a, {\pi\over2} - \theta_a)$
ambiguity~\cite{bmw, fogli}, where $(\theta_a, \delta, \theta_x)$
gives the same values for the oscillation probabilities as
$({\pi\over2}-\theta_a, \delta^\prime, \theta_x^\prime)$. This
ambiguity exists because the channel used to determine $\theta_a$,
$\nu_\mu$ survival, only measures $\sin^22\theta_a$. The ambiguity
vanishes at the experimentally preferred value for $\theta_a$
($={\pi\over4}$).

\end{enumerate}

\noindent For each of these degeneracies, a duplicity in inferred values of
$\delta$ and $\theta_x$ is possible; thus each of these degeneracies
can confuse $CP$-violating parameter sets with $CP$-conserving
ones, and {\it vice versa}. In many cases these degeneracies persist for all experimentally
allowed values of $\theta_x$. An overview of these parameter degeneracies
can be found in Refs.~\cite{bmw,minakata4}.

\begin{figure}[h]
\centering\leavevmode
\includegraphics[width=2.1in]{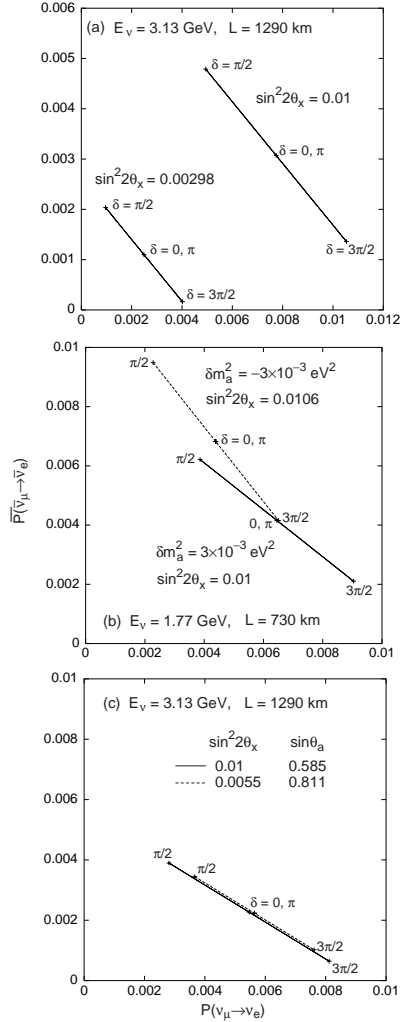}
\caption[]{Remaining degeneracies when $\Delta_a = {\pi\over2}$ for (a)
the ($\delta, \theta_x$) ambiguity, (b) the sgn($\delta m^2_a$) ambiguity,
and (c) the $(\theta_a, {\pi\over2} - \theta_a)$ ambiguity. In (a), each
value of $\sin^22\theta_x$ describes a distinct line in probability space.
In (b), the ambiguity in $\sin^22\theta_x$ is small, but in the overlap
region there is still an ambiguity in sgn($\delta m^2_a$) and a
corresponding large uncertainty in $\delta$. In (c), the ambiguity in
$\delta$ is small, but there may be a large uncertainty in\
$\sin^22\theta_x$ when $\theta_a \ne {\pi\over4}$. In all cases there
remains a $(\delta, \pi - \delta)$ ambiguity since only $\sin\delta$ is
being measured. Adapted from Ref.~\cite{bmw}.
\label{fig:degeneracies}}
\end{figure}

There are two ``magic'' baselines that are valuable to resolve some of these
parameter degeneracies:

\begin{enumerate}

\item[(i)] The detector is located at a distance that corresponds to
the first peak of the leading oscillation ($\Delta_a = {\pi\over2}$):
\begin{equation}
L \simeq 620\ {\rm km} \left( E\over 1\ {\rm GeV}\right)
\left(2.0\times10^{-3}\,{\rm eV}^2 \over
\delta m_a^2\right) \,.
\label{eq:L}
\end{equation}
Then the $\nu_\mu \to \nu_e$ probability depends only on $\sin\delta$
and not $\cos\delta$ (see Eq.~\ref{eq:Pme}), the $(\delta, \theta_x)$
degeneracy is broken, and $\theta_x$ is uniquely determined for a
given sgn($\delta m^2_a$) and $\theta_a$~\cite{bmw,peak}. There is a
residual $(\delta, \pi-\delta)$ degeneracy, but this degeneracy does
not mix $CP$ violating and $CP$ conserving solutions.
Figure~\ref{fig:degeneracies} shows the remaining degeneracies when
$L/E_\nu$ is chosen to be at the first peak of the
oscillation. Furthermore, if $L$ is taken to be very long, large
matter effects will break the sgn($\delta m^2_a$) ambiguity. The
minimum distance needed depends on the size of $\theta_x$ and
$\delta m^2_s$, but generally $L \ge 1000$~km is required.

\item[(ii)] The detector is located at a distance such that $\hat A
\Delta_a = G_F N_e L/\sqrt2 \simeq \pi$, which for the Earth's density
profile implies $L \simeq 7600$~km. Then only the leading oscillation
term survives in Eqs.~(\ref{eq:Pme}, \ref{eq:Pmbeb}, \ref{eq:Pme2}), and
(\ref{eq:Pmbeb2}), and the oscillation probabilities for $\nu_e$
appearance are independent of $\delta$ {\it and} $\delta
m^2_s$~\cite{bmw} . This allows an unambiguous measurement of
$\theta_x$~\cite{magic,magic2} (modulo the $(\theta_a, {\pi\over2} -
\theta_a)$ ambiguity).

\end{enumerate}

\noindent For each of these magic baselines, additional
measurements at different $L$ and/or $E_\nu$ values would be necessary
to break the remaining degeneracies and determine the precise values
of $\delta$ and $\theta_a$.

Precision measurements of $\delta$ and $\theta_x$ can be made at future
off-axis~\cite{bnl-e889} neutrino beam experiments proposed for the
Main Injector at Fermilab (NuMI)~\cite{para,NuMI,barenboim,ayres} and
the Japan Hadron Facility (JHF), also called the Japan Proton
Accelerator Research Complex (J-PARC)~\cite{jhfsk}. 
Compared to on-axis
beams, off-axis beams have a much narrower energy spectrum, smaller
beam contamination, and a suppression of the high-energy tail that
results in lower backgrounds to $\nu_e$ events in the detector.  They
also are well-suited for multiple detectors (a detector
cluster~\cite{bmw2}) that allows more than one measurement to be made
simultaneously.

Strategies for the future include superbeams~\cite{superbeams,
upgradedconv}, with upgrades of the neutrino flux
by a factor of 4 for SuperNuMI (SNuMI)~\cite{NuMIupgrade}, a factor of
5 for SuperJHF (SJHF)~\cite{jhfsk}, and a factor of 5 for a low-energy
option for CNGS~\cite{rubbia}, which will allow smaller values
of $\theta_x$ to be probed. 
A high-intensity neutrino beam at Brookhaven is also
being considered~\cite{wide-band}. Many different superbeam scenarios
have been considered:

\begin{enumerate}

\item[(i)] Having detectors at different off-axis angles allows one to
modify both the baseline {\it and} neutrino energy.  Multiple
detectors utilizing an off-axis beam such as the one at Fermilab can
shift degeneracies to $\sin^22\theta_x \le 0.01 -
0.001$~\cite{bmw2}. Alternatively, having detectors at two different
on-axis distances from the same superbeam can provide multiple
measurements that help to remove parameter degeneracies. One such
possibility is to combine a moderate distance such as JHF to SuperK
($L=295$~km), with a much longer distance such as JHF to a detector
near Beijing ($L\simeq 2100$~km)~\cite{hagiwara}. One specific
proposal is for a SJHF to HyperK
experiment to run 2 years with neutrinos and 5 years with
antineutrinos from, and then run with 5 years with neutrinos from SJHF
to a very large Water Cherenkov detector near Beijing. This very
ambitious scenario would require a separate beamline for the 2100~km
measurement, but could resolve the $(\delta, \theta_x)$ ambiguity down
to $\sin^22\theta_x = 0.005$~\cite{wyy}. Finally, one can vary the
beam energy at the same baseline~\cite{bmw2,hagiwara2}.

\item[(ii)] Two superbeam experiments do significantly better than one
in parameter determinations~\cite{bmw3,synergies}. For example, the
combination of data from a SJHF to SuperK experiment
($\theta_{\mbox{\scriptsize off-axis}} = 2$~deg, $E_\nu \simeq
0.6$~GeV, $L=295$~km, 22.5~kt water Cherenkov) and a SNuMI to
southwestern Ontario experiment (1~deg, $E_\nu = 1.8$~GeV, $L \simeq
900$~km, 20~kt low-Z calorimeter), both with 2 years $\nu_\mu$ running
and 6 years $\bar\nu_\mu$ running, would be sensitive to the sign of
$\delta m_a^2$ and to $CP$ violation for $\sin^22\theta_x
\gsim 0.03$~\cite{bmw3}. Running with only $\nu_\mu$ at SNuMI and SJHF may
allow one to determine sgn($\delta m^2_a$) if $\theta_x$ is not too
small~\cite{minakata3}, although $\theta_x$ and $\delta$ will not be
well-measured without $\bar\nu_\mu$ data.

\item[(iii)] Another approach is to use a wide-band superbeam, for
example from Brook-haven to a National Underground Science
Laboratory~\cite{wide-band}. The measurement of quasi-elastic events allows
a determination of the neutrino energy with reasonable precision. The
lower-energy events are more sensitive to the $\delta$ terms in the
oscillation probability, while the higher-energy events are more
sensitive to the sign of $\delta m^2_a$. Binning the quasi-elastic
events is roughly equivalent to running many narrow-band beams
simultaneously, which can help to resolve neutrino parameter
degeneracies. In principle only a neutrino beam is required.
However, running with an antineutrino beam would provide essential confirmation
(especially of $CP$ violation) and probe lower in $\theta_x$ (if $\delta m^2_a
< 0$).

\end{enumerate}

In any long-baseline neutrino experiment there is a trade-off between
detector size and the ratio of the $\nu_e$ CC signal events to
background: generally speaking, detector technologies that allow a
bigger reduction in background cannot be built as
large~\cite{superbeams}.  Four types of detectors that have been
studied for $\nu_\mu \to \nu_e$ detection are (i) water Cherenkov
(backgrounds of order $10^{-2}$ of the number of unoscillated CC
events, maximum fiducial volume of order 500~kt)~\cite{jhfsk, uno},
(ii) iron scintillator ($3\times10^{-3}$, 50~kt)~\cite{barenboim,
fermi-iron, ino}, (iii) liquid argon ($3\times10^{-3}$,
50~kt)~\cite{ar} and (iv) low-Z calorimetric~\cite{ayres}.
There is also an additional background of order $3\times10^{-3}$ due
to $\nu_e$ contamination in the beam. The larger water Cherenkov
detectors generally do better when the neutrino flux is less (such as
for a conventional beam before superbeam upgrade or for baselines $\ge
4000$~km where there is a large $1/R^2$ fall-off of the flux), whereas
the smaller detectors that can measure $e^\pm$ positions on a finer
scale generally do better when there is more flux (such as with
superbeams or for baselines below 4000~km).

Another option being considered is an intense $\nu_e$ beam (with negligible
contamination) from radioactive ion decay leading to a
so-called beta-beam~\cite{betabeams} 
in a proposed experiment from CERN to Frejus with a baseline of 130~km.

\subsection{Future reactor experiments}

\begin{figure}[ht]
\centering\leavevmode
\includegraphics[width=2.5in,angle=-90]{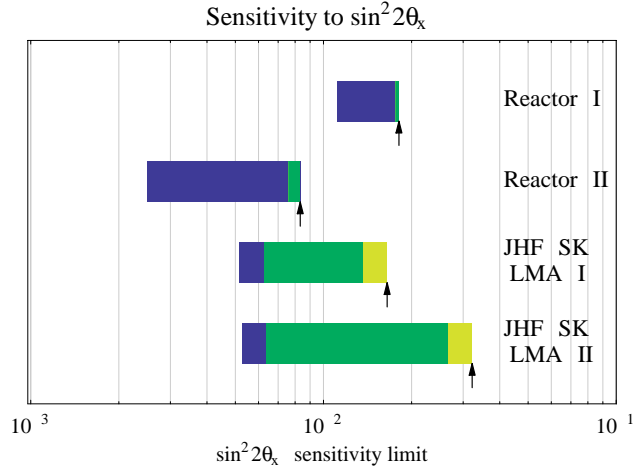}
\caption[]{Sensitivity to $\sin^22\theta_x$ at 90\%~C.L. in reactor
experiments with 400 t-GW-yr (Reactor-I) and 8000~t-GW-yr (Reactor-II),
and for a JHF-SuperK experiment assuming $\delta m^2_s =
7\times10^{-5}$~eV$^2$ (LMA-I) and $\delta m^2_s =
1.4\times10^{-4}$~eV$^2$ (LMA-II). The left edges of the bars show the
sensitivities assuming statistical uncertainties only, while the right
edges of the bars show the sensitivities as the uncertainties from
systematics, parameter correlations, and
parameter degeneracies are progressively included.
  From Ref.~\cite{huber3}.
\label{fig:lindner}}
\end{figure}

A different approach to determining $\theta_x$ without the
complication of parameter degeneracies is to measure $P(\bar\nu_e \to
\bar\nu_e)$ at a reactor experiment using two larger versions of the
CHOOZ detector, one near and one more distant ($\lsim 10$ km) from the
reactor~\cite{kozlov, yasuda, freedman}. Table~\ref{tab:reactors}
lists the detector distances for three proposals. Ignoring terms cubic or
higher in the small parameters $\theta_x$ and $\Delta_s$, the
oscillation probability is given approximately by
\begin{equation}
P(\bar\nu_e \to \bar\nu_e) =
1 - \sin^22\theta_x\sin^2\Delta_a - c_x^4\sin^22\theta_s\sin^2\Delta_s \,.
\label{eq:Pebeb}
\end{equation}
It is independent of $\delta$, $\theta_a$, and sgn($\delta
m^2_a$). Discovery down to
$\sin^22\theta_x = 0.01$ at 90\%~C.~L. may be possible~\cite{kozlov,
yasuda, freedman, huber3} for a reactor experiment with exposure of
about 400~t-GW-yr, where the tonnage refers to the detector size and
GW to the reactor power; an actual measurement of $\sin^22\theta_x$ is
possible for $\sin^22\theta_x \simeq 0.05$. A comparison of the
sensitivity to $\sin^22\theta_x$ in superbeam and reactor experiments
is shown in Fig.~\ref{fig:lindner}. When a reactor measurement is
combined with results from long-baseline experiments, it may also be
possible to determine $\delta$ and $\theta_a$.

\begin{table}[h]
\caption[]{Proposed reactor neutrino experiments for measuring $\theta_x$
using two detectors.}
\label{tab:reactors}
\medskip
\centering\leavevmode
\begin{tabular}{l l l}
\hline
Site & $L_1$ (km) & $L_2$ (km)\\
\hline
Krasnoyarsk~\cite{kozlov} & 0.1 & 1.0\\
Kashiwazaki~\cite{yasuda} & 0.3 & 1.7\\
Diablo Canyon~\cite{freedman} & 0.15 & 1.2\\
\hline
\end{tabular}
\end{table}

\subsection{Neutrino factories}

A neutrino factory (NuFact)~\cite{geer} is the ultimate technology for
neutrino oscillation studies.  Muons would be stored in a flat
oval ring. Their decays will give neutrino beams in the directions
of the straight sections of the storage ring.  Stored muons of
energies 20 GeV and above are needed; energies as high as 50 GeV
have been considered in design studies. A decaying $\mu^+$ in the ring
yields both $\bar\nu_\mu$ and $\nu_e$; detection of the charge of the
final state lepton in a charge-current event allows one to determine
the initial neutrino flavor. Table~\ref{tab:channels} shows the six
oscillation channels possible with stored $\mu^+$; the six
charge-conjugate channels can be tested using $\mu^-$ decays.

\begin{table}[h]
\caption[]{Signals for oscillation channels assuming a decaying $\mu^-$
in a neutrino factory.}
\label{tab:channels}
\medskip
\centering\leavevmode
\begin{tabular}{c c l}
\hline
Channel & Detect & Nomenclature\\
\hline
$\nu_\mu \to \nu_\mu$  & $\mu^-$  & right--sign $\mu$ survival\\
$\nu_\mu \to \nu_e$    & $e^-$    & right--sign $e$ appearance\\
$\nu_\mu \to \nu_\tau$ & $\tau^-$ & right--sign $\tau$ appearance\\
$\bar\nu_e \to \bar\nu_e$     & $e^+$     & wrong--sign $e$ survival\\
$\bar\nu_e \to \bar\nu_\mu$   & $\mu^+$   & wrong--sign $\mu$ appearance\\
$\bar\nu_e \to \bar\nu_\tau$  & $\tau^+$  & wrong--sign $\tau$ appearance\\
\hline
\end{tabular}
\end{table}

The neutrino spectrum from a NuFact ranges from zero to the stored
muon energy, $E_\mu$, with a broad peak at $0.7 E_\mu$ ($0.6 E_\mu$)
for muon (electron) neutrinos when the muons are not polarized. The
neutrino flux in the forward direction is approximately $n_0
\gamma^2/(\pi L^2)$, where $n_0$ is the number of decaying muons in
the straight section of the ring, $\gamma = E_\mu/m_\mu$, and $L$ is
the baseline. An entry-level NuFact produces a time integrated $n_0
\sim 10^{20}$ and a high-performance NuFact has $n_0 \sim
10^{21}$. Early studies of the capabilities of a NuFact can be found
in Refs.~\cite{autin, blondel2}. See also more recent study
group reports in Ref.~\cite{adams} and the
reviews in Ref.~\cite{gomez}.

\subsubsection*{Golden channel: $\nu_e \to \nu_\mu$}

Since the sign of the detected lepton is critical for determining the
oscillation channel being observed, most studies have focused on the
$\nu_e \to \nu_\mu$ and $\bar\nu_e \to \bar\nu_\mu$ oscillation
channels with final state muon detection. Employing a magnetized iron detector
allows one to determine the sign of the detected charged lepton, and
the backgrounds are quite small, of order $3\times10^{-5}$. This small
background compared to $\nu_e$ detection in a superbeam, plus the fact
that NuFact neutrino fluxes can be one or two orders of magnitude
greater than that of a superbeam, are the two main reasons a NuFact is clearly
superior.

\begin{figure}[ht]
\centering\leavevmode
\includegraphics[width=3.6in]{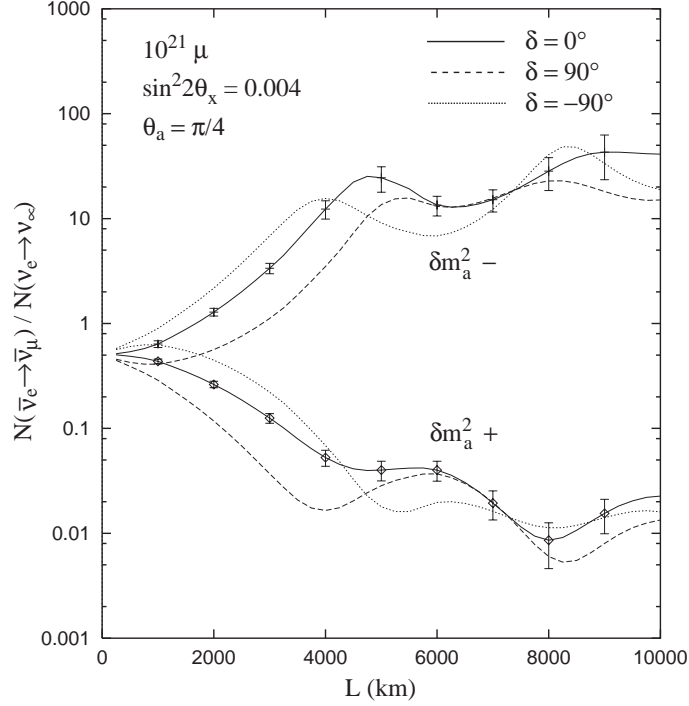}
\caption[]{Ratio of antineutrino to neutrino appearance events versus baseline
in a neutrino factory for $\sin^22\theta_x = 0.004$ and several values
of $\delta$. Both $\delta m^2_a > 0$ and $\delta m^2_a < 0$ cases are shown.
  From Ref.~\cite{entry-level}.
\label{fig:bifurcate}}
\end{figure}

There have been many many studies of the physics capabilities of a
NuFact using muon appearance~\cite{b-c,leading, cervera, entry-level, 
freund5, bueno2} (for a
discussion of the physics that can be done with electron appearance,
see Ref.~\cite{bueno}). By comparing $\nu_e \to \nu_\mu$ and
$\bar\nu_e \to \bar\nu_\mu$ event rates, factoring in Earth-matter
effects, very precise determinations of the oscillation
parameters $\theta_x$, $\delta$, and sgn($\delta
m^2_a$) can be made. Figure~\ref{fig:bifurcate} shows the ratio of antineutrino to
neutrino muon appearance events versus baseline for $\delta m^2_a > 0$
and $\delta m^2_a < 0$ for $\sin^22\theta_x = 0.004$ and several
values of $\delta$. The different signs of $\delta m^2_a$ are clearly
distinguishable when $L \ge 2000$~km, and these measurements are
especially sensitive to the amount of $CP$ violation when $L \sim
3000$~km. In practice, if a NuFact is run with $\mu^-$ about twice as
long as with $\mu^+$, then the total number of CC events in the
neutrino and antineutrino channels will be about the same since
$\bar\nu_e$ cross section is about half that of the $\nu_e$ cross
section.

One obstacle to making precise measurements of $\delta$ and $\theta_x$
and determining the sign of $\delta m^2_a$ is that the other neutrino
mass and mixing parameters, {\it i.e.}, $\theta_a$, $\theta_s$, $\delta
m^2_s$, and $\delta m^2_a$, may not be precisely known. Measurements of
$\nu_\mu$ survival in a superbeam or NuFact will reduce current
uncertainties in $\theta_s$ and $\delta m^2_a$, while future KamLAND and SNO
measurements will reduce the uncertainties in
$\theta_s$ and especially $\delta m^2_s$.

As with superbeams, there is the possibility of having an eight-fold
parameter degeneracy using neutrino and antineutrino event rates in a
NuFact. The sgn($\delta m^2_a$) ambiguity can be
resolved by choosing a baseline $\ge 2000$~km. The
ambiguity is easier to resolve for large $\theta_x$ (due to the larger
matter effect) and small $\delta m^2_s$ (due to the smaller size of
the $CP$ violating term in the oscillation probability). There have
been a number of different proposals for resolving the $(\delta,
\theta_x)$ ambiguity in the golden channel:

\begin{enumerate}

\item[(i)] Since a NuFact has a broad spectrum of neutrino energies,
measuring the energy of the detected muon gives information about the
modulation of the oscillation probability with $E_\nu$. A 10\% muon
energy resolution is sufficient to remove the $(\delta, \theta_x)$
ambiguity~\cite{freund5}. A combined fit with the $\nu_\mu$ survival
channel can also improve the measurement of $\delta$, $\theta_x$, and
sgn($\delta m^2_a$). An understanding of parameter correlations and
degeneracies is essential for extracting meaningful constraints from
the data~\cite{huber,freund5}.

\item[(ii)] Another idea is to combine $\nu_e \to \nu_\mu$ and
$\bar\nu_e \to \bar\nu_\mu$ measurements from neutrino factory
experiments at two baselines. Having one detector at $L \simeq
3000$~km and another at $L \simeq 7300$~km would provide good
discrimination between degenerate solutions for a wide range of
$\delta$ and $\theta_x$~\cite{b-c} (see Fig.~\ref{fig:b-c}).  
Measurements at $7300$~km, near the magic baseline, where the
$\delta$ and $\delta m^2_s$ dependence is minimal~\cite{bmw}, would 
provide an unambiguous measurement of the leading oscillation
amplitude $\sin^2\theta_a \sin^22\theta_x$~\cite{magic}. With an
oval design this scenario would require two separate runs for both
stored $\mu^+$ and $\mu^-$, whereas a triangular configuration could
allow data to be taken simultaneously at two baselines.

\begin{figure}[t]
\centering\leavevmode
\includegraphics[width=3in]{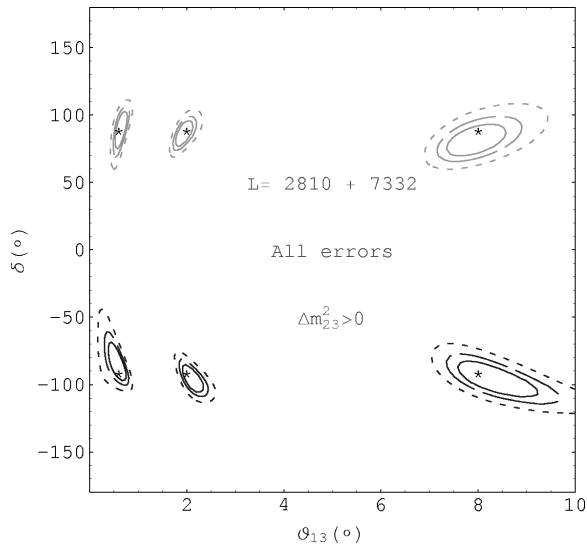}
\caption[]{Fits to $\delta$ and $\theta_x$($\equiv \theta_{13}$)
at a neutrino factory using hypothetical results from two baselines,
2810~km and 7332~km, for several input values of $\delta$ and $\theta_x$
(and $\theta_a=\pi/4$).
The three curves in each case represent the 90\%, 95\%, and 99\%~C.~L.
ranges of allowed parameters. Expected uncertainties in the 
oscillation parameters have been included, in addition to a 1\%
uncertainty in the matter density. From Ref.~\cite{b-c}, in whose notation
$\Delta m^2_{23}>0$ implies $\delta m^2_a>0$.
\label{fig:b-c}}
\end{figure}

\item[(iii)] Since a superbeam will most certainly be a precursor to a
NuFact, it is quite natural to combine data from a superbeam and a
NuFact to help resolve the $(\delta, \theta_x)$ ambiguity. Studies
show that it is possible to remove this ambiguity for $\sin^22\theta_x
\ge 0.0005$~\cite{b-c2}. In this sense, superbeams and neutrino
factories are complementary.

\end{enumerate}

\noindent Some comparisons of superbeam versus NuFact performance are
given in Refs.~\cite{superbeams,huber}. It should be noted that
even if $\theta_x = 0$, subleading terms in the oscillation
probability associated with $\delta m^2_s$ can lead to observable
effects in appearance experiments~\cite{entry-level}.

A summary of typical capabilities of future long-baseline experiments is
given in Table~\ref{tab:limits}. A neutrino factory would have
sensitivity down to $5\times10^{-4}$ and possibly lower in
$\sin^22\theta_x$ for both sgn($\delta m_a^2$) and $CP$-violation
determinations. A 1\% determination of $\delta m_a^2$ should be possible at a 
NuFact. Thus, it appears that precision reconstruction of the
neutrino mixing matrix would be possible with a neutrino factory even
for very small values of $\theta_x$.

\begin{table}[h]
\caption[]{Approximate 3$\sigma$ reaches in $\sin^22\theta_x$ of future
neutrino oscillation experiments.\label{tab:limits}}
\smallskip
\centering\leavevmode
\begin{tabular}{lrrr}
\hline
&\multicolumn{3}{c}{Reach in $\sin^22\theta_x$}\\
Experiment &  Discovery & sgn($\delta m^2_a$) & $CP$ violation\\
\hline
Reactor & $0.01$ & $-$ & $-$ \\
Conventional $\nu$ beam & $0.01$ & $-$ & $-$ \\
superbeam & $0.003$ & $0.003$ & $0.03$ \\
Entry-level NuFact & $0.0005$ & $0.0003$ & $0.002$ \\
High-performance NuFact & $0.00005$ & $0.0001$ & $0.0005$ \\
\hline
\end{tabular}
\end{table}

\subsubsection*{Silver channel: $\nu_e \to \nu_\tau$}

Another advantage of a NuFact over a superbeam is that it has
available the oscillation channel $\nu_e \to \nu_\tau$. The
oscillation probabilities for $\nu_e \to \nu_\tau$ and $\bar\nu_e \to
\bar\nu_\tau$ can be obtained from those for $\nu_e \to \nu_\mu$ and
$\bar\nu_e \to \bar\nu_\tau$ by the transformations $\sin\theta_a
\leftrightarrow \cos\theta_a$ and $\delta \to -\delta$. Comparing the muon and
tau channels in a NuFact therefore allows one to (i) help resolve the
$(\theta_a, {\pi\over2} - \theta_a)$ ambiguity since the leading term
in the oscillation probability is proportional to $\sin^2\theta_a$ for $\nu_e
\to \nu_\mu$ and $\cos^2\theta_a$ for $\nu_e \to \nu_\tau$~\cite{bmw}, and (ii) help
resolve the $(\delta, \theta_x)$ ambiguity since the $\delta$
dependence is different in the two channels~\cite{donini2}.

\subsection{$T$ and $CPT$ symmetries}

If $CPT$ is conserved, then $CP$ violation implies $T$ violation,
which can be measured, {\it e.g.}, by comparing $\nu_e \to \nu_\mu$ to
$\nu_\mu \to \nu_e$. There have been many phenomenological studies of
$T$ violation in the literature~\cite{minakata4, bueno2, arafune}.  
Unlike $CP$ violation, matter does not induce $T$ violation
in a long-baseline neutrino oscillation experiment, due to the
symmetric matter distribution ({\it i.e.}, the matter distribution is the
same from the detector to the source as for the source to the
detector), although matter can modify the amount of $T$ violation.

If $CPT$ is not conserved, then $P(\nu_\alpha \to \nu_\beta)$ is not
necessarily equal to $P(\bar\nu_\beta \to \bar\nu_\alpha)$ in a
vacuum; matter can  induce fake $CPT$ violation 
(for a study of matter-induced $CPT$ violation 
see Ref.~\cite{fakecpt}). In a NuFact, $CPT$
violation can be tested down to very low levels by comparing the
survival channels $\nu_\mu \to \nu_\mu$ and $\bar\nu_\mu \to
\bar\nu_\mu$. There have been a number of studies of possible 
tests of $CPT$ violation in the neutrino sector~\cite{cpt}.


\section{The outlier: LSND \label{sec:LSND}}

\begin{figure}[h]
\centering\leavevmode
\includegraphics[width=2.4in]{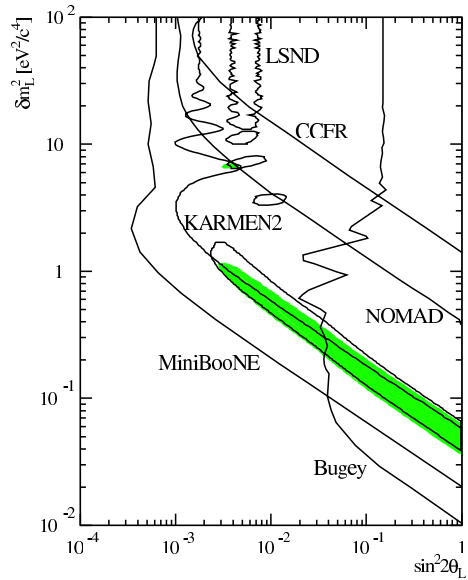}
\caption[]{The 90\% C.~L. allowed region (shaded) from a combined fit to 
$\bar\nu_\mu \to \bar\nu_e$ data from LSND~\cite{lsnd-dar} and KARMEN~\cite{karmen}.
The 90\% C.~L. allowed region from LSND alone is unshaded.
The 90\%~C.~L. exclusion regions from KARMEN,
Bugey ($\bar\nu_e \to \bar\nu_x$)~\cite{bugey}, CCFR ($\nu_\mu \to \nu_e$)~\cite{ccfr} 
and NOMAD ($\nu_\mu \to \nu_e$)~\cite{nomad} and the expected
90\%~C.~L. sensitivity of the MiniBooNE experiment~\cite{miniboone}
are also shown. From Ref.~\cite{church}.
\label{fig:lsnd}}
\end{figure}

The focus of this review so far has been on three-neutrino phenomenology.
However, the results of the LSND experiment may cast some doubt
on the three-neutrino picture. The LSND experiment found evidence for
$\bar\nu_\mu \to \bar\nu_e$ oscillations at $3.3\sigma$ significance
(oscillation probability $(2.64 \pm 0.67 \pm 0.45) \times
10^{-3}$~\cite{lsnd-dar} in data on $\mu^-$ decays at rest taken from
1993-1998. Evidence for $\nu_\mu \to \nu_e$ oscillations was found at
lesser significance from $\pi^+$ decay in flight, with oscillation
probabilities $(2.6 \pm 1.0 \pm 0.5) \times 10^{-3}$ in the 1993-1995
data~\cite{lsnd-dif} and $(1.0 \pm 1.6 \pm 0.4) \times 10^{-3}$ in
the 1996-1998 data (see the last paper of Ref.~\cite{lsnd-dar}). There
was a significant difference in the analysis of the decay in flight
data in the two time periods due to changes in the neutrino production
target, so the two $\nu_\mu \to \nu_e$ samples were not combined.
The KARMEN
experiment~\cite{karmen} also searched for $\bar\nu_\mu \to \bar\nu_e$
oscillations with a null result; the KARMEN data rule out a large
fraction of the LSND allowed region, but still allow a limited region
of oscillation parameters~\cite{church}. The Bugey reactor
experiment~\cite{bugey}, which tests the oscillation channel
$\bar\nu_e \to \bar\nu_e$, excludes the part of the LSND region with
$\sin^22\theta_L \gsim 0.04$.
In a two-neutrino parameter space, the
indicated oscillation parameters from a combination of LSND and KARMEN data that 
are consistent with the constraint from Bugey are
$\delta m^2_L \sim 0.2 - 1$ eV$^2$, $\sin^22\theta_L \sim 0.003- 0.04$ and 
$\delta m^2_L \sim 7$~eV$^2$, $\sin^22\theta_L \sim 0.004$ at the 90\% C.~L. 
(see Fig.~\ref{fig:lsnd}). The bulk of the allowed region is a narrow band
in $(\sin^22\theta_L, \delta m^2_L)$ plane lying along the line
described approximately by $\sin^22\theta_L (\delta m^2_L)^{1.64} =
0.0025$, between $\delta m^2_L = 0.2$ and 1~eV$^2$. The MiniBooNE
experiment~\cite{miniboone} will search for $\nu_\mu \to \nu_e$
oscillations over the entire parameter space allowed by the LSND
results.

\subsection{Four neutrinos?}

The LSND
parameters are very different from the oscillation parameters that
explain the solar and atmospheric neutrino data; in particular,
$|\delta m^2_L| \gg |\delta m^2_a|, |\delta m^2_s|$.
Since a theory with three neutrinos has at most two independent
mass-squared differences, a third $\delta m^2$ scale suggests that
there may be a fourth light neutrino participating in neutrino
oscillations~\cite{caldwell}. Because neutrino counting
experiments indicate that there are only three light active
neutrinos~\cite{pdg}, a fourth light neutrino must be sterile, {\it i.e.},
it does not participate in the weak interactions~\cite{leveille}.

In Section~2, we saw that an extra fully thermalized neutrino is strongly
disfavored by BBN~\cite{nupap}. The only way the LSND sterile 
neutrino can be reconciled
with BBN is if its mixing with the active neutrinos occurs only after
the active neutrinos have decoupled from the $e^{\pm}-\gamma$ plasma. Then,
the LSND neutrino is not thermalized and the effective number of neutrinos remains
equal to 3. An electron neutrino asymmetry,
\begin{equation}
L_e\equiv {n_{\nu_{e}}-n_{\bar\nu_{e}} \over n_\gamma} \approx 0.01 - 0.1\,,
\end{equation}
accomplishes this~\cite{volkas} and simultaneously improves 
the agreement between the BBN prediction 
for the primordial $^4$He abundance and the observationally inferred 
value~\cite{bari}. 
Also note that since the production of the LSND neutrino is suppressed, 
cosmological bounds on
$\sum m_\nu$ pertain only to the active neutrinos. A 
confirmation of the LSND signal by MiniBooNE
could be interpreted as a hint for a large neutrino asymmetry in the universe.

\begin{figure}[ht]
\centering\leavevmode
\includegraphics[width=4in]{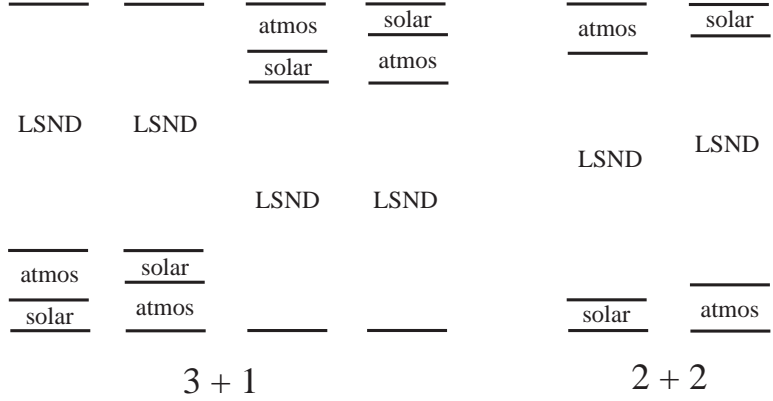}
\caption[]{The six possible mass spectra in four-neutrino models.
\label{fig:four-nu}}
\end{figure}

There are two types of mass spectra possible in four-neutrino
models~\cite{okada,bpww}. In 3~+~1 models, one mass eigenstate
is separated from a nearly degenerate triplet of mass eigenstates by
$\delta m^2_L$; the triplet has a mass ordering like that of a
three-neutrino model. The well-separated mass eigenstate can be either
lighter or heavier than the other three, and the triplet can have a
normal or inverted hierarchy, so there are four possible variations of
3~+~1 models. In 2~+~2 models there is one pair of
closely-spaced mass eigenstates separated from another closely-spaced
pair by $\delta m^2_L$; one pair has a mass-squared difference of
$\delta m^2_s$ and the other $\delta m^2_a$, and the solar $\delta
m^2$ can be in either the upper or lower
pair. Figure~\ref{fig:four-nu} shows the six possible mass spectra
with four neutrinos.

\subsection{Four neutrino models}

The 3 + 1 models are a straightforward extension of a three-neutrino
model: the three active neutrinos have mass-squared differences and
mixings similar to those in a three-neutrino model, and the sterile
neutrino state has only small mixing with active neutrinos. However,
3~+~1 models have trouble accounting for the LSND results and
simultaneously obeying the constraints of earlier accelerator and
reactor experiments~\cite{okada,bpww}. This can be
demonstrated as follows.

Assume a neutrino mass spectrum such that the nearly degenerate
triplet of mass eigenstate is lighter than the remaining state and
exhibits a normal hierarchy (the first spectrum shown in
Fig.~\ref{fig:four-nu}); similar conclusions can be drawn for the other
three 3~+~1 spectra. Then $\delta m^2_{43} \simeq \delta m^2_{42}
\simeq \delta m^2_{41} = \delta m^2_L \gg \delta m^2_{32} \simeq
\delta m^2_{31} = \delta m^2_a \gg \delta m^2_{21} = \delta m^2_s$,
and the oscillation probabilities for the leading oscillation, due to
$\delta m^2_L$, are
\begin{eqnarray}
P(\bar\nu_\mu \to \bar\nu_e) &\simeq&
4 |V_{\mu4}|^2 |V_{e4}|^2 \sin^2\Delta_L \,,
\label{eq:Pme4L} \\
P(\nu_\mu \to \nu_\mu) &\simeq&
1 - 4 |V_{\mu4}|^2 (1 - |V_{\mu4}|^2) \sin^2\Delta_L \,,
\label{eq:Pmm4L} \\
P(\bar\nu_e \to \bar\nu_e) &\simeq&
1 - 4 |V_{e4}|^2 (1 - |V_{e4}|^2) \sin^2\Delta_L \,,
\label{eq:Pee4L}
\end{eqnarray}
where $\Delta_L \equiv \delta m^2_L L/(4E_\nu)$, analogous to
Eq.~(\ref{eq:delta}). At $L/E_\nu$ values appropriate for atmospheric
neutrinos,
\begin{equation}
P(\nu_\mu \to \nu_\mu) \simeq
1 - 4 |V_{\mu3}|^2 (1 - |V_{\mu3}|^2 - |V_{\mu4}|^2) \sin^2\Delta_a \,,
\label{eq:Pmm4a}
\end{equation}
and for solar neutrinos
\begin{equation}
P(\nu_e \to \nu_e) \simeq
1 - 4 |V_{e1}|^2 |V_{e2}|^2 \sin^2\Delta_s \,.
\label{eq:Pee4s}
\end{equation}

There are very stringent limits on $\nu_\mu$ disappearance from the 
CCFR~\cite{ccfr}, NOMAD~\cite{nomad} and
CDHS~\cite{cdhs} accelerator experiments that
constrain $|V_{\mu4}|$ to be very small or very close to unity via
Eq.~(\ref{eq:Pmm4L}). However, if $|V_{\mu4}|$ is close to unity then
the amplitude of atmospheric neutrino oscillations cannot be as large
as required by observation (see Eq.~\ref{eq:Pmm4a}); therefore,
$|V_{\mu4}|^2 \ll 1$. Similarly, the Bugey reactor
experiment~\cite{bugey} puts severe limits on $\bar\nu_e$
disappearance that constrain $|V_{e4}|$ to be very small or very close
to unity, and the observation of large angle mixing in solar neutrino
oscillations therefore implies $|V_{e4}|$ cannot be close to unity
(see Eq.~\ref{eq:Pee4s}), so $|V_{e4}|^2 \ll 1$.

\begin{figure}[ht]
\centering\leavevmode
\includegraphics[width=3.75in]{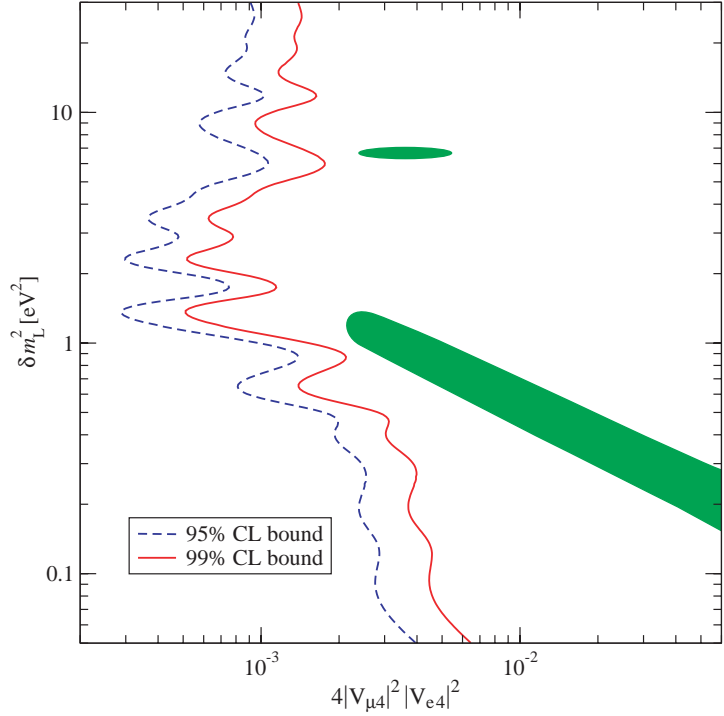}
\caption[]{Upper bounds on the LSND amplitude in 3~+~1 models 
($ 4|V_{\mu4}|^2|V_{e4}|^2$) 
from accelerator and reactor data~\cite{schwetz}. The shaded 
regions are allowed by LSND and KARMEN at the 95\%~C.~L.~\cite{church}.
\label{fig:3+1}}
\end{figure}

Since the oscillation amplitude for the LSND experiment is $4
|V_{\mu4}|^2 |V_{e4}|^2$, it has an upper limit of approximately
one-fourth of the product of the CDHS and Bugey oscillation amplitude
bounds (see Eqs.~\ref{eq:Pme4L}-\ref{eq:Pee4L}). In practice, the
limits on $|V_{\mu4}|$ and $|V_{e4}|$ depend on $\delta m^2_L$, as
does the allowed oscillation amplitude from LSND, so a comparison must
be made for each value of $\delta m^2_L$. Early analyses~\cite{okada,bpww} 
concluded that the 3~+~1 model could not consistently
explain the LSND, accelerator, and reactor data. Upon the release of
the final data analysis by the LSND collaboration (the last paper of
Ref.~\cite{lsnd-dar}, in which the central value of the average
oscillation probability decreased from 0.0031 to 0.0026), it was
thought that perhaps the 3~+~1 models would be revived~\cite{bklww,peres}. 
Subsequently, a Bayesian analysis of all relevant
data was made~\cite{schwetz}. 
Figure~\ref{fig:3+1} shows the incompatibility of the combined
accelerator and reactor upper limit on the oscillation 
amplitude in 3~+~1 models  
and the region allowed by LSND and KARMEN. 
An analysis of all relevant data yielded the best overall goodness of fit to be
$5.6\times10^{-3}$~\cite{tortola}. Thus, the viability of 3~+~1 models
is tenuous.

The 2~+~2 models are not a simple extension of a three-neutrino model,
since the removal of the sterile neutrino does not leave the standard
three-neutrino mass spectrum. If the lower pair of neutrino mass
eigenstates are primarily responsible for solar neutrino oscillations,
then we have $\delta m^2_{42} \simeq \delta m^2_{32} \simeq \delta
m^2_{31} \simeq \delta m^2_{41} = \delta m^2_L \gg \delta m^2_{43} =
\delta m^2_a \gg \delta m^2_{21} = \delta m^2_s$ (the case where the
lower pair of mass eigenstates are primarily responsible for
atmospheric neutrino oscillations leads to similar conclusions).  Then
the oscillation probabilities for the leading oscillation are
\begin{eqnarray}
P(\bar\nu_\mu \to \bar\nu_e) &\simeq&
4 |V_{\mu3}V_{e3}^*+V_{\mu4}V_{e4}^*|^2 \sin^2\Delta_L \,,
\label{eq:Pme4L2} \\
P(\nu_\mu \to \nu_\mu) &\simeq&
1 - 4 (|V_{\mu3}|^2+|V_{\mu4}|^2) (1 - |V_{\mu3}|^2 - |V_{\mu4}|^2) \sin^2\Delta_L \,,
\label{eq:Pmm4L2} \\
P(\bar\nu_e \to \bar\nu_e) &\simeq&
1 - 4 (|V_{e3}|^2+|V_{e4}|^2) (1 - |V_{e3}|^2 - |V_{e4}|^2) \sin^2\Delta_L \,,
\label{eq:Pee4L2}
\end{eqnarray}
At an $L/E_\nu$ appropriate for atmospheric neutrinos,
\begin{equation}
P(\nu_\mu \to \nu_\mu) \simeq
1 - 4 |V_{\mu3}|^2 |V_{\mu4}|^2 \sin^2\Delta_a \,,
\label{eq:Pmm4a2}
\end{equation}
and the oscillation probability for solar neutrinos in the 2~+~2 case
is the same as in the 3~+~1 case (Eq.~\ref{eq:Pee4s}).

If $\nu_e$ is primarily connected to $\nu_1$ and $\nu_2$, and
$\nu_\mu$ is primarily connected to $\nu_3$ and $\nu_4$, {\it i.e.},
$|V_{e1}|^2 + |V_{e2}|^2, |V_{\mu3}|^2 + |V_{\mu4}|^2 \simeq 1 \gg
|V_{e3}|^2, |V_{e4}|^2, |V_{\mu1}|^2, |V_{\mu2}|^2$, then it is
possible to simultaneously fit the solar, atmospheric, and LSND data
in 2~+~2 models~\cite{okada,bpww}. For example, if
$|V_{\mu3}|^2 + |V_{\mu4}|^2 = 1$ and $|V_{e1}|^2 + |V_{e2}|^2 \simeq
1$, then it is easy to have large mixings of solar and atmospheric
neutrinos while suppressing oscillations in the sensitivity regions of
CDHS and Bugey.  Since $P(\bar\nu_\mu \to \bar\nu_e)$ and $P(\bar\nu_e
\to \bar\nu_e)$ are both proportional to the second power of small
parameters ($V_{e3}$ and $V_{e4}$), the LSND oscillation amplitude is
approximately constrained by the Bugey bound in the 2~+~2 case (in
contrast to one-fourth of the product of the CDHS and Bugey bounds in
the 3~+~1 case). This less severe constraint is readily satisfied for a
wide range of $\delta m^2_L$; in fact, if either $V_{e3}$ or $V_{e4}$
is zero, then the constraint exactly reduces to the simple
two-neutrino Bugey constraint shown in Fig.~\ref{fig:lsnd}. The
situation with the roles of $\nu_e$ and $\nu_\mu$ reversed, {\it i.e.},
$\nu_e$ primarily connected to $\nu_3$ and $\nu_4$, and $\nu_\mu$
primarily connected to $\nu_1$ and $\nu_2$, gives similar results.
Some examples of explicit 2~+~2 models are given in Refs.~\cite{bpww,
bww-four}.

The preceding arguments relied upon the assumption that large mixing
solutions for solar or atmospheric neutrinos are satisfactory for
oscillations to sterile neutrinos. In 2~+~2 scenarios, solar neutrinos
oscillate to a linear combination of $\nu_\tau$ and $\nu_s$, and
atmospheric neutrinos oscillate to the orthogonal combination~\cite{bpww}:
\begin{eqnarray}
\nu_e &\to& -\sin\alpha\,\nu_\tau + \cos\alpha\,\nu_s \,,
\label{eq:nue}\\
\nu_\mu &\to& \cos\alpha\,\nu_\tau + \sin\alpha\,\nu_s \,.
\label{eq:num}
\end{eqnarray}
The oscillation probabilities in SuperK and SNO are~\cite{unknowns}
\begin{eqnarray}
R_{SK} &=& \beta P + r \beta \sin^2\alpha (1 - P) \,,
\label{eq:rsk}\\
R_{SNO}^{CC} &=& \beta P \,,
\label{eq:rsnocc}\\ 
R_{SNO}^{NC} &=& \beta P + \beta \sin^2\alpha (1 - P) \,,
\label{eq:rsnonc}
\end{eqnarray}
where $R$ is the ratio of observed to expected rates in a given
experiment, $\beta$ is the ratio of the actual $^8$B neutrino flux to
the SSM prediction, $P$ is the average oscillation probability for
$^8$B neutrinos, and $r = \sigma_{\nu_\mu,\nu_\tau}/\sigma_{\nu_e}
\simeq 1/6.48$ is the ratio of $\nu_\mu,\nu_\tau$ to $\nu_e$ elastic
scattering cross sections on electrons in the energy range of the
SuperK experiment.\footnote{SNO can also measure $\nu e$ scattering,
which is equivalent to $R_{SK}$.}  The ratio of the SNO NC to CC data
is then
\begin{equation}
{R_{SNO}^{NC} \over R_{SNO}^{CC}} =
1 + \sin^2\alpha \left( {1\over P} - 1 \right) \,.
\end{equation}
If the $^8$B neutrino flux is assumed to be 
known from the SSM ({\it i.e.}, $\beta
\equiv 1$), then $P$ is determined from SNO CC data and the sterile
fraction can be deduced from the SNO NC/CC measurement. However, if
$\beta$ is allowed to be free, $\sin^2\alpha$ cannot be determined
since $R_{SNO}^{CC}$ measures only the combination $\beta P$ and not
$P$ itself~\cite{unknowns}.

Early analyses of the solar neutrino data showed that solar solutions
with pure $\nu_e \to \nu_s$ ($\alpha=0$) did not provide as good a fit
to the solar neutrino data (see the second paper of Ref.~\cite{hata}). 
The main difficulty with pure
sterile neutrino solutions is that they give similar
predictions for the Chlorine and SuperK experiments, whereas active
neutrino solutions give a larger value for SuperK due to
neutral current $\nu_\mu e$ and/or $\nu_\tau e$ interactions in the
detector, which is in better agreement with the experimental
data. Oscillations to sterile neutrinos have different matter
effects since the value of $\delta m^2_s$ that gives resonant
oscillations is $\delta m^2_s = 2\sqrt2 G_F E_\nu (N_e -
{1\over2}N_n)/\cos2\theta_s$ instead of the value $2\sqrt2 G_F E_\nu
N_e/\cos2\theta_s$ for oscillations to active neutrinos (see
Eq.~\ref{eq:sterile}).

Imposing the SSM uncertainties on $\beta$,
 the pure sterile solution is excluded at the $7.6\sigma$ level, although a
large sterile fraction could still be allowed
even after the SNO CC and NC data is included~\cite{ourfit,unknowns,
bahcall-concha2}. The SuperK data (or, equivalently, the SNO $\nu e$
scattering data) do not give an independent constraint; in fact,
there is a sum rule~\cite{unknowns,piecing}
\begin{equation}
R_{SNO}^{NC} = [R_{SK} - (1-r)R_{SNO}^{CC}]/r \,,
\label{eq:sumrule}
\end{equation}
that must be obeyed for any value of $\sin^2\alpha$. A future test of
$\sin^2\alpha$ that is independent of the SSM $^8$B neutrino
flux predictions would be a neutrino-nucleon NC measurement of
intermediate energy solar neutrinos~\cite{unknowns}, or the independent
measurement of $P$, such as in the KamLAND reactor
experiment~\cite{bahcall-concha2}.

The opposite extreme is to have pure sterile solutions to
the atmospheric neutrino data ($\alpha = {\pi\over2}$).  Here there
are strong matter effects due to coherent forward scattering in the
Earth that is present for $\nu_\mu$ but not $\nu_s$ (see
Eq.~\ref{eq:sterile}), with $N_{eff} = -{1\over2}N_n$; for pure
$\nu_\mu \to \nu_\tau$ oscillations, matter effects are small. The
SuperK atmospheric data rule out pure $\nu_\mu \to \nu_s$
oscillations at the 99\% C.~L.~\cite{toshito,superK-tau}, and require
$\sin^2 \alpha$ to be smaller than 0.2 at the 90\% C.~L.~\cite{hayato,nakaya}. 
Other analyses also
found that a substantial sterile component is allowed by
the atmospheric neutrino data~\cite{yasuda-atmos}.

Since pure $\nu_e \to \nu_s$ solar and pure $\nu_\mu \to \nu_s$
atmospheric neutrino oscillations are ruled out, but partial sterile
solutions are allowed in each case, the only remaining option in the
2~+~2 scenario is to have an active-sterile mixture for both solar and
atmospheric neutrinos ($0 < \alpha < {\pi\over2}$). 
The analysis of Ref.~\cite{tortola} 
indicates $\sin^2\alpha \ge 0.55$ from pre-SNO salt phase solar neutrino 
data and
$\sin^2\alpha \le 0.35$ from atmospheric neutrino data (both at
99\%~C.~L.), and that the overall goodness of fit of 2~+~2 models
is only $1.6\times10^{-6}$, much worse than that of the
3~+~1 models.

To summarize the situation for four neutrinos, the most recent
analysis~\cite{tortola} shows that 3~+~1 scenarios provide a
better fit than 2~+~2 scenarios, but neither scenario gives a good description
of the combined solar, atmospheric, accelerator, and reactor
data. However, the inclusion of additional sterile neutrinos can
enhance the size of the LSND amplitude allowed by CDHS and Bugey in
the 3~+~1 scenario~\cite{peres}, and a recent study found that a
3~+~2 model does significantly better than the 3~+~1 models in fitting
the data~\cite{sorel}. Also, studies using a more complete set of
four-neutrino parameters indicate there may still be room for the 2~+~2
models~\cite{pas-song-weiler}. Therefore, although the
positive appearance result in LSND remains puzzling, a consistent
four-neutrino explanation may still be possible.

\subsection{Three-neutrino models with $CPT$ violation}

It has been suggested that if $CPT$ were not conserved, then
oscillations of three active neutrinos could describe the solar,
atmospheric {\it and} LSND data simultaneously~\cite{murayama}. 
In this proposal, the mass matrices (and hence
mass-squared differences) for neutrinos and antineutrinos are
different, which violates $CPT$ (while preserving Lorentz invariance)
\footnote{Whether such a model can be constructed using nonlocality
of the interactions is still a matter of debate~\cite{greenberg}.}.

In the original versions of $CPT$-violating models, the neutrino
sector had the usual three-neutrino mass spectrum that can account for
the oscillation of solar and atmospheric neutrinos, while in the
antineutrino sector the mass-squared differences account for the
oscillation of antineutrinos in the atmospheric and LSND experiments
(the weak indication for $\nu_\mu \to \nu_e$ oscillations in LSND must
be ignored). KamLAND data, consistent with oscillations of $\bar\nu_e$
at the $\delta m^2_s$ scale, forced a modification of the antineutrino
spectrum, so that it describes the oscillation of antineutrinos in
LSND and KamLAND (but not in the atmosphere)~\cite{borissov2}. Since
the atmospheric data does not distinguish between neutrinos and
antineutrinos, this latter scenario was not in obvious contradiction
to the data. However, one analysis of atmospheric data indicates that
$CPT$-violating scenarios are not in good agreement with the
atmospheric data~\cite{strumia}.  Furthermore, global analyses of all
data including KamLAND excludes these $CPT$-violating scenarios at the
3$\sigma$ level~\cite{strumia, cptreject}. 

We await the MiniBooNE results on $\nu_\mu \to \nu_e$ oscillations
to confirm or reject the LSND effect. A positive signal will rule out
current $CPT$-violating models, and force a reconsideration of the
disfavored four-neutrino models, possibly by extending them to
include more than one sterile state. A negative result will rule out
the standard four-neutrino scenario (which predicts the same
oscillations for neutrinos and antineutrinos in LSND), but does not
completely extinguish the LSND flame since it does not test the
primary LSND $\bar\nu_\mu \to \bar\nu_e$ channel. In either case, it will be 
difficult to exclude the very speculative possibility of the LSND anomaly
arising from $CPT$-violation in four-neutrino models~\cite{cptus}.

\section{Summary and outlook}

Great advances have been made over the past five years in our
understanding of the neutrino sector. The field is now poised for
further breakthroughs. In this article we have strived to summarize
the present state of the field and the ongoing experimental,
phenomenological, and theoretical efforts towards a bottom-up
reconstruction of the fundamental properties and theory of neutrinos.
In this concluding section we very briefly recapitulate the highlights of
these recent accomplishments and the major planned directions for
future progress.

The standard formalism of three-neutrino mixing and oscillation
probabilities was recounted. The modifications from $\nu_e$ scattering
in matter were quantified as appropriate for solar neutrinos and for
long-baseline accelerator neutrino experiments. The experimental
resolution of the solar neutrino problem, particularly by the SNO
experiment, as the large mixing angle oscillation solution, was
discussed along with its validation by the KamLAND reactor and other
neutrino experiments. The evidence for $\nu_\mu\to\nu_\tau$
oscillations, from the dependence on the $\nu_\mu \to \nu_\mu$
oscillation probability versus zenith angle in the Super-Kamiokande
atmospheric neutrino experiment, was summarized.

In the framework of three-neutrino mixing, the major physics goal is
to determine the six neutrino oscillation parameters: two
mass-squared differences ($\delta m_s^2,\ \delta m_a^2$), 
three mixing angles ($\theta_s,\ \theta_a,\ \theta_x$), and a $CP$ phase
$\delta$. From present data the approximate values are $\delta m_s^2
\approx 7 \times 10^{-5}\rm\,eV^2$, $\left|\delta m_a^2\right| \approx
2\times 10^{-3}\rm\,eV^2$, $\theta_s \approx 33^\circ$, $\theta_a
\approx 45^\circ$. The unknown parameter $\theta_x$ (with present
limit $\theta_x \leq 13^\circ$ for $\delta m^2_a=2.0\times 10^{-3}$ eV$^2$ at the 95\% C.~L.) is critical to further understanding; it
can be probed in $\nu_\mu \to \nu_e$ or $\nu_e \to\nu_\mu$ appearance
channels. Both off-axis and broad-band beams are being considered for
its measurement. More intense neutrino beams (superbeams or neutrino
factories) and longer baselines ($\gtrsim 1000$~km) are essential to
determine the sign of $\delta m_a^2$ from matter effects and have
sensitivity to $CP$-violation. Expected sensitivities of proposed future
oscillation experiments to $\theta_x$, sgn$(\delta m_a^2)$, and
$\delta$ are compared; see Table~\ref{tab:limits}. 
Reactor experiments at baselines $\lesssim
10$~km could also determine $\theta_x$ if $\sin^22\theta_x \gtrsim 0.05$, 
but only at about the 90\% C.~L.

The evidence for $\bar\nu_\mu \to \bar\nu_e$ oscillations at a
higher $\delta m^2$ ($\approx 0.1 - 1$~eV$^2$) may imply the existence of
 a sterile
neutrino if $CPT$ is conserved. Global fits to data now reject this
possibility at more than the 99\%~C.~L. The ongoing MiniBooNE experiment
is designed to confirm or exclude the LSND effect. Big Bang Nucleosynthesis,
combined with cosmic microwave background measurements by WMAP,
strongly disfavors more than three thermalized neutrinos, but a large
$\nu_e$-asymmetry allows an escape from this constraint.

The search for the absolute scale of neutrino mass in beta decay,
neutrinoless double-beta decay, and large scale structure in the
universe was discussed. The best present constraint comes from
cosmology and bounds the sum of neutrino masses to be less than about
1~eV.

Galactic supernova explosions produce high fluxes of neutrinos and
antineutrinos that can determine the neutrino mass hierarchy (normal
or inverted) if the mixing angle $\theta_x$ is known from accelerator
experiments.

Models of neutrino mass are largely of two generic types, Grand
Unification with a seesaw mechanism and radiative mass
generation. Unified models can accommodate all present knowledge of
quark and lepton masses and mixing, with different models giving
different predictions for the unknown neutrino parameters. Successful
Grand Unified models predict $\delta m_a^2>0$, for which
there would be one heavier neutrino mass and two lighter masses. Also
in unified models, the neutrino masses are hierarchical: $m_3 \gg m_2
\gg m_1$. Radiative mass generation involves lepton number violating
interactions that  have testable implications for collider physics
experiments.

Unified models offer a possible explanation of the baryon asymmetry of
the universe in terms of a neutrino-antineutrino asymmetry. In such
models the $CP$-violating phase in heavy Majorana neutrino decays that
generate leptogenesis may be related to the $CP$-violating phase of
neutrino oscillations at low energy.

The ways by which the unknown neutrino oscillation parameters can be
determined by an ambitious future experimental program with baselines
at long distances from intense sources was detailed. Conventional
neutrino beam experiments are now under construction. Later these may
be upgraded with more intense superbeams. Eventually neutrino
factories will provide the ultimate sensitivity and precision in
determining neutrino mixing and mass-squared difference parameters.

\section*{Acknowledgments}

We thank E. Kearns for updated  (preliminary) results and figures
from the Super-Kamiokande experiment. This research was supported
in part by the U.S. Department of Energy under Grants
No.~DE-FG02-95ER40896, DE-FG02-01ER41155, and DE-FG02-91ER40676, by
the NSF under Grant No.~PHY99-07949, and in part by the Wisconsin
Alumni Research Foundation. We also thank the Kavli Institute for
Theoretical Physics at the University of California, Santa Barbara for its support and
hospitality while this work was in its initial stages. VB thanks the Aspen
Center for Physics for its hospitality.


\end{document}